# A review on the advancements in the characterization of the high-pressure properties of iodates

Akun Liang, Robin Turnbull, Daniel Errandonea*

*Departamento de Física Aplicada-ICMUV-MALTA Consolider Team, Universitat de València, c/Dr. Moliner 50, Burjassot (Valencia) 46100, Spain.*

* Corresponding author.

E-mail address: daniel.errandonea@uv.es

# Abstract

The goal of this work is to report a systematic and balanced review of the progress made in recent years on the high-pressure behavior of iodates, a group of materials with multiple technological applications and peculiar behaviors under external compression. This review article presents results obtained from multiple characterization techniques which include: X-ray diffraction, Raman and infrared spectroscopy, optical-absorption, resistivity, and second-harmonic generation measurements. The discussion of the results from experiments will be combined with density-functional theory calculations which have been shown to be a very useful tool for the interpretation of experimental data. Throughout the manuscript many of the phenomena observed will be connected to the presence of a lone electron pairs of the iodine atoms in the studied iodates. The presence of the lone electron pairs plays a crucial role in the high-pressure behavior of iodates and it is associated with many of the phenomena discussed here, in particular with the pressure-induced changes in the character of iodine-oxygen bonds, which causes many physical properties to behave nonlinearly. Towards the end of this review, a discussion of current problems that remain unsolved is presented as well as proposals for possible avenues for future studies.

# Contents

# Glossary of abbreviations

**BM EOS** – Birch-Murnaghan equation of state

**CBM** – Conduction band minimum

**COHP** – Crystal orbital Hamilton populations

**COOP** – Crystal orbital overlap population

**DAC** – Diamond-anvil cell

**DFG** – Different frequency generation

**DFT** – Density-functional theory

**DOS** – Density of states

**FWHM** – Full-width at half maximum

**GGA** – Generalized gradient approximation

**HP** – High-pressure

**HP-HT** – High-pressure high-temperature

**ICSD** – Inorganic Crystal Structure Database

**LEP** – Lone electron pair

**LP** – Low-pressure

**MO** – Molecular orbital

**NLO** – Nonlinear optical

**P-T** – Pressure-temperature

**PBE** – Perdew-Burke-Ernzerhof

**PDOS** – Projected density of states

**RT** – Room temperature

**SFG** – Sum frequency generation

**SHG** – Second-harmonic generation

**VBM** – Valence band maximum

**XRD** – X-ray diffraction

# 1. Introduction

Iodine is an element with large electronegativity which, in its pentavalent state, forms multiple stable iodate compounds. There are at least 450 entries for pentavalent iodine iodates in the Inorganic Crystal Structure Database (ICSD). The list of iodates contains compounds with different stoichiometries, including: $AgIO_3$, $KIO_3$, $LiIO_3$, $Mg(IO_3)_2$, $Zn(IO_3)_2$, $Fe(IO_3)_3$, etc. Triangular pyramids of $[IO_3]^-$, where iodine has a lone electron pairs (LEP), is a common building block of the crystal structure of these iodates. Many of the crystal structures are non-centrosymmetric, resulting in materials with nonlinear optical (NLO) properties, for instance second-harmonic generation (SHG) [1,2]. Consequently, metal iodate compounds were the focus of multiple studies in the 1970s. The main goal of those studies was to investigate their NLO, ferroelectric, and piezoelectric properties [3–5]. At the beginning of the 21st century there was a resurgence of the interest in metal iodates. Great progress was achieved in the development of infrared applications and in the improvement of synthesis and crystal growth techniques [6–13].

Metal iodates are also very interesting for fundamental research. This is because of the iodine LEP which favors the formation of halogen bonds, an electrostatic interaction between the LEP and a nearby nucleophilic region [14]. Therefore, in metal iodates, in addition to the three short I-O covalent bonds forming the $IO_3$ pyramid (~1.8 Å) there are three additional weak I···O halogen bonds with bond distances ~ 2.5 – 2.8 Å. These bonds are not formal chemical bonds, but rather a strong electrostatic attraction, however, they play an important role in stabilizing iodate compounds [15]. The presence of a LEP and the existence of halogen bonds make metals iodates very attractive to be studied under high-pressure (HP) conditions [16]. It is very well-known that HP changes interatomic bonds, inducing changes in the physical and chemical properties of materials [17]. The existence of the LEP and the halogen bonds also makes metal iodates highly sensitive to external pressure, therefore their properties can be easily tuned under a compression much smaller than that needed to modify most oxides.

Nowadays experimental pressures of several hundreds of gigapascals (GPa) can be achieved in laboratories using diamond-anvil cells (DAC), with some reports even reaching thousands of GPa [18]. Under such conditions of compression inter-atomic distances can usually be shortened up to an order of magnitude, for example in rare gas solids. Consequently, electronic, magnetic, vibrational, and elastic properties can be dramatically altered, existing bonds can be broken, new bonds can be formed, or phase transitions can be triggered. As a result, diverse and interesting phenomena have been found in HP studies in the last decades. The list of pressure-driven phenomena includes: pressure-driven phase transitions [19]; metallization [20]; amorphization [21]; superconductivity [22]; formation of new compounds [23], and so forth. Thus, HP studies provide unique opportunities to control the crystal structure and emergent physicochemical properties of materials without altering their chemical composition. It is also a useful tool for crystal engineering, allowing the development of novel materials with interesting properties, which could be potentially valuable for technological developments.

At the dawn of the twentieth century, Percy Bridgman created experimental devices that radically extended the HP range covered by experiments. His innovative work transformed HP research [24]. The next great advance in the field of HP research was the development of the DAC at the end of the 1950s [25]. Since then, the DAC has become a prevalent HP device. Pressures greater than 600 GPa can be reached in a controlled and precise way using contemporary DAC techniques. The highest pressure limit of experiments has been recently extended to terapascals using a double-stage DAC [18]. The combined use of the DAC with the use of synchrotron sources and other modern characterization techniques has boosted exponentially the number of discoveries in several research fields, which ranges from inorganic/organic solids to liquids and gases [17]. In the context of metal iodates, it is useful to note that the majority of the interesting pressure-induced phenomena are observed at pressures below 30 GPa. A pressure transmitting medium, such as neon or methanol:ethanol, is used to maximize the hydrostaticity of the sample [26].

The goal of this article is to review the effect of compression on iodates. HP research is a useful tool to help improve the understanding of the physical properties of iodates. Phase transitions, and singular phenomena like phonon softening and the occurrence of symmetry-preserving phase transitions, have been discovered. Most of these phenomena have been related to the existence of the LEP in iodine. Since the beginning of the present decade, several studies have been carried out by various laboratories across the globe to understand the physics behind the HP properties of iodates [27–32]. However, the interesting results in the literature are dispersed and fragmented in various journals. Therefore, this review presents and discusses all of the results from HP studies on pentavalent iodine iodates published so far in the form of a single review. This should be beneficial for the scientific community, in general, and the HP community, specifically. This review includes discussions pertaining to results from: powder X-ray diffraction (XRD); vibrational spectroscopy; optical absorption spectroscopy; and electrical resistivity measurements. Results of density-functional theory (DFT) calculations will be also discussed. Such calculations combined with experiments played an important role for establishing systematics in the HP behavior of iodates.

The aim of the article is discussing HP properties of iodates. Therefore, a thorough description of HP techniques, experimental methods, or computer simulation procedures are not provided in this review because extensive reviews can be found elsewhere the literature [33–36]. This review article is organized as follows. It begins with a concise historical background summarizing previous HP studies on pentavalent iodine iodates. This is followed by descriptions of the commonly adopted crystal structures of iodates at ambient pressure. The review then presents descriptions of the iodate Raman and infrared vibrational modes as well as the electronic band structure and optical properties at ambient pressure. These sections are necessary for the reader to follow the subsequent sections which focus on HP studies. Results from HP studies are be presented beginning with **Section 6**. In the subsequent sections, the observed phase transitions and phase transition mechanisms are discussed together with the pressure-induced changes in electronic, optical, and vibrational properties. A discussion of the systematics of the HP behavior of iodates together with various equations of state, coordination changes around the different polyhedral units, axial, bulk, and polyhedral compressibility follows from the initial discussion of phase transitions in iodates. Before concluding, a perspective for future research topics for HP studies of pentavalent iodine iodates is presented, discussing open questions and possible avenues to follow in order to solve the contemporary questions. Finally, a summary is presented which highlights the impact of the review on HP research in particular and on wider research fields in general.

## 2. Historical background

The first study of iodates involving HP conditions was reported in the year 1969 [37]. In this work the authors studied lithium iodate ($LiIO_3$) and reached a maximum pressure of 0.45 GPa at temperatures ranging from 40 to 420 ºC. X-ray measurements were performed ex-situ in recovered samples. It was concluded that, at ambient pressure, above 75 ºC the hexagonal α-polymorph (space group $P6_3$), formed by $LiO_6$ octahedra linked by $IO_3$ triangular pyramids, undergoes a phase transition to a tetragonal β-polymorph (space group $P4_2/n$). The α-β phase boundary was reported to have a positive slope with intercept to the melting curve at a triple point at 410 ºC and 0.45 GPa. The pressure-temperature (*P-T*) phase diagram of $LiIO_3$ was updated in the 1980s [38] after a phase transition to a tetragonal ε-polymorph (space group *P*4/*nmm* or *P*4/*n*) was found at 2.5 GPa. A high pressure-high temperature (HP-HT) Raman study was subsequently performed at temperatures from 12 to 500 K and at pressures up to 0.75 GPa [39]. No pressure-induced phase transition was found, but a softening behavior of the high-frequency modes was reported, which are associated with stretching vibrations of the $IO_3$ molecule. Such phenomena are consistent with the existence of phase transitions at a higher pressure [40,41]; for instance, the transition found at 2.5 GPa by Shen *et al.* [38]. However, the existence of a pressure-induced transition has not been confirmed by more recent DAC experiments [27,42], which reported that the α-phase is stable up to at least 75 GPa at room temperature (RT). Contradictions in the literature about the HP behavior of α-$LiIO_3$ also involve the pressure-volume equation of state. The reported values of the bulk modulus are scattered from 38 to 68 GPa [27,42–44]. The tetragonal β-polymorph was also studied under compression by Raman spectroscopy and XRD up to 9.5 GPa [45]. A phase transition to a phase with monoclinic symmetry *P*2/*n* was characterized at 5 GPa.

The study of other iodates under HP conditions was not continued until nearly 15 year later. First, Shen *et al.* [46] studied potassium iodate ($KIO_3$) by Raman spectroscopy up to 35 GPa at RT. $KIO_3$ crystallizes in a triclinic structure [28], but its structural framework is qualitatively similar to that of $LiIO_3$ and many other iodates. The crystal structure is formed by highly distorted $KO_6$ octahedral units linked by typical pyramidal $IO_3$ molecules. Drastic changes observed in Raman spectra, which were interpreted as phase transitions, were observed at 5, 8.7, 15, and 22 GPa, respectively [46]. In a subsequent work, Bayanjagal *et al.* [28] studied $KIO_3$ at RT up to 30 GPa and confirmed the HP phase transition previously reported at 8.7 GPa, which they found at 7 GPa. However, they did not observe the phase transition previously reported at 5 GPa. Indeed, these authors only observed two phase transitions up to 30 GPa [28] and not four as alleged by Shen *et al.* reported [46]. The transitions take place at 7 and 14 GPa. Single-crystal XRD was used to identify the crystal structure of the first HP phase at 8.7 GPa. The structure has been identified as trigonal (space group *R*3) and it is related to a double perovskite structure [28]. In the HP structure, potassium is coordinated by twelve oxygen atoms and iodine remains in the typical pyramidal 3-fold coordination of iodates. However, because of the presence of three second neighboring oxygen atoms, the coordination polyhedron of iodine is also described in the literature as a highly distorted octahedron. In addition to the identification of phase transitions, the compressibility of the ambient-pressure and first HP polymorphs was also determined [28]. The ambient-pressure phase is highly compressible with a bulk modulus of 24.7(5) GPa.

The studies on $KIO_3$ were followed by a study on silver iodate ($AgIO_3$) [47]. This iodate has an orthorhombic crystal structure (space group $Pbc2_1$) at ambient conditions, which is constituted by $AgIO_3$ layers parallel to the (010) plane and linked by weak I···O halogen bonds along the direction of the LEP [48]. As in other iodates, the metallic cation (Ag) is coordinated by six oxygen atoms, and I is three-fold coordinated forming a trigonal pyramid. Evidence of a phase transition was found at 260 ºC and 2.7 GPa using a Belt apparatus and differential-thermal analysis [47]. The crystal structure of the HP-HT phase has been determined *ex-situ* from recovered samples. The phase transition involves a contraction of the unit-cell volume of 4.2%. The HP-HT phase is also layered and has similar coordination polyhedral units like those of the ambient-pressure phase. The compressibility of the two phases of $AgIO_3$ has not been studied yet. In parallel to the study of $AgIO_3$ there was a computer-simulation study of mercury iodate, $Hg(IO_3)_2$ [48]. By means of DFT calculations a bulk modulus of 108.4 GPa was determined for $Hg(IO_3)_2$. Such bulk modulus is unusually large in comparison with typical values reported previously for other iodates [27,42,44,46,49] and similar to the bulk modulus of inorganic compounds like vanadate and phosphates, which is non-standard [50,51].

The above-described facts indicated to researchers that the HP behavior of iodates could be extremely complex and that systematic studies were needed. Methodical research to further understand the HP behavior of iodates was started very recently. The first efforts were focused on iron iodate, $Fe(IO_3)_3$ [29]. The structure of $Fe(IO_3)_3$ resembles that of α-$LiIO_3$ and is described by the same space group. $Fe(IO_3)_3$ has been studied under HP using powder XRD, infrared spectroscopy, Raman spectroscopy, optical-absorption, resistivity, and DFT calculations [29,52,53]. Evidence of two subtle symmetry preserving phase transitions has been found and the occurrence of a first-order phase transition at 22 GPa has been reported. The behavior of phonons, optical properties, resistivity, and mechanical properties under compression have been accurately established. The phenomena observed include a collapse of the volume, the bandgap, and the resistivity at 22 GPa. DFT calculations helped for the building of a rationale to explain all observed phenomena. One of the most relevant findings was the unveiling of the key role played by LEP in the HP behavior of $Fe(IO_3)_3$. The existence of the LEP favors the formation of extra I-O bonds which has important consequences in physical properties [29,52,53] as discussed throughout this review.

After the work on $Fe(IO_3)_3$, a substantial amount of the progress was attained in understanding the experimentally observed HP behavior of double iodates. The compounds studied include $Zn(IO_3)_2$ [54–56], $Mg(IO_3)_2$ [30,56], and $Co(IO_3)_2$ [57,58]. They were studied combining powder XRD with spectroscopic and optical experiments. Several phase transitions were discovered. Experimental findings were corroborated and explained with the help of theoretical studies based on state-of-the-art computational models. One of the major contributions of these studies is the understanding of the observed phonon softening, which is a consequence of changes in the coordination polyhedron of iodine. This is one of the shared characteristics observed in all studied iodates. This phenomenon is connected to the existence of the LEP and weak I···O halogen bonds. Important progress has been also achieved on the understanding of electronic properties [56]. One interesting phenomenon reported is the non-linear pressure behavior observed for the bandgap energy which is a consequence of the interplay between the pressure-induced increase in length of the first-nearest neighbor iodine-oxygen bonds and the decrease in length of the second-nearest neighbor weak halogen iodine-oxygen bonds.

Before the end of this section, the review now briefly discusses some very recent studies on more complex iodates. The subject of these studies include $Na_3Bi(IO_3)_6$ [59,60], $BiOIO_3$ [31], $HIO_3$ [61], and $Li_2Ti(IO_3)_6$ [32]. In $Na_3Bi(IO_3)_6$, HP Raman experiments were performed and changes in the Raman spectra has been interpreted as caused by two reversible phase transitions caused by changes of the $IO_3$ vibrational modes, the subsequent single-crystal XRD reveal a pressure-induced structural phase transition from space group $P\bar{1}$ to $P1$ at around 9.5 GPa [60]. In $BiOIO_3$, a pressure-induced two-step SHG switching has been achieved. The phenomenon has been attributed to the suppression of the LEP in iodine atoms. In $HIO_3$, Raman measurements and DFT simulations provide evidence of the strengthening of I···O halogen bonds at the very low pressure of 0.5 GPa. These studies also indicate possible structural changes in the system at 5 and 15 GPa. In $Li_2Ti(IO_3)_6$ zero-linear compressibility and zero-area compressibility have been reported. These works, and those described in previous paragraphs, enrich the knowledge of the HP behavior of iodates and open new avenues for exploring novel physical and chemical phenomena under HP conditions.

## 3. Crystal structure of metal iodates at ambient conditions

This section introduces the most common crystal structures of the metal iodates reported at ambient conditions. Priority is given to structures relevant for HP studies already performed. The metal iodates contain two parts, one is the metal oxygen polyhedral unit and another is the iodate unit. There are several different types of isolated iodate polymers which have been reported in the literature. They include the $[IO_3]^-$ iodate group, and the $I_xO_y$ polyiodates with I-O-I bridges: $I_2O_5$, $[I_3O_8]^-$, $[I_4O_{11}]^{2-}$ and $[I_5O_{14}]^{3-}$. A schematic illustration of these anions can be found in **Fig. 1**. Amongst these types of iodates, $[IO_3]^-$ is the most common one reported in the literature. In this molecule, all three of the oxygen atoms are located on the same side of iodine unit due to the existence of the iodine LEP. In this arrangement the configurational energy of the $[IO_3]^-$ unit is minimised [62]. If two $[IO_3]^-$ are connected by sharing one oxygen atom between them, the resulting unit is the $I_2O_5$ polyiodate. This configuration has been reported in compounds like $HBa_{2.5}(IO_3)(I_2O_5)$ [63] and $Rb_3(IO_3)_3(I_2O_5)(HIO_3)_4(H_2O)$ [64]. If two $[IO_3]^-$ units are linked by an $[IO_4]^{3-}$ unit, the result is the $[I_3O_8]^-$ polyiodate. This configuration has been reported in $NaI_3O_8$ [9], and α, β-$AgI_3O_8$ [65]. If one more $[IO_4]^{3-}$ unit is added in the center of the $[I_3O_8]^-$ chain, the result is $[I_4O_{11}]^{2-}$, and this configuration has been reported in $HBa(IO_3)(I_4O_{11})$ [63] and α, β-$CsI_4O_{11}$ [66].

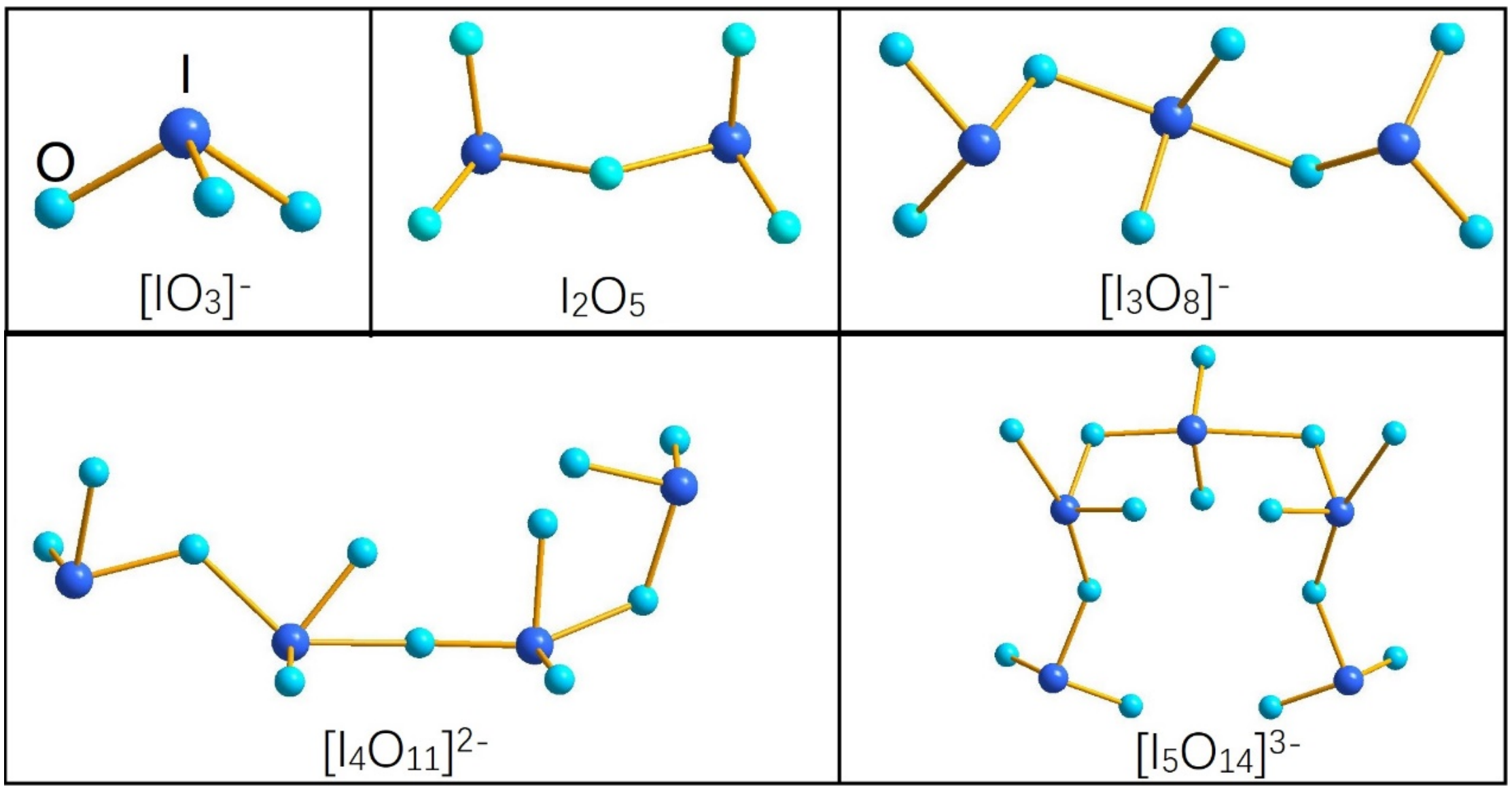

***Fig. 1**. A schematic representation of iodate $[IO_3]^-$ and polyiodates $I_2O_5$, $[I_3O_8]^-$, and $[I_4O_{11}]^{2-}$ extracted from the crystal structure of $Mg(IO_3)_2$* [67], *$HBa_{2.5}(IO_3)(I_2O_5)$* [63], *$NaI_3O_8$ [9], and $CsI_4O_{11}$* [66].

**Table 1** provides a list the crystal structures of most metal iodates reported in the literature. A large number of them crystallize in the monoclinic crystal system. The crystal structures of these monoclinic materials is described by the space groups: $P2_1$ (No. 4), *Cc* (No. 9), $P2_1$/c or P$2_1$/n (No. 14), *C*2 (No. 5), and *C*2/c (No.15). The crystal structures of other metal iodates are represented by the lowest symmetry triclinic crystal system. Space groups *P*1 and $P\bar{1}$ have been used to describe the crystal structures of $KIO_3$ [68] and $Tl_4(IO_3)_6$ [69]. $BiOIO_3$ [70] and α-$AgIO_3$ [71] are described by the space group $Pca2_1$ (No. 29) in the orthorhombic crystal system at ambient conditions, while the earlier reported iodic acid, α-$HIO_3$, is reported in the space group $P2_12_12_1$ in the same crystal system. Many of the metal iodates are also described by the space group $R\bar{3}$ (No. 148) in the trigonal crystal system, especially when the chemical formula contains six iodate molecules. Additionally, there are a few iodates which can be described by another space group *R*3m (No. 160) in the trigonal system. There are also some metal iodates whose crystal structures are described by the space group $P6_3$ (No. 173) in the hexagonal symmetry crystal system. For example, the first metal iodate studied under HP, and one of the most studied as a potential NLO material, α-$LiIO_3$ [27,38,42,72], crystalizes in this space group. Its structure is represented in **Fig. 2d.** All of the crystal structures shown in **Fig. 2**. share the common characteristic of the presence of pentavalent iodine $[IO_3]^-$ molecules, which have a stereochemically active LEP. In the $[IO_3]^-$ unit the three oxygen atoms are located on the same side of iodine because the electrostatic repulsion of the LEP and thus have the minimized configurational energy compared to other arrangements [62].

***Table 1.*** *Summary of most common metal iodates reported in the literature including the space group of the crystal structure. References to articles describing the crystal structure are provided.*

| Compound | Space Group | Ref. | Compound | Space Group | Ref. |
|---|---|---|---|---|---|
| α-$LiIO_3$ | $P6_3$ | [72] | α-$NaAu(IO_3)_4$ | *P*1 | [73] |
| α-$HIO_3$ | $P2_12_12_1$ | [74] | β-$NaAu(IO_3)_4$ | $P2_1$/c | |
| $KIO_3$ | *P*1 | [68] | β-$CsAu(IO_3)_4$ | $P\bar{1}$ | |
| $CsIO_3$ | *R*3m | [75] | $AgAu(IO_3)_4$ | $P2_1$/c | |
| $RbIO_3$ | *R*3m | [76] | $RbAu(IO_3)_4$ | *C*2 | |
| $BiO(IO_3)$ | *P*ca$2_1$ | [70] | α-$CsAu(IO_3)_4$ | *C*2 | |
| α-$AgIO_3$ | $Pbc2_1$ | [71] | $Sn(IO_3)_4$ | $P\bar{1}$ | [77] |
| $Zn(IO_3)_2$ | $P2_1$ | [78] | $NaBi(IO_3)_4$ | *Cc* | [11] |
| $Mg(IO_3)_2$ | $P2_1$ | [67] | $NaY(IO_3)_4$ | *Cc* | [8] |
| $Co(IO_3)_2$ | $P2_1$ | | $NaLa(IO_3)_4$ | *Cc* | [80] |

| | | |
|---|---|---|
| $Ca(IO_3)_2$ | $P2_1/n$ | [81] |
| $Ba(IO_3)_2$ | $C2/c$ | [82] |
| α-$Cu(IO_3)_2$ | $P2_1$ | [17] |
| β-$Ni(IO_3)_2$ | $P2_1$ | [83] |
| $Mn(IO_3)_2$ | $P2_1$ | [67] |
| $Pb(IO_3)_2$ | $Pbcn$ | [85] |
| $Hg(IO_3)_2$ | $P2_1$ | [87] |
| $Fe(IO_3)_3$ | $P6_3$ | [88] |
| $Tl(IO_3)_3$ | $R\bar{3}$ | [69] |
| $Nd(IO_3)_3$ | $P2_1/n$ | [90] |
| $Al(IO_3)_3$ | $P6_3$ | |
| $Bi(IO_3)_3$ | $P2_1/n$ | [26] |
| $^{248}Cm(IO_3)_3$ | $P2_1/c$ | [87] |
| α-$A(IO_3)_3$, A=Y and Dy | $P2_1/c$ | [28] |
| β-$A(IO_3)_3$ A=Y, Ce, Pr, Eu, Gd, Tb, Dy, Ho, Er | $P2_1/n$ | |
| $A(IO_3)_3$, A=Yb and Lu | $P2_1/n$ | [93] |
| α-$Am(IO_3)_3$ | $P2_1/n$ | [30] |
| β-$Am(IO_3)_3$ | $P2_1/c$ | |
| $La(IO_3)_3$ | $Cc$ | [8] |
| β-$In(IO_3)_3$ | $R\bar{3}$ | [96] |
| α-$In(IO_3)_3$ | $P6_3$ | |
| $LiZn(IO_3)_3$ | $P6_3$ | [13] |
| $LiCd(IO_3)_3$ | $P6_3$ | |
| $KLi_2(IO_3)_3$ | $P2_1/n$ | [99] |
| $LiMg(IO_3)_3$ | $P6_3$ | [38] |
| $NaCe(IO_3)_4$ | $Cc$ | |
| $NaSm(IO_3)_4$ | $Cc$ | |
| $NaEu(IO_3)_4$ | $Cc$ | |
| $BaNbO(IO_3)_5$ | $Cc$ | [11] |
| $LaVO(IO_3)_5$ | $P2_1/n$ | [84] |
| $Ag_3In(IO_3)_6$ | $P\bar{1}$ | [86] |
| $Tl_4(IO_3)_6$ | $P\bar{1}$ | [69] |
| $Cs_2Sn(IO_3)_6$ | $R\bar{3}$ | [89] |
| $Li_2Sn(IO_3)_6$ | $P6_3$ | |
| $K_2Sn(IO_3)_6$ | $R\bar{3}$ | |
| $Na_2Sn(IO_3)_6$ | $P6_3$ | |
| $Rb_2Sn(IO_3)_6$ | $R\bar{3}$ | |
| $Li_2Ti(IO_3)_6$ | $P6_3$ | [66] |
| $Na_2Ti(IO_3)_6$ | $P6_3$ | |
| $Rb_2Ti(IO_3)_6$ | $R\bar{3}$ | |
| $Cs_2Ti(IO_3)_6$ | $R\bar{3}$ | |
| $Tl_2Ti(IO_3)_6$ | $R\bar{3}$ | |
| $K_2Ti(IO_3)_6$ | $R\bar{3}$ | |
| $SrSn(IO_3)_6$ | $R\bar{3}$ | [31] |
| $Ag_2Zr(IO_3)_6$ | $P2_1/c$ | [97] |
| $Na_3Bi(IO_3)_6$ | $P\bar{1}$ | [34] |
| $Na_2Pt(IO_3)_6$ | $R\bar{3}$ | [98] |
| $K_2Pt(IO_3)_6$ | $R\bar{3}$ | |
| $Rb_2Pt(IO_3)_6$ | $R\bar{3}$ | |
| $Cs_2Pt(IO_3)_6$ | $R\bar{3}$ | |

The triclinic crystal structure of $KIO_3$ at ambient condition was reported by two groups in 1973 [68] and 1978 [101], based on the results from single-crystal XRD experiments. However, there is a disagreement about the inner atomic bond distance in those two works. In 1984 [102], neutron powder diffraction was performed to reinvestigate the crystal structure of $KIO_3$. By the means of a Rietveld refinement on the diffraction data, it confirmed the structure reported in 1978. The related crystal structure is shown at the top of **Fig. 2a** as a representation of metal iodates whose crystal structure can be described in the space group *P*1 (No. 1). In the crystal structure the iodine atoms are located at four non-equivalent Wyckoff positions. They are in the corners and the faces of the structure. Each iodine atom is bonded with three oxygen atoms and the average bond distance is around 1.78 Å. The three I-O bonds are slightly different in length. In the $KIO_3$ structure, the potassium atoms are in the edges and the center of the structure. They occupy four different Wyckoff positions, forming distorted $KO_6$ octahedra. However, due to the presence of second neighbor oxygen atoms close to the first coordination sphere, the coordination polyhedra of potassium have been also described as $KO_7$ or $KO_8$ polyhedra. Perpendicular to the *b*-axis, the crystal structure of $KIO_3$ exhibits a layered structure, each layer contains both $IO_3$ and $KO_n$ (n = 7, 8) units. The $IO_3$ and $KO_n$ units are corner-sharing or edge-sharing connected.

Another common space group for metal iodates in the triclinic crystal system is $P\bar{1}$ (No. 2). Here $Tl_4(IO_3)_6$ is used as an example to discuss this group of iodates (see **Fig. 2a**). The crystal structure of $Tl_4(IO_3)_6$ was solved by single-crystal XRD [69]. The iodine and thallium atoms are in three Wyckoff positions. Each iodine atom is bonded to three oxygen atoms and formed a distorted $IO_3$ pyramid. The three I-O bond distances range from 1.79 to 1.87 Å, thereby showing a large distortion compared with the I-O bonds in the crystal structure of $KIO_3$. The thallium atoms are bonded with six or eight oxygen atoms, forming either $TlO_6$ or $TlO_8$ polyhedra. This is because in this compound Tl has two difference valence states $Tl^{3+}$ and $Tl^{+}$. $TlO_6$ and $TiO_8$ units are edge-sharing connected and $TlO_8$ (or $TlO_6$ units) units are bridged by $IO_3$ units. Perpendicular to the *b*-axis, the material exhibits a layered structure, where $TlO_6$ and $TlO_8$ are alternating in successive layers. On the other hand, the $IO_3$ units are present in every layer.

In the monoclinic crystal system, three commonly found space groups for metal iodates are identified in this review. The first discussed here is space group $P2_1$ (No. 4), which is represented by the crystal structure of α-$Cu(IO_3)_2$ [4] in **Fig. 2b**. As **Table 1** summarizes, many metal iodates with two $[IO_3]^-$ molecules in the chemical formula are described by this space group. In particular, the structures of $Zn(IO_3)_2$, $Co(IO_3)_2$, and $Mg(IO_3)_2$ are intimately related to that of α-$Cu(IO_3)_2$. In this structure the copper atom is in one Wyckoff position and iodine is in two different Wyckoff positions. Along the *b*-axis, each copper is bonded with six oxygen atoms, and each iodine atom is bonded with three oxygen atoms, the $CuO_6$ and $IO_3$ polyhedra are connected by sharing one oxygen atom between them. The $CuO_6$ octahedra are bridged by two $IO_3$ units. Perpendicular to the *b*-axis, the crystal structure shows a layered structure formed by $IO_3$ units. The layers are connected to each other by $CuO_6$ octahedra. In addition, all the $IO_3$ pyramids are aligned to the same direction. This structure described here is similar to that of α-$LiIO_3$ in the bottom of **Fig. 2d**, which crystallizes in the hexagonal crystal system and is described by the space group $P6_3$ (No. 173). The main difference between the two structures is that the metal cations in space group $P2_1$ are in the interior of the structure, but in space group $P6_3$, the metal cations are in the corners of the structure.

In **Table 1**, there are multiple metal iodates with three $[IO_3]^-$ molecules in the chemical formula. The structures are described by space group $P2_1$/c or $P2_1$/n (No. 14). One example of them is the crystal structure of β-$Am(IO_3)_3$ [94] shown in **Fig. 2b**. The americium atom with a 3+ valence state has an eight-fold coordination. On the other hand, each iodine atom is bonded with three oxygen atoms. Perpendicular to the *c*-axis, there is a layered structure and each layer contain three $IO_3$ pyramids and one $AmO_8$ polyhedra, which is corner-sharing connected within the layer.

The crystal structure of $NaY(IO_3)_4$ (also shown in **Fig. 2b**) at ambient conditions is described by the space group *Cc* (No. 9) [8]. It contains $YO_8$ polyhedra, and asymmetric $IO_3$ pyramids. The $YO_8$ and $IO_3$ units are connected by sharing corners or edges. Perpendicular to the *a*-axis, the material exhibits a layered structure, formed by $YO_8$ and $IO_3$ units. Within the layer, there are channels formed by $YO_8$ polyhedra, which are bridged by $IO_3$ pyramids. The sodium atoms are located inside the channels.

It can be seen from **Table 1** that the crystal structures of some metal iodates is described by the space group $R\overline{3}$. Such structures are common in metal iodates with six $[IO_3]^-$ units in the chemical formula. Here the crystal structure of $Sc(IO_3)_3$ is chosen to represent the metal iodates in this space group (**Fig. 2c**). In the structure, the scandium atom is six-coordinated and iodine is three-fold coordinated. In the *ab* plane, the $ScO_6$ polyhedra are bridged by $IO_3$ pyramids. In the structure, the scandium atoms are in the four corners and inside of the unit cell, while the $IO_3$ pyramids are aligned along the *c*-axis at the edge and inside the unit cell. However, all the $IO_3$ pyramids do not point to the same direction. Perpendicular to the *c*-axis, there is a layered structure consisting of $ScO_6$ and $IO_3$ units, each layer is isolated from each other.

$LiIO_3$ crystallizes in two polymorphs. The standard polymorph, α-$LiIO_3$, is a uniaxial nonlinear crystal with high nonlinear coefficients and a wide optical transparent range. The crystal structure of α-$LiO_3$, shown in **Fig. 2d**, is described by space group $P6_3$. As with the rest of structures described in this section, α-$LiO_3$ is built up from discrete $IO_3$ and $LiO_6$ octahedra. The average I-O bond distance within the $IO_3$ pyramid is 1.801 Å. In addition, each iodine atom is linked by halogen bonds to three other oxygen atoms at distances in the range 2.8 to 3.2 Å. The average Li-O bond distance within the $LiO_6$ octahedron is 2.130 Å. $IO_3$ units and $LiO_6$ octahedra are connected by corner sharing oxygen atoms. $LiO_6$ octahedra are also connected to other $LiO_6$ octahedra sharing edges and forming a chain along the *c*-axis. The structure of other compounds described by space group $P6_3$, like $Fe(IO_3)_3$, have many similarities with α-$LiIO_3$.

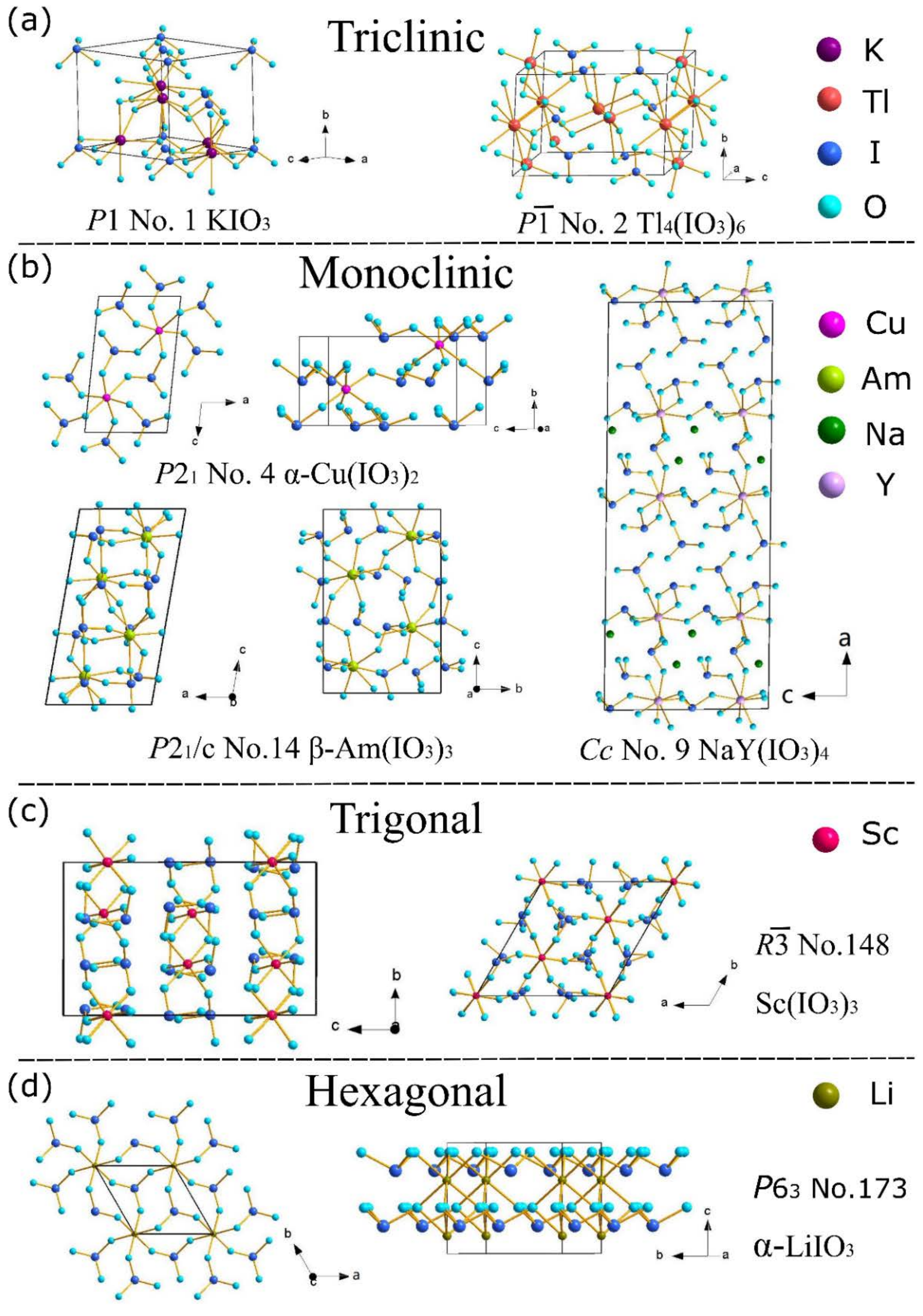

***Fig. 2***. *Projections of the seven most common crystal structures of metal iodates. (a) The triclinic structures of $KIO_3$* [83] *and $Tl_4(IO_3)_6$* [69]. *(b) The monoclinic structures of α-$Cu(IO_3)_2$ [4], $NaY(IO_3)_4$ [8] and β-$Am(IO_3)_3$* [94]. *(c) The trigonal structure of $Sc(IO_3)_3$* [103]. *(d) The hexagonal structure of α-$LiIO_3$* [72]. *These structures are representative examples of metal iodate structures described by the space groups $P1$, $P\overline{1}$, $P2_1$, $Cc$, $P2_1/c$, $R\overline{3}$, and $P6_3$. The color code used to represent atoms is given in the figure.*

# 4. Vibrational spectra of iodate materials at ambient conditions

## 4. 1 Complementarity of Raman and infrared spectroscopy

Raman and infrared spectroscopies are experimental diagnostics which utilize different phenomena associated with interactions of electromagnetic radiation with matter to probe molecular dynamics. For example, Raman spectroscopy utilizes the phenomenon of inelastic scattering of electromagnetic radiation from matter to probe the relative energies of allowed vibrational and rotational energy levels in the scattering molecules. In contrast, infrared spectroscopy utilizes the phenomenon of photo-absorption to probe the absolute energies of molecular vibrations and rotations. The two techniques are also differentiated by their gross selection rules. For example, in infrared spectroscopy, the gross selection rule for photo-absorption is that the molecular dipole moment must change during the molecular vibration. Therefore, symmetrical molecules with no dipole moments, such as the homonuclear diatomics: $H_2$, $N_2$ and $O_2$, do not exhibit infrared absorption. In contrast, the gross selection rule for Raman spectroscopy is that the molecular polarizability must change with molecular vibration. Therefore, the fundamental vibrational energy levels of symmetrical molecules can be probed via Raman spectroscopy. In fact, in centrosymmetric materials, any fundamental vibrations which are not active in infrared are active in Raman spectroscopy, and vice versa, making Raman and infrared spectroscopy complementary experimental diagnostics.

It should be emphasized that Raman and infrared spectroscopies are equally complementary to XRD as discussed in **Section 6.1**. For example, imagine a crystal composed of $N_2$ molecules. XRD data can tell you the spatial arrangement of $N_2$ molecules in the crystal structure, but it cannot tell you about the strength or order of the $N_2$ triple bond. The reason is that XRD data provide structural information based on time-averaged data. This is because molecular vibrations occur on a much shorter time scale than the XRD data acquisition. Therefore, XRD alone provides little information regarding lattice dynamics, (although it is possible to identify, for example, freely rotating molecules occupying specific lattice sites). On the other hand, the information encoded into Raman and infrared spectra is fundamental to the crystal dynamics, as discussed in the above paragraph in the context of the spectral line energies originating from fundamental vibrational modes. Additionally, spectral line intensities are determined by many factors such as: transition probability; population of states (*e.g.* stokes vs. anti-stokes); and optical path-length. The spectral line widths are also encoded with information, for example, a distribution of vibrational energies for the same fundamental mode would cause linewidth broadening. In the case of HP Raman spectroscopy, a distribution of energies such as this could be caused by a pressure gradient across the scattering medium thereby indicating strain in the sample.

## 4.2 Vibrational characteristics of the $[IO_3]^-$ group at ambient conditions

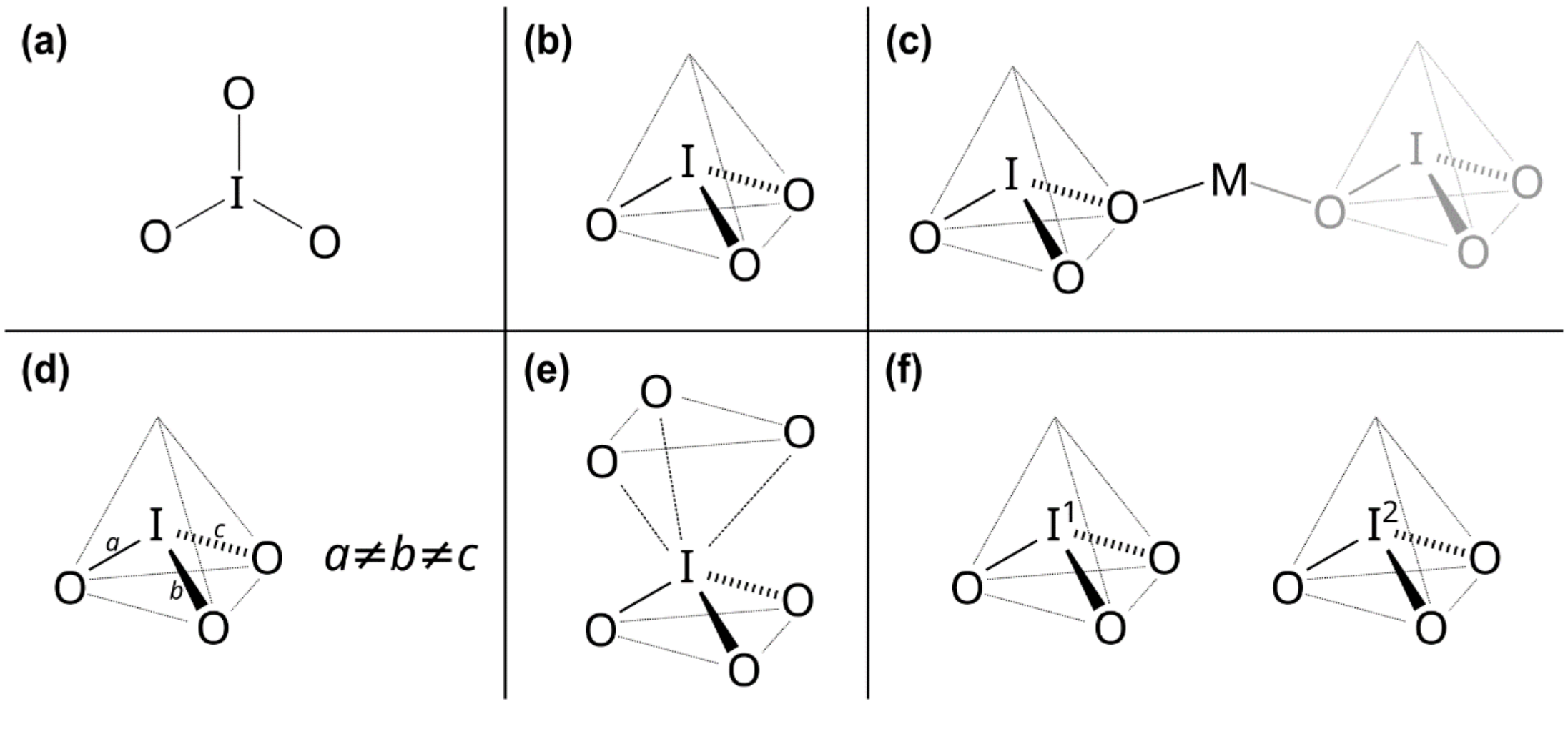

***Fig. 3.*** *Six different idealised $[IO_3]^-$ configurations corresponding to the vibrational modes described in* **Table 2**. ***(a)*** *Trigonal planar;* ***(b)*** *Trigonal pyramidal;* ***(c)*** *Covalently bonded iodate group;* ***(d)*** *Distorted trigonal pyramidal;* ***(e)*** *Octahedral; and* ***(f)*** *crystallographically unique iodine sites.*

***Table 2** Description of the predicted vibrational modes for the independent [$IO_3$]$^-$ configurations (a-f) show in **Fig. 3**. Whether the vibrational modes is Raman (R) or infrared (IR) active have been indicated.*

| | **Point group** | | | | | | |
|---|---|---|---|---|---|---|---|
| **(a) Trigonal planar** | $D_{3h}$ | $\nu_1$ $A_1$'(R) symmetric I-O stretch | $\nu_2$ $A_2$''(IR) symmetric I-O bend | $\nu_3$ E'(R, IR) antisymmetric I-O stretch | | $\nu_4$ E'(R, IR) antisymmetric I-O bend | |
| **(b) Trigonal pyramidal** | $C_{3v}$ | $\nu_1$ $A_1$(R, IR) symmetric I-O stretch | $\nu_2$ $A_1$(R, IR) symmetric I-O bend | $\nu_3$ E'(R, IR) antisymmetric I-O stretch Doubly degenerate | | $\nu_4$ E'(R, IR) antisymmetric I-O bend Doubly degenerate | |
| **(c) Covalently bonded iodate group: $O_2IO$-M** | $C_s$ | $\nu_1$ (R, IR) symmetric stretch O-$IO_2$ | $\nu_2$ (R, IR) symmetric bend O-$IO_2$ | $\nu_{3a}$ (R, IR) antisymmetric $IO_2$ stretch | $\nu_{3b}$ (R, IR) symmetric $IO_2$ stretch | $\nu_{4a}$ (R, IR) antisymmetric $IO_2$ bending | $\nu_{4b}$ (R, IR) symmetric $IO_2$ bending |
| **(d) Distorted trigonal pyramidal** | $C_s$ | $\nu_1$ (R, IR) symmetric stretch I-O | $\nu_2$ (R, IR) symmetric bend I-O | $\nu_{3a}$ (R, IR) antisymmetric I-O stretching | $\nu_{3b}$ (R, IR) antisymmetric I-O stretching | $\nu_{4a}$ (R, IR) antisymmetric I-O bend | $\nu_{4b}$ (R, IR) antisymmetric I-O bend |
| **(e) Octahedral** | $O_h$ | $\nu_1$ $A_{1g}$ (R) symmetric stretch | $\nu_2$ $E_g$ (R) stretch Doubly degenerate | $\nu_3$ $F_{1u}$ (IR) bend Triply degenerate | $\nu_4$ $F_{1u}$ (IR) stretch Triply degenerate | $\nu_5$ $F_{2g}$ (R) bend Triply degenerate | $\nu_6$ $F_{2u\ (Forbidden)}$ bend Triply degenerate |
| **(f) Unique independent I sites: $I^1$ and $I^2$** | $C_{3v}$ | $\nu_{1,1}$ and $\nu_{1,2}$ $A_1$'(R) symmetric I-O stretch non-degenerate | $\nu_{2,1}$ and $\nu_{2,2}$ $A_2$''(IR) symmetric I-O bend non-degenerate | $\nu_{3,1}$ and $\nu_{3,2}$ E'(R, IR) antisymmetric I-O stretch non-degenerate | | $\nu_{4,1}$ and $\nu_{4,2}$ E'(R, IR) antisymmetric I-O bend non-degenerate | |

All the iodate materials studied at HP and discussed in this review contain the iodate [$IO_3$]$^-$ ion, therefore, in this section the vibrational spectra of iodate materials in general will be discussed and interpreted based on the vibrational modes of these [$IO_3$]$^-$ units. Specifically, the discussion begins with a brief description of the vibrational spectrum of an idealized symmetrical pyramidal iodate [$IO_3$]$^-$ ion. Subsequently, the symmetry of the iodate [$IO_3$]$^-$ ion is lowered, following symmetry reductions observed in real iodate materials, and the resultant increasingly complex vibrational spectra are discussed. Some experimental vibrational spectra are presented to emphasize the complexity and to illustrate the similarities between the vibrational spectra of different iodate materials regardless of the underlying crystal structure. The pressure induced evolution of the vibrational properties of iodate materials is discussed in **Section 6.2**. The reader is directed to the following resources for a more fundamental and in-depth discussion of vibrational spectroscopy [104–109].

**Fig. 3** shows six independent $[IO_3]^-$ iodate molecular configurations. Iodate materials can exhibit a combination of all these configurations to varying degrees which makes the experimental spectra extremely complicated. In this discussion each of them is considered separately, followed by a discussion of actual experimental vibrational spectra.

(a) The iodate $[IO_3]^-$ ion was originally believed to exhibit trigonal planar symmetry (**Fig. 3a**), similar to that of the carbonate $CO_3^{2-}$ ion [110]. As shown in row (a) of **Table 2**, a trigonal planar $[IO_3]^-$ ion would belong to the point group $D3_h$, at it would possess four vibrational modes in total: three of them being Raman-active ($\nu_1$, $\nu_3$ and $\nu_4$) and three of them infrared-active ($\nu_2$-$\nu_4$). However, the iodate ion possesses a LEP which repels the three oxygen atoms, resulting in a trigonal pyramidal geometry where the LEP occupies one of the pyramidal vertices.

(b) The iodate $[IO_3]^-$ ion was first observed to exhibit a trigonal pyramidal geometry (**Fig. 3b**), similar to the stereochemically similar chlorate $ClO_3^-$ and bromate $BrO_3^-$ ions, in 1937 in spectroscopic studies of iodate containing solutions [111]. The trigonal pyramidal geometry was confirmed due to the observation of four vibrational modes, $\nu_1$-$\nu_4$, all of which are both Raman- and infrared-active (see row (b) of **Table 2**), with the antisymmetric modes $\nu_3$ and $\nu_4$ both being doubly degenerated. The pyramidal iodate ion was first observed in a crystalline iodate in ceric iodate, $Ce(IO_3)_4$ [112], wherein the authors note three different I-O bond lengths of: 1.78, 1.84 and 1.83 Å. The effect of different bond lengths on the vibrational spectrum is discussed later in this section.

The symmetric and asymmetric stretching modes of the symmetrical trigonal pyramidal $[IO_3]^-$ iodate ion, $\nu_1$ and $\nu_3$ respectively, exhibit vibrational frequencies between 750 and 850 $cm^{-1}$ with $\nu_1$ mode being lower in frequency but more intense. This mode has been observed in Raman spectra at: 777 $cm^{-1}$ in $Mg(IO_3)_2$ (see **Fig. 5**) [30]; 782 $cm^{-1}$ in $Zn(IO_3)_2$ [54]; 790 $cm^{-1}$ in $Fe(IO_3)_3$ [52]; 765 $cm^{-1}$ in $Co(IO_3)_2$ [57];

774 $cm^{-1}$ in $Mn(IO_3)_2$ [113]; and at 750 $cm^{-1}$ in $KIO_3$ [114]. The typical frequency separation between $\nu_1$ and $\nu_3$ is typically less than 100 $cm^{-1}$ and $\nu_3$ may appear as a shoulder on $\nu_1$. $\nu_1$ and $\nu_3$ are separated from the symmetric and asymmetric stretching bending modes, $\nu_2$ and $\nu_4$ respectively, which typically occur between 300 and 400 $cm^{-1}$. Between the bending modes region, 300-400 $cm^{-1}$, and the stretching mode region, 750-850 $cm^{-1}$, the vibrational spectra are silent (see **Fig. 5**). The vibrational spectra of iodate materials may be complicated by the occurrence of the $\nu_2$ overtone, $2\nu_2$, which occurs in the higher energy stretching frequency region 750-850 $cm^{-1}$ and is amplified due to Fermi resonance with the $\nu_1$ mode which has similar frequency and the same symmetry. Therefore, even in the idealized case of symmetrical trigonal pyramidal $[IO_3]^-$ iodate ion it is possible to observe five peaks in the experimental spectra.

(c) In iodate materials, the environment of the $[IO_3]^-$ ion can deviate substantially from the idealized case, thereby leading to marked differences in the vibrational spectra such as mode splitting. For example, Dasent *et al.* studied the vibrational characteristics of the $[IO_3]^-$ ion when covalently bonded to another chemical species (see **Fig. 3c**), showing that the iodate symmetry is reduced from $C_{3v}$ to $Cs$, resulting in a lifting of the degeneracy of the antisymmetric stretching and bending modes $\nu_3$ and $\nu_4$ (see row (c) of **Table 2**) [115]. In this case, for example, the doubly degenerate antisymmetric I-O stretch, $\nu_3$, splits into $\nu_{3a}$ and $\nu_{3b}$, which respectively correspond to the symmetric and antisymmetric $IO_2$ stretches and are still found in the high frequency stretching region between 750 and 850 $cm^{-1}$.

(d) Furthermore, in iodate materials the three I-O covalent bonds may not necessarily have the same bond length (see **Fig. 3d**). For example, in our recent powder XRD study on $Fe(IO_3)_3$, the three shortest I-O bond distance were found to be 1.88(3), 1.924(17) and 1.943(19) Å at ambient conditions [29]. The observation of different I-O bond lengths is consistent throughout the literature [30,54,57]. In this geometrical configuration shown in **Fig. 3d**, the point group of the iodate $[IO_3]^-$ ion is $C_s$, which is same as previously discussed for the case of the iodate $[IO_3]^-$ ion covalently bonded to another atom via one of its oxygen atoms. Therefore, (as shown in row (d) of **Table 2**) a lifting of the degeneracy of the antisymmetric stretching and bending modes $\nu_3$ and $\nu_4$ and a peak splitting the vibrational spectrum would be expected.

(e) As described in **Section 8.2** about the oxygen coordination increase of iodine in iodate materials, the iodate atoms of the iodate $[IO_3]^-$ units tend be coordinated firstly by three first nearest neighbor oxygen atoms, and then by three additional next nearest neighbour oxygen atoms (see **Fig. 3e**). For example, in our $Fe(IO_3)_3$ study the three next nearest oxygen atoms showed I···O halogen bond distances of: 2.574(18), 2.800(20) and 2.932(18) Å. These atoms are probably not negligible since the sum of the van der Waals radio of iodine and oxygen is around 3.5 Å [116]. The addition of these three far away oxygen atoms to the coordination sphere of the iodine atom means that the coordination can be thought of as that of approaching octahedral configuration. Indeed, as discussed in **Section 8.2**, with increasing pressure the weak I···O halogen bonds decrease in length as the oxygen atoms approach of central iodine atom. Additionally, the initial shorter I-O covalent bonds increase in length at the same time, meaning a pressure-induced symmetrization of the octahedral coordination is observed which can make the experimental vibrational spectrum extremely complex wherein the vibrational modes cannot easily be unambiguously assigned due to the mixed character of the system. The vibrational modes for a symmetrical octahedral coordination are shown in row (e) of **Table 2**.

(f) Another cause of complexity in the vibrational spectra is the fact that not all $[IO_3]^-$ units in the crystallographic unit cell are equivalent (see **Fig. 3f**). Therefore, it may be that case that, for example, the I-O bond distances in one $[IO_3]^-$ unit may not be equivalent to those in a different $[IO_3]^-$ unit within the same crystal structure. Equivalently, one $[IO_3]^-$ unit may be covalently bonded one other atom whereas a different $[IO_3]^-$ unit may be bonded to two. Even two identical $[IO_3]^-$ units located in different crystal fields would be expected to exhibit different vibrational mode energies (as shown in row (f) of **Table 2**).

The above discussion considered several different $[IO_3]^-$ configurations independently and briefly discussed the resulting vibrational modes. In reality, the $[IO_3]^-$ units in iodate materials may exhibit all of the above characteristics, albeit to varying degrees. Additionally, so far, only the internal stretching and bending modes have been discussed. The vibrational spectra are complicated further still by external modes, (also called lattice modes,) which arise from vibrational coupling between individual $[IO_3]^-$ units. The external modes can further be classified into rotational and translational modes, and in iodate materials they are typically observed below 200 $cm^{-1}$. Therefore, the vibrational spectra can be divided into three separate regions, as shown for the example iodate $Mg(IO_3)_2$ in **Fig. 4** [30]. Firstly, the external mode region below 200 $cm^{-1}$; secondly, the bending mode region between 300 and 450 $cm^{-1}$; and finally, the stretching mode region between 750 and 850 $cm^{-1}$. A total of 22 Raman modes are observed in the $Mg(IO_3)_2$ spectrum shown in **Fig. 4**. To illustrate the fact that the underlying iodate crystal structure does not greatly affect the vibrational spectra, as long as the crystal structure contains iodate $[IO_3]^-$ units, Raman spectra from seven different iodates are shown in **Fig. 5**, where it can be seen that the external, bending and stretching modes are located in the same frequency region regardless of crystal structure.

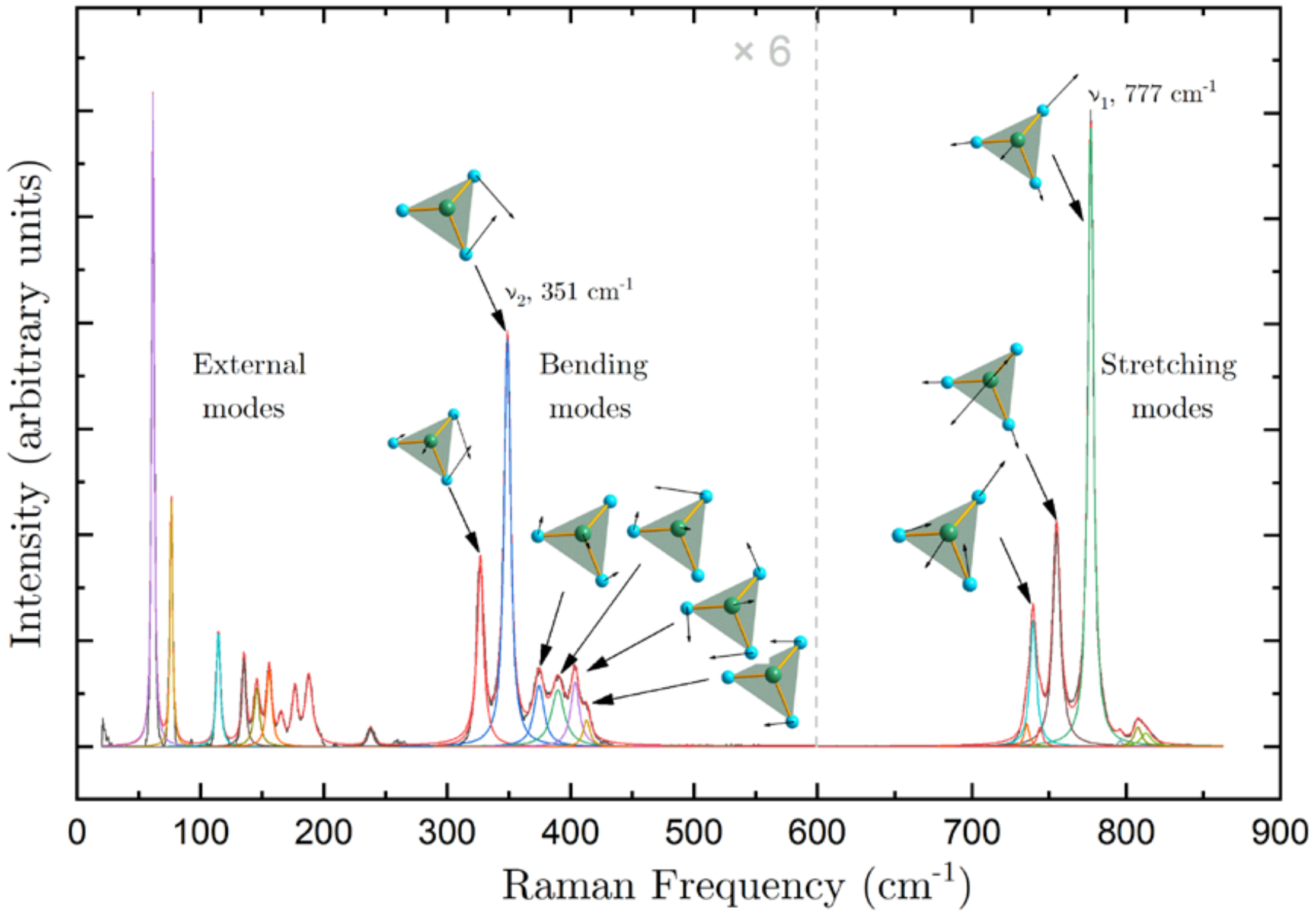

***Fig. 4.*** *Raman spectrum of $Mg(IO_3)_2$ at close to ambient conditions (0.2 GPa). The experimental data (grey) is fitted with Voight profiles (various colours). As indicated in the figure, the intensity of the data the lower frequency region between 0-600 $cm^{-1}$ has been multiplied by a factor of 6 to improve clarity.*

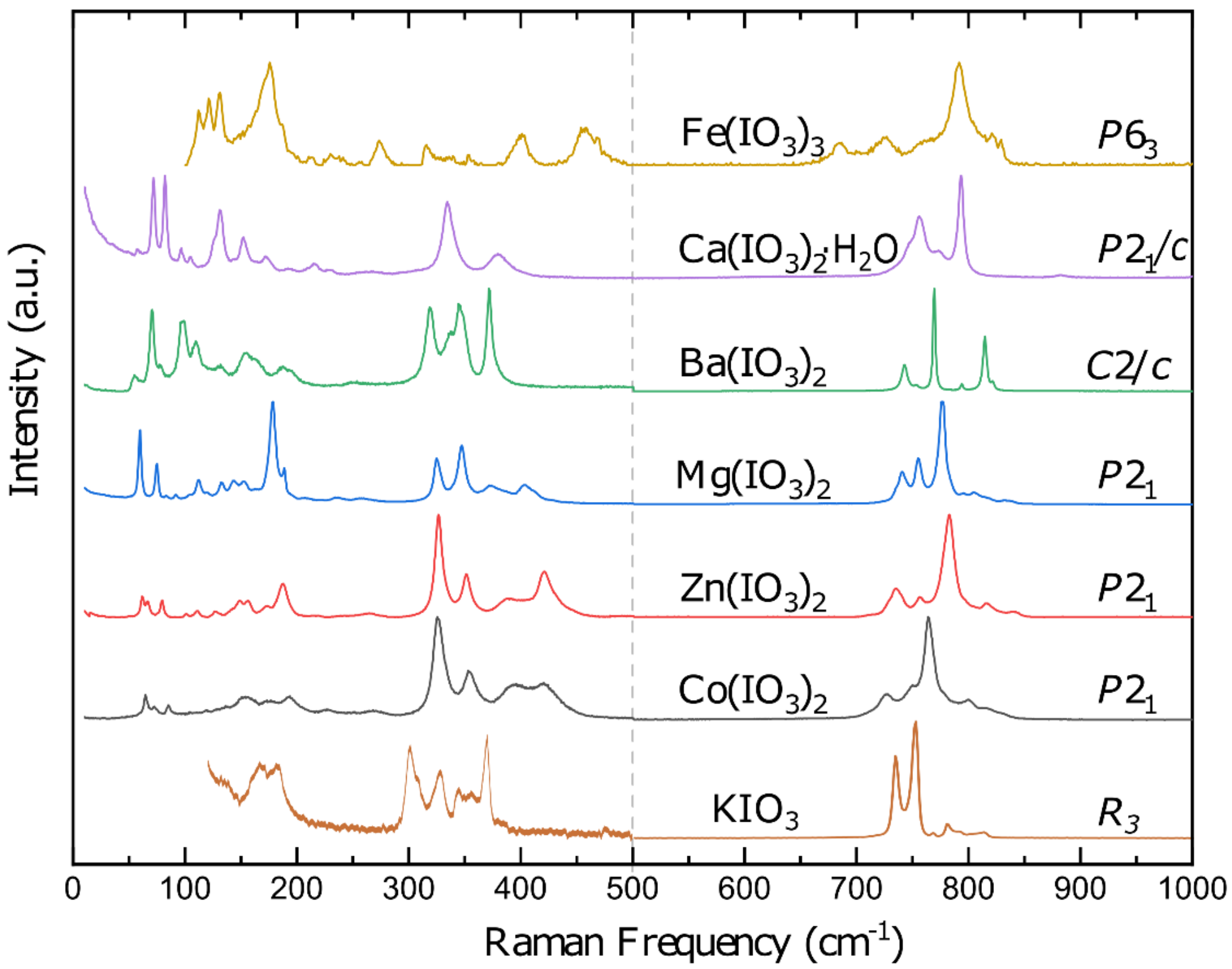

***Fig. 5*** *Representative ambient pressure Raman spectra from seven different iodate materials. The intensity of the data below 500 cm$^{-1}$ has been normalized to match that of the most intense stretching mode between 750-800 cm$^{-1}$. The space group of the iodate crystal structure is written at the right hand side of the figure. The data for $KIO_3$ are taken from Ref.* [28].

# 5. Bandgap energy of metal iodates at ambient conditions

A wide bandgap is one of the essential requirements for metal iodates to be used as NLO materials. Such materials can be used to modify the output frequency of lasers via several different processes, including: second-harmonic generation (SHG); different frequency generation (DFG); or sum frequency generation (SFG). Most of the metal iodates have a large SHG response and can employed to double the frequency (energy) of an input/pump laser. In order to optimize this process, the energy of the input laser must be lower than the bandgap, so that the photons are not absorbed promoting electrons from the valence band maximum (VBM) to the conduction band minimum (CBM) in the electronic band structure. To achieve this optimization, it is desirable to engineer the iodate bandgap to be as large as possible to prevent absorption of light from the input or output laser.

The most commonly used method for measuring the bandgap energy of metal iodates is the measurement of the optical-absorption spectrum or absorbance [117–122]. The optical-absorption spectra of several metal iodates collected at ambient conditions are plotted in **Fig. 6**. The absorption edge of metal iodates is distributed in a wide range of energies from 2.0 eV to 5.0 eV. The bandgap energy of different metal iodates can be obtained from the Tauc plot of the absorption spectrum [123], in which the linear fit of the high-energy part of the spectrum is extrapolated to zero energy. The Tauc plot shows $(\alpha h\upsilon)^x$ vs $h\upsilon$, where x = ½ if the bandgap is indirect, and x = 2 if the bandgap is direct. In the previous expression $\alpha$, $h$, and $\upsilon$ are absorption coefficient, Plank constant, and photon frequency, respectively. In addition to the absorption edge caused by the fundamental band gap, the optical absorption spectra of $Co(IO_3)_2$ also exhibits absorption peaks at the energy range from 2.0 to 2.7 eV, which are associated with the internal *d-d* electron transitions. This part will be explained in detail in **Section 6.3.2** in this review.

The experimental and calculated bandgaps of most the reported metal iodates at ambient conditions are summarized in **Table 3**, as well as the bandgap nature (direct or indirect), and dominating features in the VBM and CBM in the electronic band structures which have been determined from theoretical calculations. In **Table 3**, most of the metal iodates are of the $[IO_3]^-$ type, however other compounds containing iodine oxide molecules are included, such as $I_2O_5$ [64], $[I_3O_8]^-$ [65], and $[I_5O_{14}]^{3-}$ [124].

There are some common features can be established from **Table 3**. The first feature is that the majority of the metal iodates which exhibit an indirect bandgap also crystallize in low-symmetry crystal structures as summarized in **Table 1** and **Fig. 2**.

The second feature is that the experimentally determined bandgaps of the metal iodates are reported in a very wide range. The metal iodate with the lowest bandgap (2.1 eV) reported at ambient conditions is $Fe(IO_3)_3$ [53], while the metal iodate with the highest bandgap (5.1 eV) reported at ambient conditions is $Rb_3(IO_3)_3(I_2O_5)(HIO_3)_4(H_2O)$ [64], which crystallizes in the centrosymmetric $P2_1/c$ (No. 14) space group [64]. There are discrepancies between the experiments and the calculated bandgap energy for most metal iodate. In general, calculations underestimate the bandgap energy. This is related to the fact that most computer simulations calculate the band structure using the generalized gradient approximation (GGA) with the Perdew-Burke-Ernzerhof (PBE) for solids prescription (PBEsol) functionals [125]. On the other hand, there are also a few cases where the theoretically calculated bandgap overestimates the bandgap determined from experiments. A few examples of them are $Fe(IO_3)_3$ [53], $Mg(IO_3)_2$ [56], and $RbIO_2F_2$ [76]. The reasons related to these overestimations of the bandgap energy are still under debate [126] and out of the scope of this paper, so it will not be discussed here.

Thirdly, in the electronic band structure, the VBM is dominated by the O-2*p* orbitals for the metal iodates which only contain non-transition metals or closed-shell transition metals, while the CBM for metal iodates is usually dominated by the I-5*p* states or have an almost equal contribution from I-5*p* and O-2*p* orbitals. This is the case for instance of $Mg(IO_3)_2$ and $Zn(IO_3)_2$ [56]. On the other hand, the partially filled *d* orbitals for the transition metal will usually contribute to either the VBM or the CBM. For instance, the VBM of both α- and β-$AgI_3O_8$ is dominated by O-2*p* and Ag-4*d* orbitals [65], the CBM of α-$NaAu(IO_3)_4$, $RbAu(IO_3)_4$ and α-$CsAu(IO_3)_4$ are dominated by Au-5*d* orbitals [73], and the CBM of $Li_2Ti(IO_3)_6$ are dominated by Ti-3*d* orbitals [10].

Fourthly, when the chemical formulae are similar, the bandgaps of the metal iodates containing the partially filled *d* orbitals tend to be smaller than those of metal iodates which only contain non-transition or closed-shell transition metals. This is caused by the fact that the partially filled *d* orbitals will contribute to either the VBM or the CBM and then narrow the bandgap energy. For example, both the experimentally and theoretically determined bandgap energy of $NaI_3O_8$ is higher than that in α- and β-$AgI_3O_8$ [65]. The bandgap energy of $NaBi(IO_3)_4$ [79] is 1.0 eV higher than that of $RbAu(IO_3)_4$ [73] in experiments, and the bandgap energy of $Na_2Ge(IO_3)_6$ [127] is 1.3 eV higher than that of $Na_2Ti(IO_3)_6$ [62] in experiments.

By analyzing the theoretical calculated density of state (DOS), projected density of state (PDOS), and crystal orbital overlap population (COOP) or crystal orbital Hamilton populations (COHP), it is possible to determine the molecular orbital diagram (MO) for solids [128]. For non-transition and closed-shell transition metal iodates, the MO diagram is shown in **Fig. 7**. Although there are some minor changes from one to one, the overall features are the same, as well as the MO diagram around the Fermi level [56,129]. For example, in the MO diagram of $Ca(IO_3)_2{\cdot}H_2O$, there is an extra non-bonding state between the bonding and anti-bonding state of the *s*-*s* interaction between iodine and oxygen and an extra Ca-3*d* state at the topmost of conduction band [129]. Therefore, the bandgap for non-transition and closed-shell transition metal iodates is determined by the non-bonding states of O-2*p* and the anti-bonding states of the *p*-*p* interaction between iodine and oxygen atoms. For the metal iodates which contain transition metals, the partially filled *d* states insert to either the VBM or CBM have strong effects on the bandgap energy of metal iodates.

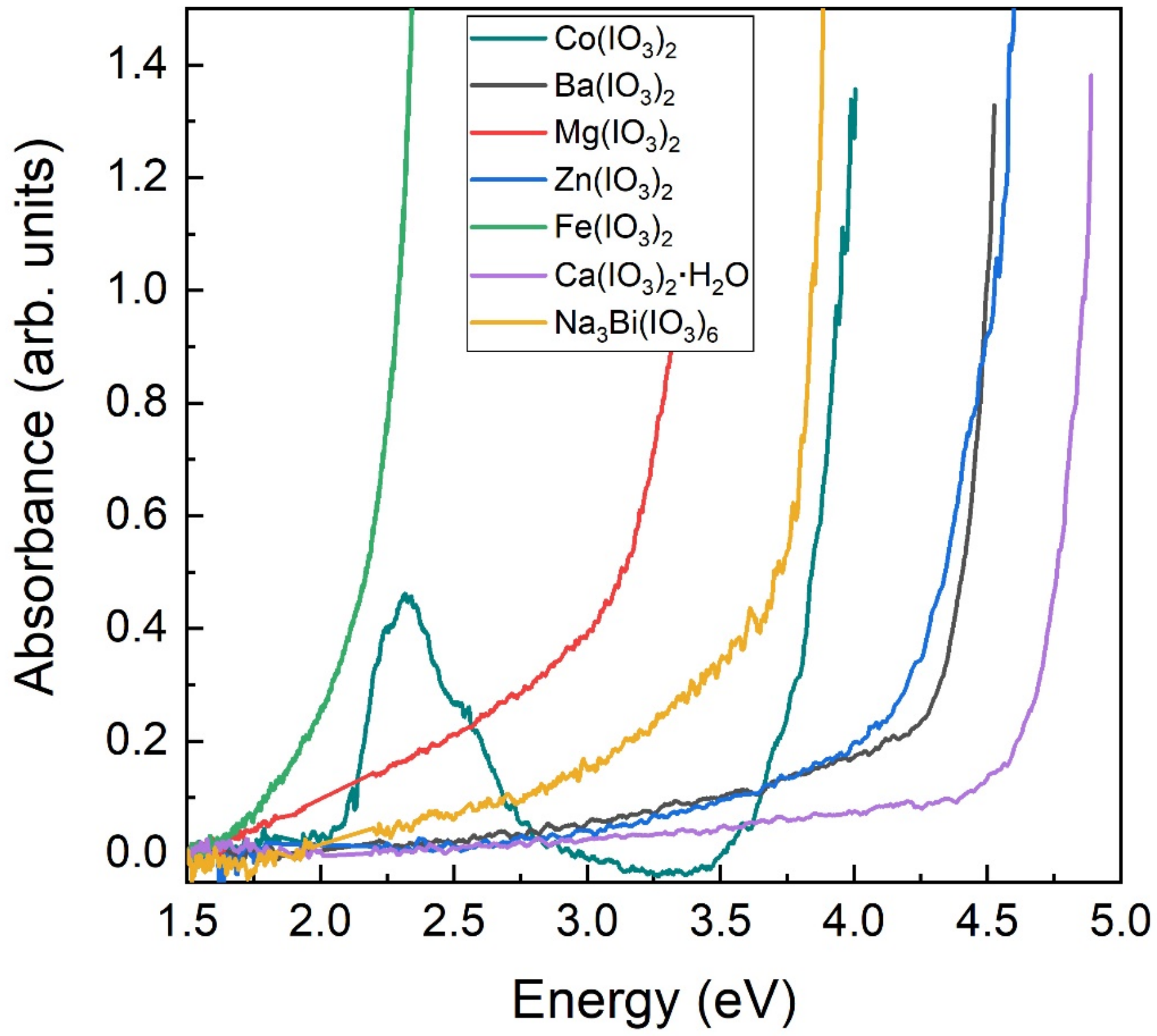

***Fig. 6.*** *Optical absorption spectra of several metal iodates collected at ambient conditions. Including* $Co(IO_3)_2$ [58], $Ba(IO_3)_2$, $Mg(IO_3)_2$, $Zn(IO_3)_2$ [56], $Fe(IO_3)_3$ [53], $Ca(IO_3)_2{\cdot}H_2O$ [129], *and* $Na_3Bi(IO_3)_6$ [60].

| Compounds | Band gap type | Band-gap energy (eV) | | Valence band maxima (VBM) | | Conduction band minima (CBM) | | Ref. |
|---|---|---|---|---|---|---|---|---|
| | | Exp. | Cal. | dominated | other | dominated | other | |
| $Fe(IO_3)_3$ | In | 2.1 | 2.50 | O(2*p*) | I(5*s*,5*p*) | Fe(3*d*), I(5*p*) | O(2*p*) | [53] |
| $Co(IO_3)_2$ | In | 3.7 | 2.26 | O(2p) Co(3d) | I(5*p*) | I(5*p*) | Co(3d) O(2p) | [58] |
| $Zn(IO_3)_2$ | In | 3.9 | 2.96 | O(2*p*) | I(5*p*) | I(5*p*), O(2*p*) | O(2*s*) | [56] |
| $Mg(IO_3)_2$ | In | 3.0 | 3.40 | O(2*p*) | I(5*p*) | I(5*p*), O(2*p*) | × | |
| $Ca(IO_3)_2 \cdot H_2O$ | In | 4.5 | 3.59 | O(2*p*) | I(5*p*) | I(5*p*), O(2*p*) | × | [129] |
| $CsIO_3$ | D | 4.2 | 3.25 | O(2*p*) | I(5*s*) | I(5*p*) | O(2*p*) | [75] |
| $CsIO_2F_2$ | In | 4.5 | 4.15 | O(2*p*) | F(2*p*) | I(5*p*) | O(2*p*) | |
| $Ce_2I_6O_{18}$ | × | 2.5 | × | O(2*p*) | × | I(5*p*) | Ce(5*d*) | [2] |
| $Sn(IO_3)_4$ | In | 4.0 | 2.75 | O(2*p*) | I(5*s*,5*p*) | I(5*p*) | Sn(5*s*) | [77] |
| $RbIO_3$ | × | 4.0 | 2.79 | O(2*p*) | × | I(5*p*), O(2*p*) | × | [76] |
| $RbIO_2F_2$ | × | 4.2 | 4.3 | O(2*p*) | F(2*p*) | I(5*p*) | O(2*p*) | |
| α-$AgI_3O_8$ | In | 3.8 | 2.43 | O(2*p*), Ag(4*d*) | I(5*s*,5*p*) | I(5*p*) | O(2*p*) | [65] |
| β-$AgI_3O_8$ | In | 3.6 | 2.46 | O(2*p*), Ag(4*d*) | I(5*s*,5*p*) | I(5*p*) | O(2*p*) | |
| $NaI_3O_8$ | In | 3.9 | 3.11 | O(2*p*) | I(5*s*,5*p*) | I(5*p*) | O(2*p*) | |
| α-$LiIO_3$ | In | 4.0 | 2.81 | O(2*p*) | × | I(5*p*) O(2*p*) | × | [44,130] |
| $Tl(IO_3)_3$ | × | 3.4 | 2.20 | O(2*p*) | Tl(6*s*) | O(2*s*,2*p*), I(5*s*,5*p*) | × | [69] |
| $Tl_4(IO_3)_6$ | × | 2.9 | 1.80 | O(2*p*) | I(5*s*,5*p*) | O(2*s*,2*p*),Tl(6*s*) | × | |
| $YI_5O_{14}$ | D | 3.8 | 2.17 | O(2*p*) | I(5*s*,5*p*) | I(5*p*), O(2*p*) | × | [124] |
| $GdI_5O_{14}$ | D | 4.1 | 2.24 | O(2*p*) | I(5*s*,5*p*) | I(5*p*), O(2*p*) | × | |
| $Na_3Bi(IO_3)_6$ | In | 3.2 | 3.34 | O(2*p*) | I(5*s*,5*p*), Bi(6*s*) | I(5*p*), O(2*p*) | I(5*s*), Bi(6*p*) | [59] |
| $NaBi(IO_3)_4$ | In | 3.5 | 3.0 | O(2*p*) | I(5*s*,5*p*), Bi(6*s*) | I(5*p*) | O(2*p*) | [79] |
| $K_2BiI_5O_{15}$ | In | 3.5 | 2.51 | O(2*p*) | × | I(5*p*) | O(2*p*) | [131] |
| $Rb_2BiI_5O_{15}$ | In | 3.5 | 2.47 | O(2*p*) | × | I(5*p*) | O(2*p*) | |
| $K_8Ce_2I_{18}O_{53}$ | In | 2.3 | 1.13 | O(2*p*) | I(5*s*,5*p*) | I(5*p*) | O(2*p*) | [132] |
| α-$NaAu(IO_3)_4$ | In | 2.6 | 2.14 | O(2*p*) | Au(5*d*),I (5*s*,5*p*) | Au(5*d*) | O(2*p*), I(5*p*) | [73] |
| $RbAu(IO_3)_4$ | In | 2.5 | 2.31 | O(2*p*) | Au(5*d*),I | Au(5*d*) | O(2*p*), | |
| | | | | | (5*s*,5*p*) | | I(5*p*) | |
| α-$CsAu(IO_3)_4$ | In | 2.6 | 2.30 | O(2*p*) | Au(5*d*),I (5*s*,5*p*) | Au(5*d*) | O(2*p*), I(5*p*) | |
| $AgGa(IO_3)_4$ | × | 4.0 | 2.91 | O(2*p*), Ag(4*d*) | × | I(5*p*) | O(2*p*) | [86] |
| $Ag_3In(IO_3)_6$ | × | 3.8 | 2.40 | | × | I(5*p*) | O(2*p*) | |
| $Cs_2Sn(IO_3)_6$ | × | 4.1 | × | × | × | × | × | [89] |
| $Li_2Sn(IO_3)_6$ | × | 3.9 | × | × | × | × | × | |
| $K_2Sn(IO_3)_6$ | × | 4.0 | × | × | × | × | × | |
| $Na_2Sn(IO_3)_6$ | × | 4.0 | × | × | × | × | × | |
| $Rb_2Sn(IO_3)_6$ | × | 4.1 | × | × | × | × | × | |
| $LiMg(IO_3)_3$ | In | 4.3 | 3.35 | O(2*p*) | I(5*s*, 5*p*) | I(5*p*) | O(2*p*) | [100] |
| $Li_2Ti(IO_3)_6$ | × | 3.0 | 1.60 | O(2*p*) | I(5*s*, 5*p*) | Ti(3*d*) | O(2*s*,2*p*) | [10] |
| $Na_2Ti(IO_3)_6$ | × | 3.3 | 2.95 | O(2*p*) | × | Ti(3*d*), I(5*p*) | O(2*s*,2*p*) | [62] |
| $Rb_2Ti(IO_3)_6$ | × | 3.3 | × | × | × | × | × | |
| $Cs_2Ti(IO_3)_6$ | × | 3.2 | × | × | × | × | × | |
| $Tl_2Ti(IO_3)_6$ | × | 3.2 | 2.88 | O(2*p*) | × | Ti(3*d*), I(5*p*) | O(2*s*,2*p*) | |
| $K_2Ti(IO_3)_6$ | × | 3.3 | 3.14 | O(2*p*) | × | Ti(3*d*), I(5*p*) | O(2*s*,2*p*) | |
| $KLi_2(IO_3)_3$ | × | 4.3 | × | × | × | × | × | [99] |
| $Li_2Ge(IO_3)_6$ | In | 3.9 | 3.18 | O(2*p*) | I(5*s*, 5*p*) | I(5*p*) | O(2*p*), Ge(4*s*) | [127] |
| $Na_2Ge(IO_3)_6$ | In | 4.6 | 3.31 | × | × | × | × | |
| $Rb_2Ge(IO_3)_6$ | × | 4.1 | × | × | × | × | × | |
| $Cs_2Ge(IO_3)_6$ | × | 4.1 | × | × | × | × | × | |
| $Ba_3Ga_2(IO_3)_{12}$ | In | 3.1 | 3.09 | O(2*p*) | × | I(5*p*), O(2*p*) | × | [133] |
| $Bi(IO_3)F_2$ | In | 4.0 | 3.26 | O(2*p*) | Bi(6*s*), I(5*s*, 5*p*) | I(5*p*), Bi(6*p*) | O(2*p*) | [134] |
| $KBi_2(IO_3)_2F_5$ | In | 3.8 | 2.89 | O(2*p*) | F(2*p*), I(5*s*, 5*p*) | I(5*p*) | O(2*p*) | [135] |
| $RbBi_2(IO_3)_2F_5$ | In | 3.8 | 2.97 | O(2*p*) | F(2*p*), I(5*s*, 5*p*) | I(5*p*) | O(2*p*) | |
| $CsBi_2(IO_3)_2F_5$ | In | 3.8 | 2.99 | O(2*p*) | F(2*p*), I(5*s*, 5*p*) | I(5*p*) | O(2*p*) | |
| $LiZn(IO_3)_3$ | D | 4.2 | 4.4 | O(2*p*) | I(5*p*) | I(5*p*), O(2*p*) | × | [13] |
| $LiCd(IO_3)_3$ | D | 4.2 | 4.3 | O(2*p*) | I(5*p*) | I(5*p*), O(2*p*) | × | |
| $SrSn(IO_3)_6$ | In | 4.1 | 3.19 | O(2*p*) | I(5*p*) | I(5*p*) | O(2*p*) | [95] |
| $Ag_2Zr(IO_3)_6$ | In | 3.8 | 2.64 | O(2*p*), Ag(4*d*) | I(5*s*, 5*p*) | I(5*p*), O(2*p*) | × | [97] |
| $LaZr(IO_3)_5F_2$ | In | 4.1 | 3.29 | O(2*p*) | I(5*s*, 5*p*) | I(5*p*), Zr(4*d*) | O(2*p*) | |
| $BiO(IO_3)$ | × | 3.3 | 2.0 | O(2*p*) | I(5*s*, 5*p*) | I(5*p*), O(2*p*) | × | [70] |
| $BaNbO(IO_3)_5$ | D | 3.6 | 2.55 | O(2*p*) | I(5*p*) | Nb(4*d*), I(5*p*) | × | [11] |
| $Zn_2(VO_4)(IO_3)$ | In | 3.3 | 2.7 | O(2*p*) | V(3*d*), Zn(3*d*) | V(3*d*), I(5*p*), O(2*p*) | × | [136] |
| $LaVO(IO_3)_5$ | In | 3.6 | 1.7 | O(2*p*) | I(5*s*, 5*p*) | V(3*d*), | × | [84] |
| | | | | | | La(5*d*), I(5*p*) | | |
| $LaV_2O_6(IO_3)$ | In | 3.6 | 2.54 | O(2*p*) | I(5*s*, 5*p*) | V(3*d*), I(5*p*) | O(2*p*) | |
| $Rb_3(IO_3)_3(I_2O_5)(HIO_3)_4(H_2O)$ | × | 5.1 | × | × | × | × | × | [64] |

***Table 3***. *A summary of the experimental (Exp.) and theoretical calculated (Cal.) bandgap energies of metal iodates reported at ambient conditions, as well as the bandgap nature, and the dominated feature in the valence band maxima (VBM) and conduction band minima (CBM) in the electronic band structure. Most of them are $[IO_3]^-$ type metal iodates, it also includes some of the $[I_3O_8]^-$, $[I_3O_9]^{3-}$, $[I_5O_{14}]^{3-}$ type metal iodates. for the bandgap nature, "In" and "D" means indirect and direct bandgap, respectively. The blank cells filled with "×" means no available data was found in the literature.*

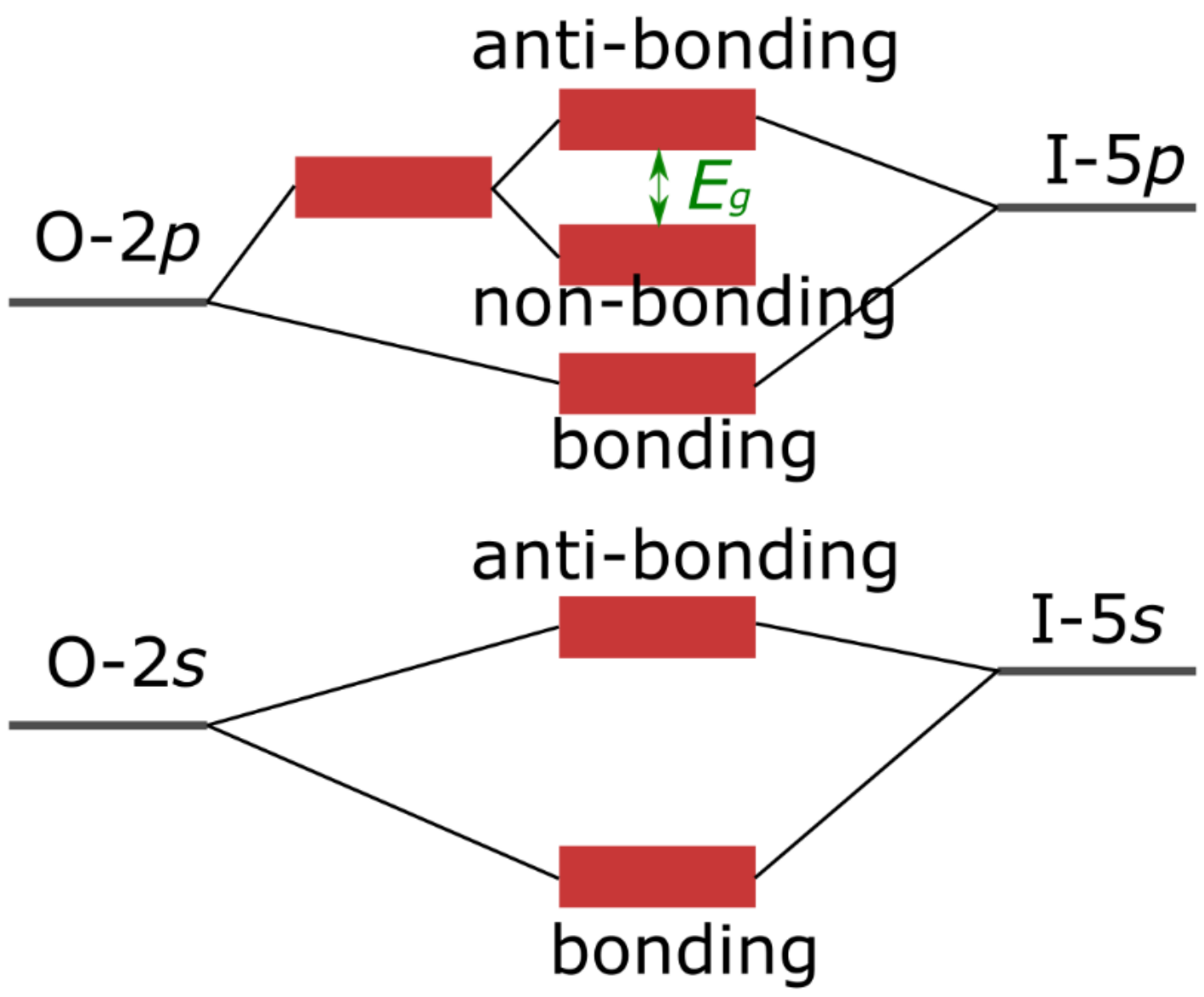

***Fig. 7***. *Molecular orbital diagram for metal iodates containing only non-transition or close-shell transition metal.*

According to the MO diagram established for non-transition and closed-shell transition metal iodates (**Fig. 7**), the bandgap energy for these kinds of metal iodates is dominated by the distance change between iodine and oxygen. If the I-O bond distance decreases, the *p*-*p* interaction will increase and enlarge the separation between bonding and anti-bonding states, in that way opens the bandgap. In the opposite case, if the I-O bond distance increase, then it will narrow the bandgap. In other words, there is an inverse proportionality between the I-O bond distance and the bandgap energy of non-transition or closed-shell transition metal iodates. The bandgap energy *vs.* average I-O bond distance of 71 different metal iodates is represented in **Fig. 8**. **Fig. 8** contains only compounds which contain non-transition metal or closed-shell transition metals, regardless of the chemical formula or crystal structure. The full list of the metal iodates for each point in **Fig. 8** can be found in Ref. [56]. Only I-O bonds shorter than 2.0 Å were considered because calculations have shown that the interaction is weak if the bond distance is longer than 2.0 Å [56]. As expected, the bandgap energy for these kinds of metal iodates exhibits an inverse proportionality with the average I-O bond distance. That is to say, longer I-O bonds result in smaller bandgap energies, and vice versa. The linear fitting yields a quantitative relationship between them as:

$$E_g = (-42.1 \pm 2.4) \times \langle d_{I-O} \rangle + 80.4 \pm 4.4,$$

$$\langle d_{I-O} \rangle = \sum_{i=1}^{k} d_{I(i)-O} / k \quad \text{where} \quad d_{I(i)-O} < 2.0$$

Here $E_g$ is the bandgap energy and $d_{I\text{-}O}$ is the I-O bond distance, and the units for them are "eV" and "Å", respectively.

This result therefore enables the realization of two criteria for designing metal iodates with wide bandgap energy, which are: i) avoiding the use of partially filled transition metal, as the partially-filled *d*-state will contribute to either the VBM of CBM in the electronic band structure and narrow the bandgap energy; ii) shortening the bond distance between iodine and oxygen atoms, both the short I-O covalent bonds and long weak I···O halogen bonds.

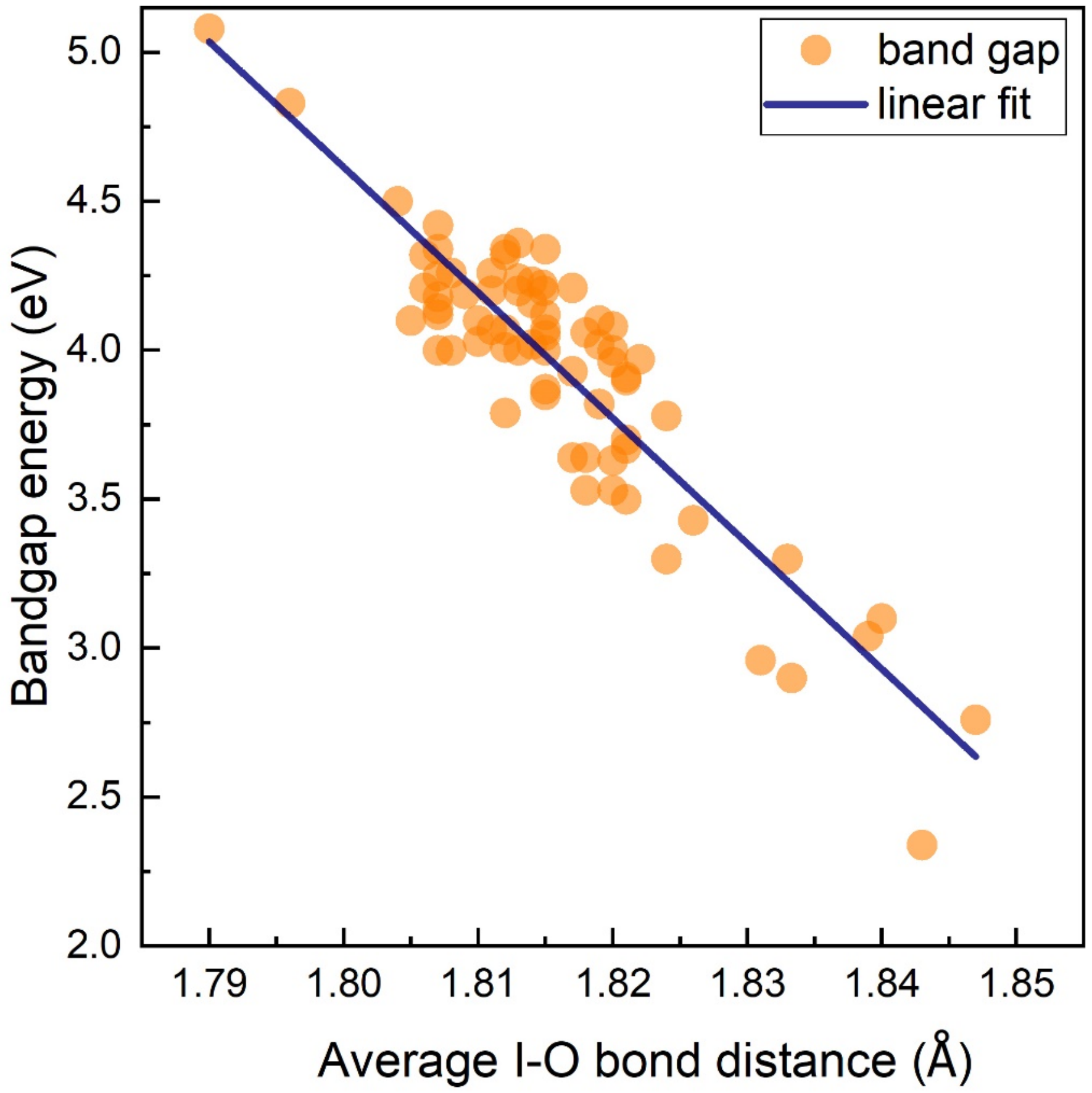

***Fig. 8.*** *Plot of the bandgap energy vs average I-O short bond distance for 71 different metal iodates reported in the literature, contain only non-transition or closed-shell transition metals. The blue line is the linear fitting of those data.*

# 6. High-pressure studies on iodates

## 6.1 X-ray diffraction experiments

The first XRD study reporting evidence of a pressure-induced phase transition was reported for $KIO_3$ [28]. Using single-crystal XRD it was found that the ambient-conditions triclinic polymorph (space group *P*1) undergoes a phase transition at 8.7(1) GPa. After the phase transition it was observed that the single crystal sample remained intact. Using synchrotron radiation and a four-circle diffractometer the crystal structure of the HP phase was solved in space group *R*3 [28]. The HP polymorph is acentric and has a strong SHG signal. The structure of this polymorph has similarities to a double-perovskite structure with the potassium and iodine atoms close to the ideal perovskite positions. As in the low-pressure phase, iodine is coordinated by three short I-O covalent bonds with a distance close to 1.8 Å forming a trigonal pyramid. There are three additional long halogen I···O bonds with distances close to 2.5 Å. The potassium atoms are in a highly distorted 12-fold cuboctahedra coordination. Powder XRD experiments have been performed up to 20 GPa [28]. A second transition has been detected at 14 GPa but the crystal structure of the second HP phase has not been solved yet. In the case of $AgIO_3$ and new polymorph has been synthesized under HP-HT conditions [47]. The HP-HT polymorph was obtained at 2.7 GPa after a first-order reconstructive transition at 260 ºC. This polymorph is quenchable as a metastable phase at ambient conditions. The crystal structure of the HP phase was determined by powder XRD and Rietveld refinement. It crystallizes in the *Pbca* centric orthorhombic space group, consequently the structural transformation quenches the SHG activity. *In-situ* HP XRD experiments have not been performed in $AgIO_3$ yet, so it is unknown if the reported HP-HT phase could be obtained at RT by applying only pressure.

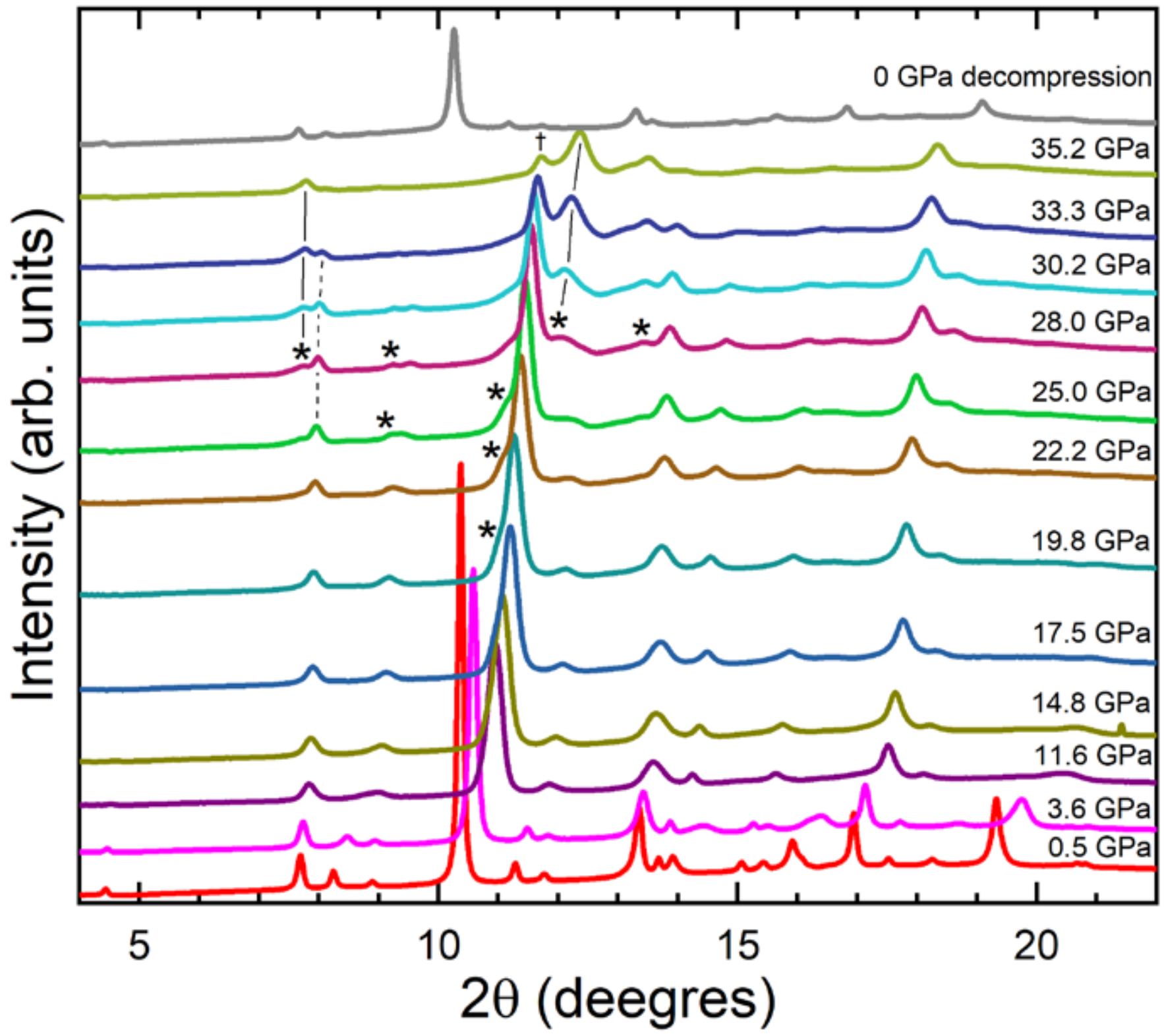

***Fig. 9.*** *Integrated XRD patterns measured in Fe(IO3)3 up to 35.3 GPa and at ambient pressure after decompression. Pressures are indicated in the figure. The asterisks identify emerging peaks assigned to the HP phase. The dagger symbol shows a peak of the low-pressure phase which can be observed at 35.2 GPa. Solid lines connect peaks of the HP phase at different pressures. The dashed lines connect peaks of the low-pressure phase at different pressures.*

An important breakthrough in the HP behavior of metal iodates came from powder XRD studies on $Fe(IO_3)_3$. *In-situ* angle dispersive powder XRD experiments up to 35.2 GPa were performed at the Shanghai Synchrotron Radiation Facility (SSRF, BL15U1 beamline) using an X-ray wavelength of 0.6199 Å [29]. A selection of XRD patterns measured in the experiment is shown in **Fig. 9**. In the figure it can be seen that the gradual appearance of extra peaks at 19.8 GPa and higher pressures, which are identified by asterisks in the figure. In parallel, the most intense peaks at low-pressure gradually lost intensity. The relative changes of intensities between the peaks identified at ambient pressure and emerging peaks can be seen in the peaks around 8º. The emerging peak (on the left) grows in intensity (see the solid line connecting peaks) while the persistent peak (on the right) becomes weaker. A similar phenomenon occurs between the strongest peak of $Fe(IO_3)_3$ – located near 10º - which can be followed up to 35.2 GPa (identified by the dagger symbol at this pressure) and the peak emerging on its right, which can be observed from 28 GPa. The described changes in the XRD patterns are a consequence of a first-order isostructural phase transition; both the low-pressure (LP) and high-pressure (HP) phases are described by the same space group, $P6_3$ [29]. Consequently, the crystal structure of $Fe(IO_3)_3$ suddenly undergoes an abrupt change of the *c/a* axial ratio and a volume collapse of nearly 5%. Interestingly, the two phases coexist over a large pressure range. The strongest peak of the low-pressure phase can be observed at the highest pressure reached by experiments (see dagger symbol in **Fig. 9**). The observed transition also involves a change in the coordination of iodine from 3-fold to 3+3-fold. All structural changes are reversible as shown by the XRD patterns measured after decompression (see top trace of **Fig. 9**). The XRD studies in $Fe(IO_3)_3$ also allow for the determination of the pressure dependence of unit-cell parameters and volume. Changes in the pressure dependence of unit-cell parameters and bond distances have been interpreted as a consequence of subtle symmetry preserving phase transition which take place at near 2 and 6 GPa [52]. Such changes affect the behavior of phonons and electronic properties of $Fe(IO_3)_3$ as shows later in **Sections 6.2** and **6.3**. Subsequent XRD measurements performed in $Co(IO_3)_2$ and $Zn(IO_3)_3$ lead to the identification of subtle symmetry preserving phase transitions [54,57,59]. These transitions were always observed below 10 GPa. Such pressure-induced isosymmetric phase transitions are characterized by increasing oxygen coordination of the iodine atoms and the probable formation of pressure-induced metavalent I-O bonds with distances smaller than 2.48 Å. These are bonds with exceptional bonding characteristics which cannot be categorized as purely covalent, purely metallic nor as intermediate between the two [137]. The existence of these transitions is a direct consequence of the existence of a stereochemically active LEP associated to the iodate ion, $[IO_3]^-$ [54,57]. A detailed analysis of the transformation from halogen I···O bond to metavalent bond is given in **Section 8.2**.

The structural behavior of $Mg(IO_3)_2$ under compression has been also studied by synchrotron-based HP powder XRD. In this material, experiments have been carried out up to 22 GPa at ALBA synchrotron using a wavelength of 0.4642 Å [30]. In this case, unambiguous evidence of a phase transition taking place at 9.7 GPa has been found. The phase transition enhances the symmetry of the crystal. The low-pressure (LP) phase is monoclinic (space group $P2_1$) and the HP phase has a trigonal (space group $P3$) structure. Both the LP and HP phases are non-centrosymmetric. In particular, the space groups of both phases belong to chiral groups. This means that the HP phase should also have SHG activity. **Fig. 10** shows XRD patterns measured for the LP and HP phases of $Mg(IO_3)_2$ with the respective Rietveld refinements. There are evident differences between XRD patterns of the two phases. In particular, the number of peaks in the HP phase is smaller to that in the LP phase. This supports the symmetry increase at the phase transition. In fact, the $P2_1$ space group predicts multiple reflections that are not present in the XRD patterns of the HP phase, as illustrated in the pattern measured at 10.5 GPa shown in **Fig. 10**. The reported phase transition is reversible and has been further confirmed by Raman, infrared and optical absorption experiments [30,56]. The reported transition involves an increase of the oxygen-iodine coordination from 3 to 6. The bond formation under compression is a consequence of the existence of LEP on the iodine cation.

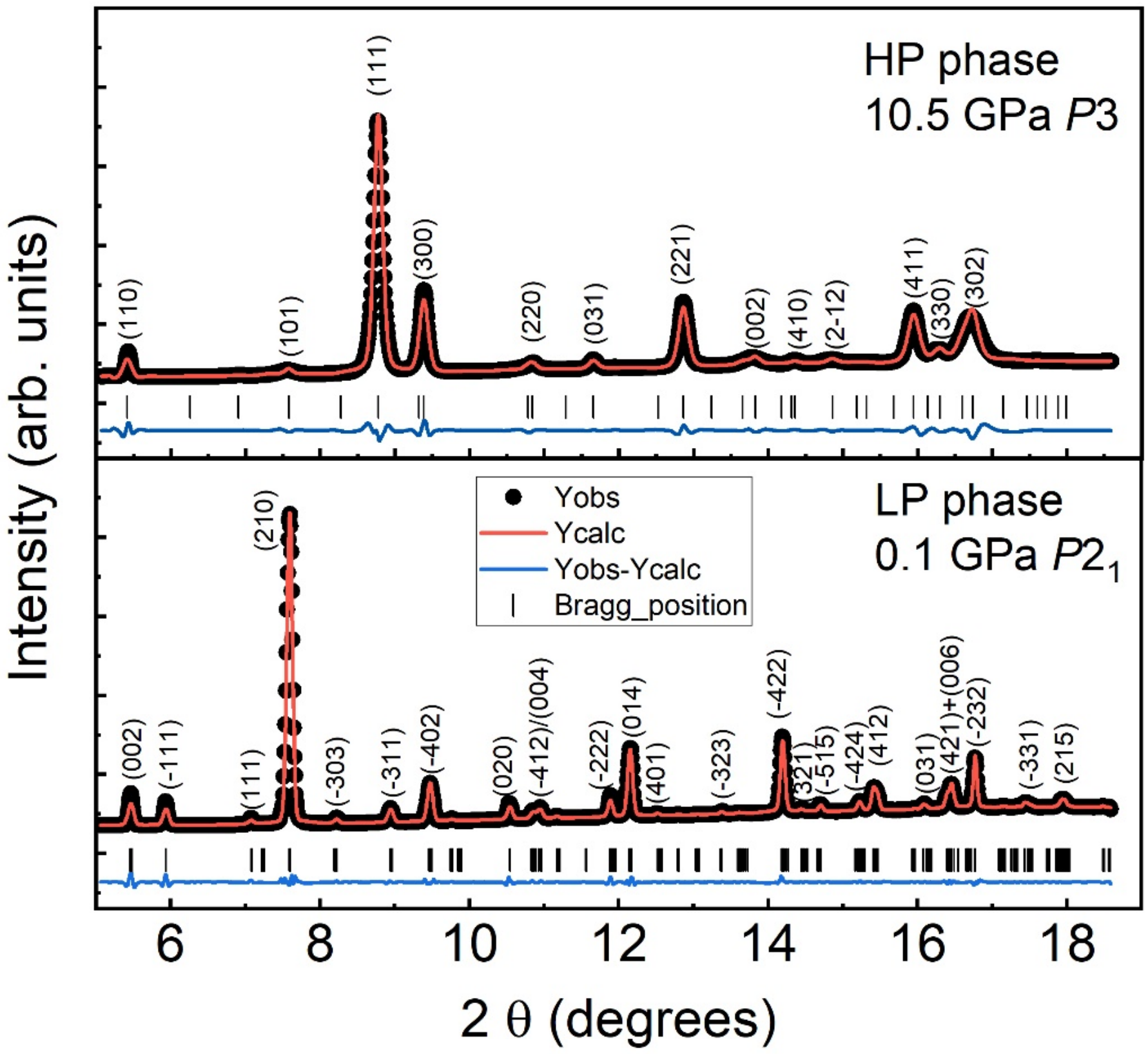

***Fig. 10.*** *XRD patterns measured in $Mg(IO_3)_3$ at 0.1 GPa (LP phase) and 10.5 GPa (HP phase). Experiments are shown with black circles and Rietveld refinements with orange solid lines. The blue solid lines are the residuals of the refinements and ticks show the positions of diffraction peaks. Most intense peaks are labeled with the corresponding Miller indices. The goodness-of-fit factors are $R_p$ = 9.23% and $R_{wp}$ = 8.85% for the LP phase and $R_p$ = 10.6% and $R_{wp}$ = 10.1% for the HP phase.*

The previously described XRD studies on metal iodates have triggered subsequent studies in more complex iodates. In 2022, two studies including XRD measurements have been published. In one study, $BiOIO_3$ was investigated up to 20.8 GPa [31]. As observed in studies of other iodates, an isostructural phase transitions occurs in $BiOIO_3$ which takes place at around 5 GPa. The transition induces a soft-mode behavior in the internal stretching mode of $IO_3$ and, as previously established in other iodates, it is related to the presence of the LEP and the existence of I···O halogen bonds. At 10 GPa a second phase transition occurs. The HP phase is denser than the LP phase and the transition causes an abrupt reduction of the unit-cell volume. Interestingly, the second transition drives a symmetry decrease, changing the space group from non-centrosymmetric space group $Pca2_1$ to centrosymmetric space group $P2/n$, to the suppression of the SHG activity. In the second study, $Li_2Ti(IO_3)_6$ was studied up to 40 GPa [32]. No phase transition has been claimed for this compound. However, a close examination of the reported results suggests that two isostructural phase transitions occur at 8 and 19 GPa. This quaternary iodate also shows a highly non-isotropic response to compression similar to that observed in the majority of metal iodates. It has been shown that $Li_2Ti(IO_3)_6$ has a zero-linear and zero-area compressibility over a wide range of pressure. Based on this it has been proposed that iodates can be used as shock-resistant materials in technological applications [32].

## 6.2 Raman and infrared spectroscopy measurements

The HP evolution of the vibrational spectra of most iodate materials exhibits two characteristic features. These are, firstly, a pressure-induced softening of the internal stretching modes of $[IO_3]^-$, and secondly, a non-linear pressure response of the vibrational modes in general. Both features are discussed in this section.

Regarding the softening of the stretching modes (the modes located in the frequency region from 600 to 900 $cm^{-1}$ in Raman spectra, **Figs. 4 and 11**), it should be noted that the archetypal response of vibrational modes under pressure is to 'harden', that is, to increase in frequency as pressure is increased due to compression of chemical bonds. Therefore, the pressure induced softening in the HP vibrational spectra of iodates is somewhat counterintuitive. This observation is explained by the fact that the initially short I-O bonds in the iodate crystal structures increase in length as pressure is increased (see **Section 8.2**). This result agrees with XRD studies and with DFT calculations. This key characteristic of iodates, the pressure-induced mode softening, leads to most observations regarding iodate physical properties, for example: phase transitions, changes in bandgap energy, changes in compressibility and changes in vibrational spectra.

This characteristic softening of the stretching vibrational modes has been observed in at least the following iodates: $KIO_3$ [28], $BiOIO_3$ [31], $HIO_3$ [61], α-$LiIO_3$ [39], $Li_2Ti(IO_3)_6$ [32], $Fe(IO_3)_3$ [52], $Zn(IO_3)_2$ [54], $Co(IO_3)_2$ [57] and $Mg(IO_3)_2$ [30]. To provide a representative example of the pressure-induced evolution of the iodate vibrational spectra, HP Raman spectra from $Mg(IO_3)_2$ are shown in **Fig. 11** [30].

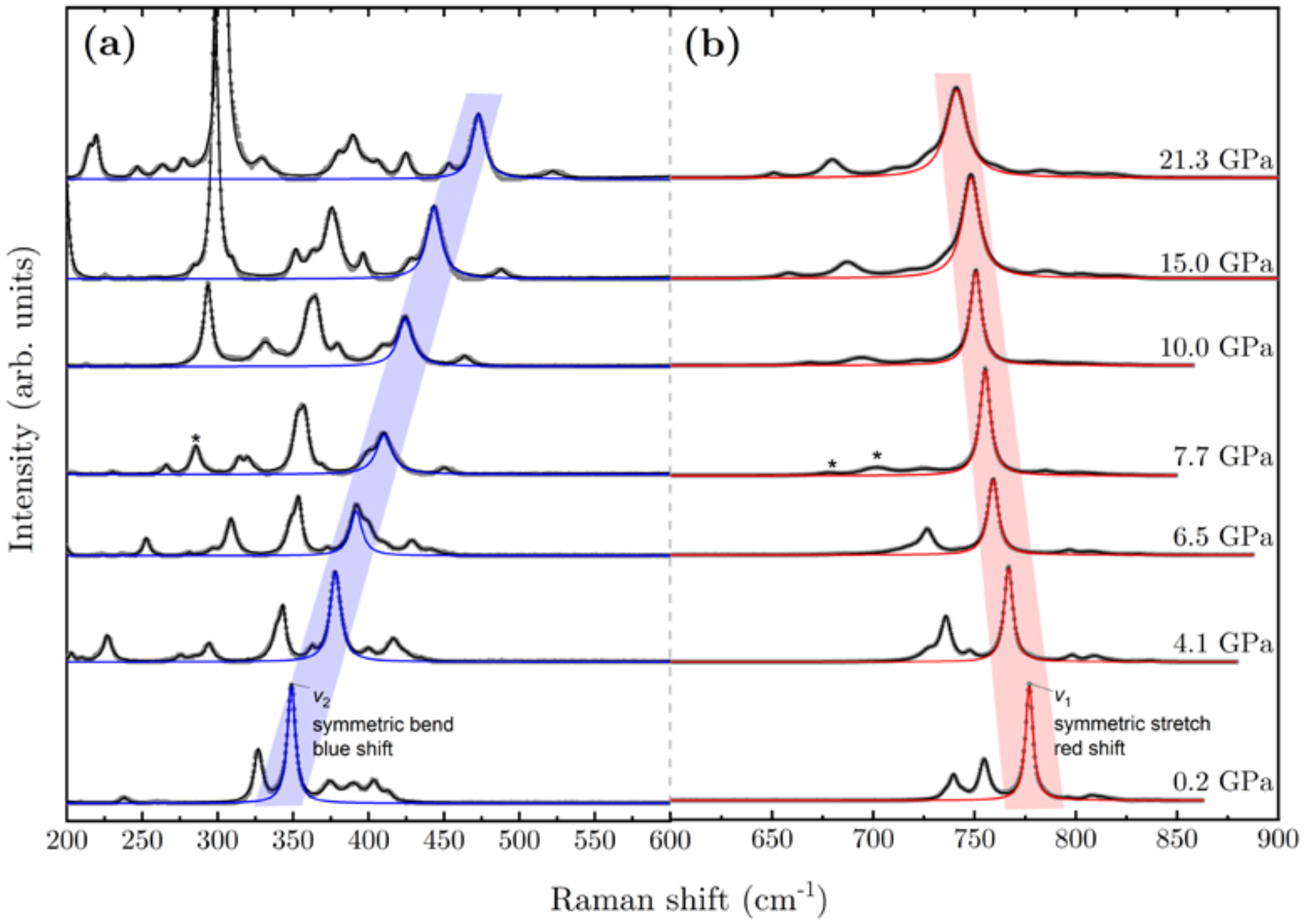

***Fig. 11.*** *Pressure evolution of the Raman spectra of $Mg(IO_3)_2$.* ***(a)*** *The mid-frequency I-O bending mode region.* ***(b)*** *The high frequency I-O stretching mode region. $\nu_1$ and $\nu_2$ respectively are the symmetric stretching and symmetric bending modes. Raw data points are shown as grey symbols and the sum of all Voight peaks fitted to the data is shown as a single black line. The Voight-profile peaks fitted to $\nu_1$ and $\nu_2$ are shown in blue and red respectively. The shaded regions are a guide to the eye to clearly illustrate the red- and blue-shifting of the labelled modes. The modes marked with asterisks appear at 7.7 GPa are associated with the phase transition observed in XRD measurements.*

The lowest pressure spectrum of **Fig. 11** is the same spectrum shown in more detail in **Fig. 4**. It is immediately clear that the vibrational spectra are simplest at low pressures. Therefore, as a general observation, increasing pressure makes the vibrational spectra considerably more complex due to the splitting of degenerate modes and the appearance of new vibrational modes. In **Fig. 11**. the symmetric stretching mode, $\nu_1$, and the symmetric bending mode, $\nu_2$, have been highlighted. These two modes exhibit opposing behaviours in terms of their pressure-induced evolution. The $\nu_2$ symmetric bending mode follows the archetypical vibrational response to compression and moves to higher energies with increasing pressure - also known as blue-shifting. Conversely, the internal I-O symmetric stretching mode, $\nu_1$, exhibits the characteristic and counterintuitive shift to lower pressures – also known as red-shifting. As aforementioned, this red-shifting is due to the pressure-induced increase in length of the three short I-O covalent bonds in the initial $[IO_3]^-$ pyramids as the three next nearest neighbor oxygen atoms are pushed closer to the central iodine atom (see **Section 8.2**). As discussed in ref. [54] for the case of $Zn(IO_3)_2$ [16] the frequency, $\omega^{-2/3}$ of the internal I-O stretching modes of the $[IO_3]^-$ iodate ion, exhibits a linear dependence of the average I-O bond distance. Therefore, the pressure-induced softening of several Raman modes in high-frequency part of the spectrum is clearly caused by the enlargement of the I-O covalent bond distance under compression.

The nonlinear pressure dependence of the Raman modes also has been observed in most of the metal iodates which have been studied under HP conditions. For example, **Fig. 12** shows experimental and calculated Raman modes of $Zn(IO_3)_2$. In this compound the crystal lattice parameters *a* and *b* increase below 9 GPa and then decrease above 9 GPa (see **Section 8.3**). Across the whole range of pressure shown in **Fig. 12** the XRD results indicate that no first-order phase transition occurs and that all diffraction peaks can be explained by the same monoclinic, $P2_1$, structure [54]. Therefore, the non-linear behavior observed in the Raman modes is not rooted in any type reconstructive or displacive phase transition. The non-linearity of the vibrational modes arises due to the changing coordination of the iodine atoms and there are two very closely related yet competing effects. That is to say that, as the three short I-O covalent bonds increase in length the immediate crystal field around the iodine atom decreases. However, the short I-O bonds increase in length due to the three approaching oxygen atoms associated with the three long I···O halogen bonds. As these long I···O halogen bonds shorten, the result in an increase in the crystal field around the iodine atoms. Therefore, there is a competition between the short and long I-O bonds resulting in a crystal field which in contributed to by two factors, one growing and one shrinking, which both change at different rates. The non-linear response of the vibrational modes to increasing pressure has been clearly observed in $Fe(IO_3)_3$ [52], $Co(IO_3)_2$ [57], $KIO_3$ [28], $HIO_3$ [61], $Li_2Ti(IO_3)_6$ [32]. The non-linear response to pressure is also observed in iodates in other properties such as the bandgap energy (see **Section 6.3.1**). This nonlinear behavior observed in most of the Raman modes in metal iodates is different from the common nonlinear behavior observed in most of the compounds under pressure, the later usually could be described by a quadratic equations and caused by the anharmonic effect between atoms. like the evolution of Raman modes at around 200 $cm^{-1}$ in $SiO_2$ [138], Raman modes at round 210 $cm^{-1}$ in $AlPO_4$ [138], or the Raman modes at around 164 $cm^{-1}$ in CuBr under pressure [139].

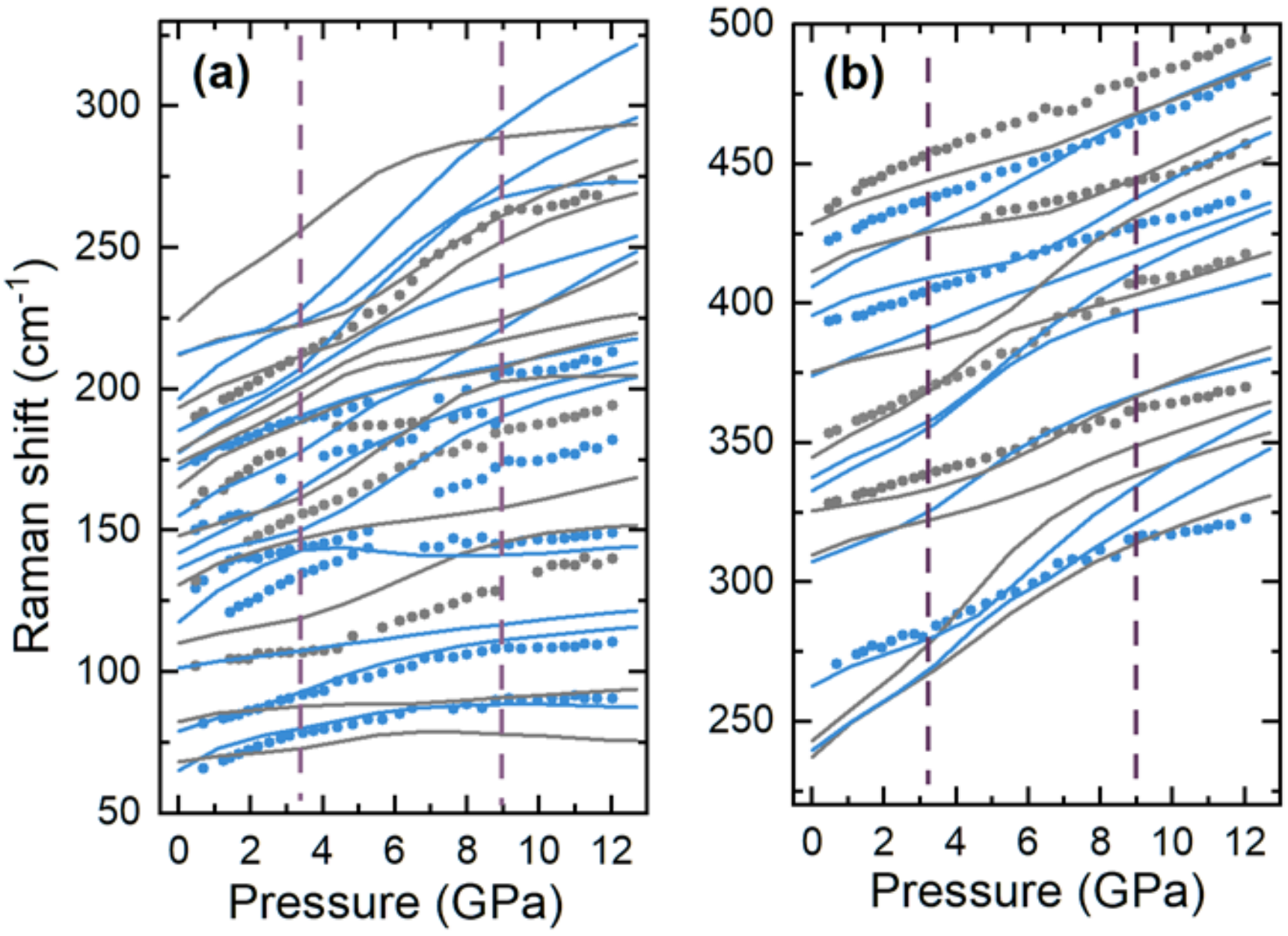

***Fig. 12.*** *Experimental (dots) and calculated (lines) Raman modes of $Zn(IO_3)_2$. (a) The low frequency external (or lattice) mode region. (b) The bending mode region. The vertical dashed lines indicate isostructural phase transitions pressure. A tentative mode assignment of the experimental data has been made based on the comparison of the modes at ambient pressure and the pressure dependence under HP conditions between experiment and calculations. The A and B modes are shown in blue and gray color, respectively.*

**Fig. 13** shows HP infrared spectra acquired from the same $Mg(IO_3)_2$ sample shown in **Fig. 11**. Notice in **Fig. 13a** that many of the infrared modes exhibit the same frequency as the Raman modes at ambient pressure because all of the vibrations modes are both Raman and Infrared-active. The IR modes also exhibit hardening with increasing pressure. **Fig. 13b** shows infrared spectra on increasing pressure, where the phase transition around 7.7 GPa is again observed via the appearance of new modes around 250 $cm^{-1}$, as indicated in the figure by the black arrows. **Fig. 13c** shows the non-linear response of the vibrational frequency to increasing pressure. For example, the pressure derivative of the frequency of the infrared mode which is initially at 250 $cm^{-1}$ changes due to the phase transition, being steeper before the transition. This characteristic non-linearity can likely be used to identify the occurrence of isostructural phase transitions in iodate materials which have yet to be studied. The symmetry of the modes in **Fig. 13c** was assigned based on comparison of the experimental data and DFT calculations. The reader is referred to Ref. [30] for further reading the symmetry assignment of the vibrational modes based on comparison with DFT calculations.

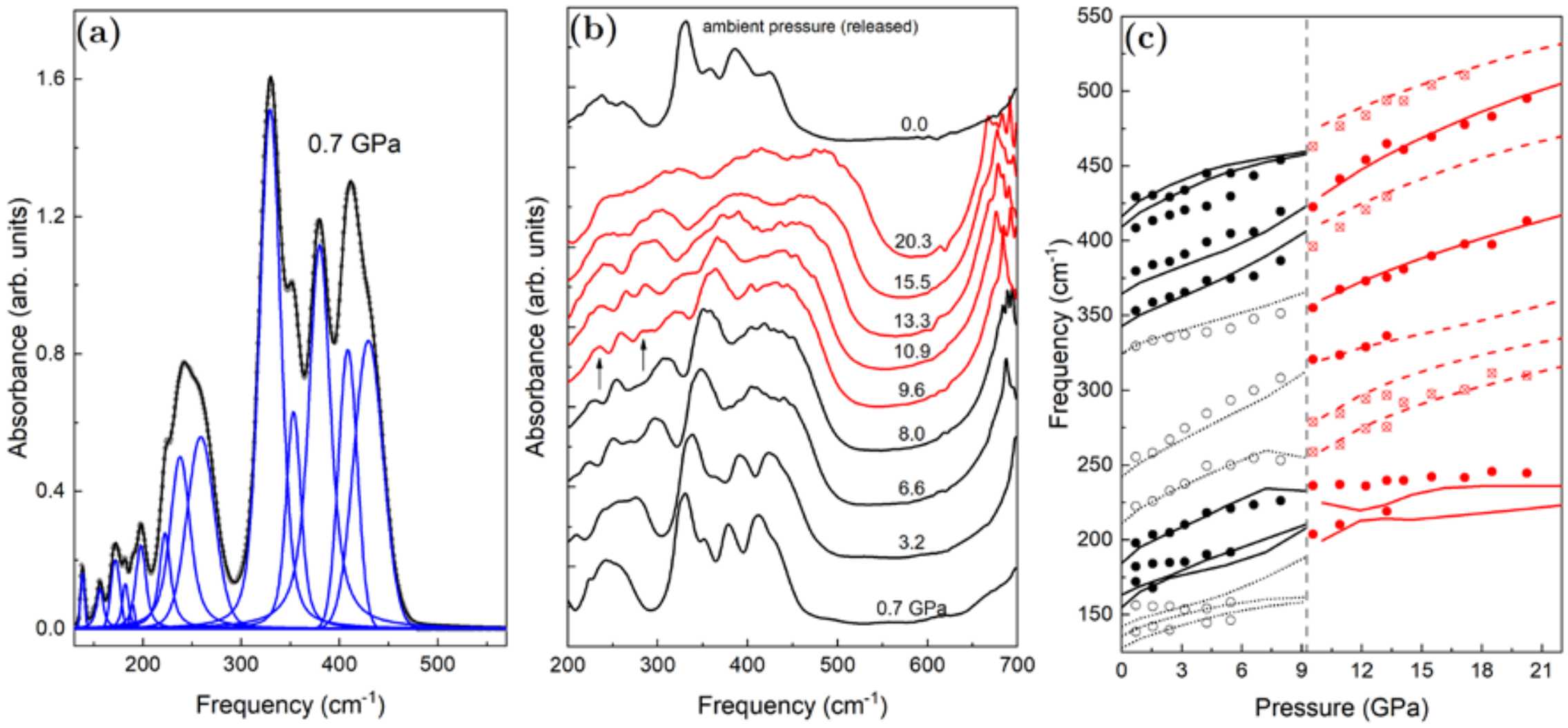

***Fig. 13.*** *HP infrared spectra of Mg(IO$_3$)$_2$. (a) Voight-peaks fitted to a near ambient pressure spectrum (0.7 GPa). The raw data are shown as grey symbols, the Voight peaks are shown in blue and the overall fit is shown in black. (b) Infrared spectra acquired on compression to 20.3 GPa. (c) The infrared mode frequencies shown as a function of pressure. Symbols correspond to the Voight peaks fitted to experimental data. Lines correspond to DFT calculations. Solid symbols and solid lines correspond to symmetric (A) modes. In the low-pressure phase (black) the open symbols and dotted lines correspond to asymmetric (B) modes. In the high pressure phase (red) crossed symbols and dashed lines correspond to E symmetry.*

## 6.3 Optical and transport measurements

### 6.3.1 Optical absorption measurements

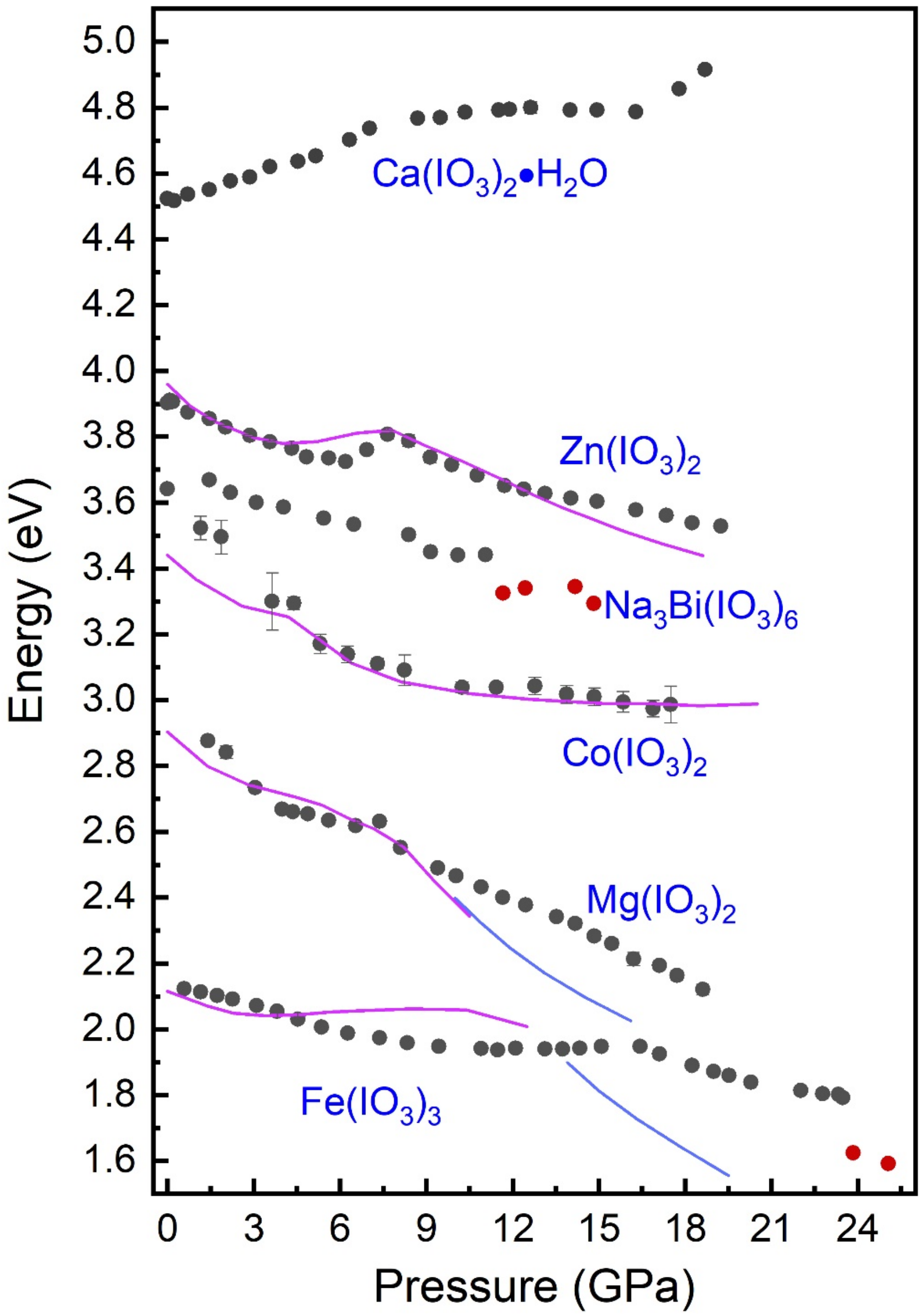

***Fig. 14.*** *Pressure dependence of the bandgap energy of several metal iodates as determined from experiment (symbols) and calculation (solid lines). For a better comparison between the experimental and calculated data, the calculated bandgap of* $Fe(IO_3)_3$ [53], $Co(IO_3)_2$ [58], $Zn(IO_3)_2$ [56], $Mg(IO_3)_2$ [56] *have been shifted by -0.4 eV, +1.2 eV, +1.0 eV and -0.5 eV, respectively.*

Only the bandgap energy of a few metal iodates has been studied and reported as a function of pressure in the literature. The reported studies include HP bandgap studies for $Fe(IO_3)_3$ [53], $Co(IO_3)_2$ [58], $Mg(IO_3)_2$ [56], $Zn(IO_3)_2$ [56], α and β phase of $Na_3Bi(IO_3)_6$ [60], $BiOIO_3$ [31] and hydrated $Ca(IO_3)_2$ [129]. The pressure dependence of the bandgap energy of several metal iodates has been summarized in **Fig. 14**, including the metal iodates with space group $P2_1$ (No. 4), $P2_1/c$ (No. 14), $Pca2_1$ (No. 29), $P\bar{1}$ (No. 2), $C2/c$ (No. 15) and $P6_3$ (No. 173).

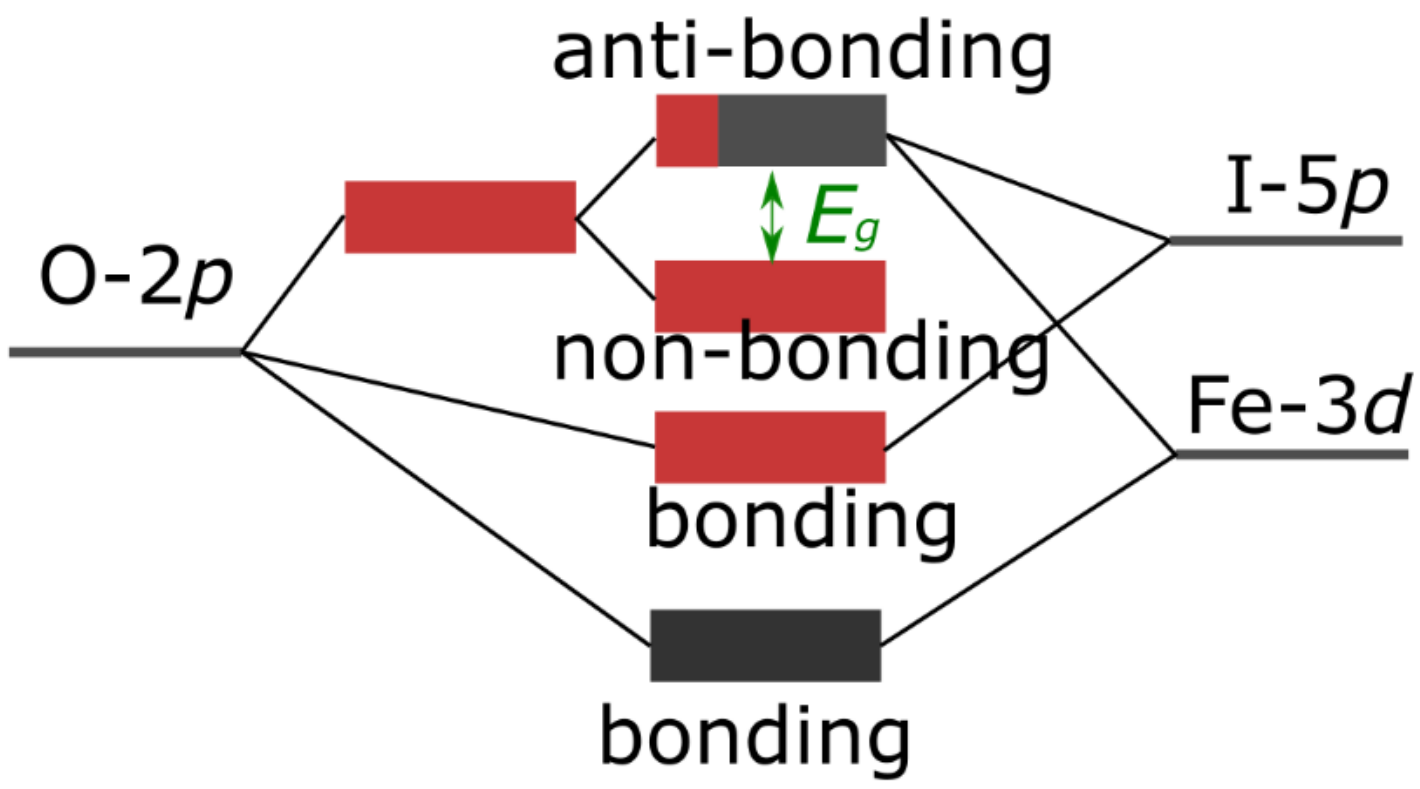

***Fig. 15.*** *Molecular orbital diagram suggested for* $Fe(IO_3)_3$.

The first reported HP bandgap study in iodates relates to $Fe(IO_3)_3$ (space group $P6_3$, No 173) [53]. In the experiment, $Fe(IO_3)_3$ exhibits the smallest bandgap energy at ambient pressure (2.1 eV) [53]. In addition, the bandgap shows a nonlinear narrowing under compression up to around 24 GPa followed by a bandgap collapse. The bandgap jump is around 0.2 eV and is caused by the pressure-induced first-order isostructural phase transition accompanied by a unit-cell volume collapse at around 21 GPa found in HP-XRD experiments [29]. In calculations, a pressure-induced nonlinear narrowing of the bandgap below 13 GPa has been observed, which is followed by a bandgap collapse upon further compression. The same pressure-induced bandgap collapse also has been found in $Na_3Bi(IO_3)_6$ at around 12 GPa [60]. It is believed to be caused by the pressure-induced phase transition and unit-cell volume collapse at the same pressure [60].

To understand the reported results it should be noted that the VBM of the electronic band structure in $Fe(IO_3)_3$ is dominated by the O-2*p* orbital, and the CBM is dominated by the Fe-3*d* orbital, with some minor contribution from I-5*p* states [53]. The MO diagram established for $Fe(IO_3)_3$ is shown in **Fig. 15**. Compared to the MO diagram for non-transition or closed-shell transition metal iodates (**Fig. 7**), the anti-bonding state of the interaction between the O-2*p* and Fe-3*d* orbitals present in the CBM dominates the behavior of the bandgap energy. Therefore, the bandgap energy is dependent on the interaction between oxygen and the cations, including both iron and iodine. The nonlinear behavior of the bandgap energy under pressure could be explained by the competition of three effects. One effect is the pressure-induced shortening of the Fe-O bond distances (*effect 1*) [53], which increase the overlap between the Fe-3*d* and O-2*p* orbitals and as a consequence open the separation between the bonding and anti-bonding states, this effect favors the opening the bandgap energy. The second effect (*effect 2*) originates from the fact that the bond distances change between iodine and the inlayer oxygen atoms (short I-O covalent bonds). As introduced in **Section 8.2**. The bond distance between iodine and inlayer oxygen atoms actually expands under compression to accommodate the extra three interlayer oxygen atoms [29,30,52,54,57], thereby decreasing the overlap between I-5*p* and O-2*p* orbitals and closing the separation between the bonding state and anti-bonding states. This effect favors a narrowing of the bandgap energy. The third effect (*effect 3*) arises from the bond distances change between iodine and interlayer oxygen atoms (long I···O halogen bonds). The pressure-induced shortening of the I-O bond distances increases the overlap between the I-5*p* and O-2*p* orbitals and enlarges the separation between the bonding and anti-bonding states [29,30,52,55,57]. Therefore, this effect also favors the opening of the bandgap. Because the iodine and inlayer oxygen atoms have the shortest bond distance (around 1.8 Å) compared to the bond distance between iron and oxygen or interlayer oxygen atoms and iodine, *effect 2* dominates the change of the bandgap energy of $Fe(IO_3)_3$. Furthermore, the three effects compete under compression causing the nonlinear behavior of the bandgap. The bandgap of a different partially filled transition metal iodate, $Co(IO_3)_2$, also shows a slightly nonlinear narrowing under compression. The MO diagram for $Co(IO_3)_2$ is similar to that of $Fe(IO_3)_3$ with the only difference being the fact that the bonding state between O-2*p* and Co-3*d* is closer to the VBM.

Here we want to explain more about the relationship between the atomic bond distance and the energy difference between the bonding and anti-bonding state we talked above. In the tight-binding approach, the energy difference between the bonding state and the original energy state of *s* orbital or *p* orbital is ***V*** for homopolar molecule, the same as the energy difference between the original s or p state and the anti-bonding state, this relationship also could be used in crystal. Here $\boldsymbol{V}$ is the matrix element of the interaction Hamiltonian between the atomic orbital, also be called as overlap parameter. This overlap parameter of the orbital between two atoms could be defined as

$$S_{AB} = \int \psi_A^* \psi_B dr$$

where $\psi_A$ and $\psi_B$ is the wave function of the orbital of atomic A and B, respectively. the overlap parameter could be derived as

$$\mathrm{S(R)} = \langle 1s_A | 1s_B \rangle = e^{-R/a_0} \left(1 + \frac{R}{a_0} + \frac{R^2}{3a_0^2}\right),$$

where R is the atomic distance, $a_0$ is Bohr radius. So normally, when $R=6a_0$, the overlap parameter is close to zero, and when the R=0, the overlap parameter is close to the maxima 1. The atomic distance and the overlap parameter have a negative relationship. When pressure shortening the atomic distance, the overlap parameter will be enlarged, therefore, the difference between the bonding and anti-bonding state is increased and vice versa.

The non-transition metal and closed-shell transition metal iodates, including $Zn(IO_3)_2$, $Mg(IO_3)_2$, $Ba(IO_3)_2$, and $Na_3Bi(IO_3)_6$, also exhibit a nonlinear narrowing of the bandgap under compression. A representative MO diagram for those metal iodates is shown in **Fig. 7**. Since there are no partially filled *d* orbitals contributing to either the VBM or CBM, there is no *effect 1* affecting the bandgap. Only *effects 2* and *3* compete under compression. As discussed in **Section 8.2**, the I-O bond distances for these iodates share the same behavior under compression. The bond distances between iodine and inlayer oxygen atoms slightly enlarge under compression. On the other hand, the bond distances between iodine and interlayer oxygen will shorten under compression. As the bond distance between iodine and inlayer oxygen atoms is shorter than the bond distance between iodine and interlayer oxygen atoms, *effect 2* dominates the change of the bandgap energy, and the competition of *effects 2* and *3* causes the nonlinear behavior.

$BiOIO_3$ was reported to undergo two pressure-induced phase transitions at around 5 and 10 GPa [31]. The first phase transition ($BiOIO_3$-I to $BiOIO_3$-II) is an isostructural phase transition, and the second phase transition is from space group $Pca2_1$ to $P2/n$ (from $BiOIO_3$-II to $BiOIO_3$-III), which is accompanied by a unit-cell volume collapse. The bandgap energy shows discontinuity at the transition pressure, and the bandgap exhibits a pressure-induced narrowing in $BiOIO_3$-I and $BiOIO_3$-III phases, showing a nonlinear behavior in $BiOIO_3$-II phase. This behavior can be explained by the same arguments used to describe the behavior of the bandgap energy in the previously discussed iodates.

The bandgap behavior of $Ca(IO_3)_2{\cdot}H_2O$ under high pressure is different from that of all other reported metal iodates [129]. Uniquely, it shows a nonlinear opening of the bandgap under compression, which is due to a different pressure-induced evolution of the three short I-O covalent bonds. For this hydrated iodate, the average bond distance of the three long I···O halogen bonds shows a linear decrease under compression up to around 13 GPa. Subsequently, it shows unchanged behavior up to the highest pressure (18 GPa). The average bond distances of the three short I-O covalent bonds first show little change under pressure below 7.5 GPa and then slight enlargement up to 13 GPa. Beyond this pressure the I-O bonds sharply decrease in length. Even though the bond distances of the short I-O covalent bonds are shorter, (and consequently have a stronger interaction,) than those of the long I···O halogen bonds, the changes in length of the short I-O covalent bonds are negligible below 13 GPa. Therefore, *effect 3* dominates the change of the bandgap below 13 GPa. Above this pressure, *effect 2* affects the bandgap in the opposite way and starts to open the bandgap. The pressure-induced bandgap opening observed in $Ca(IO_3)_2{\cdot}H_2O$ is a consequence of its crystal structure. In contrast with the layered crystal structure of $Co(IO_3)_2$, $Mg(IO_3)_2$, $Zn(IO_3)_2$ or $Fe(IO_3)_3$, $Ca(IO_3)_2{\cdot}H_2O$ does not have a layered structure formed by the aligned $[IO_3]^-$ pyramids, therefore, the inlayer oxygen atoms are not pushed far from the iodine under compression by the approaching of the interlayer oxygen atoms.

## 6.3.2 Electronic *d-d* transition in cobalt iodates under compression

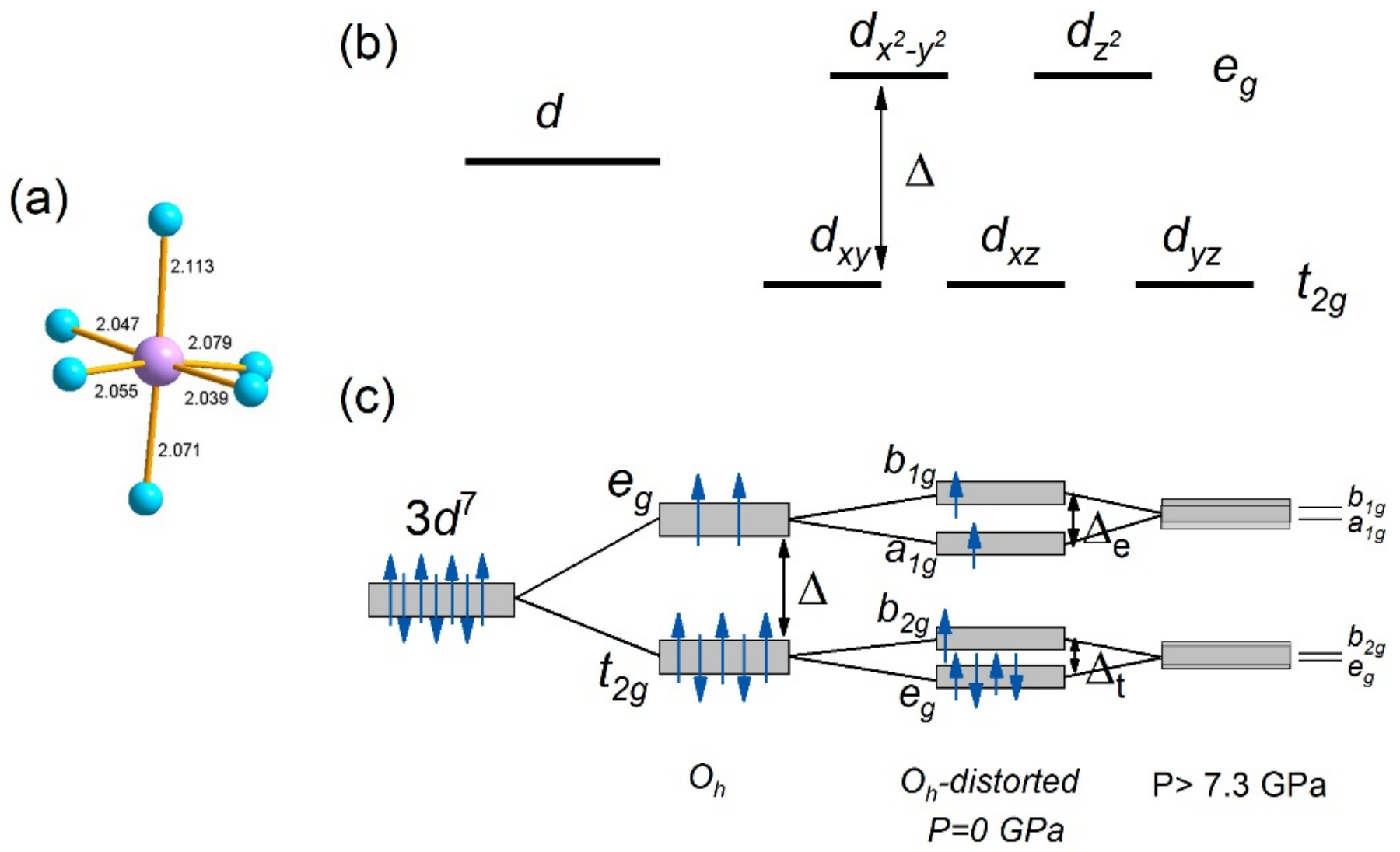

***Fig. 16.*** *(a) A representation of the $CoO_6$ octahedron at ambient conditions. The cobalt and*

*oxygen atoms are shown in pink and light blue color, respectively. The bond distances are labelled and the units are Å. (b) Scheme of the crystal field splitting and (c) crystal field splitting of distorted $CoO_6$ octahedra at ambient condition and at pressure above 7.3 GPa.*

In the case of $Co(IO_3)_2$, as the transition metal is bonded with six ligands and is in an octahedral environment, the degeneracy of the five *d*-orbitals ($d_{xy}$, $d_{xz}$, $d_{yz}$, $d_{x^2-y^2}$, and $d_{z^2}$) of the cation is lifted, resulting in a splitting into two energy levels depending on their orientation: orbital $d_{x^2-y^2}$, and $d_{z^2}$, represent by the symmetry label "$e_g$", point directly to the location of the six ligands, and the other three orbitals (represent by the symmetry label "$t_{2g}$") have an orientation which is between the six ligands. The repulsion in the first case is greater than that for the later. So, the energy for the $e_g$ state is higher than the energy of the $t_{2g}$ state as sown in **Fig. 16b**. The difference between the $e_g$ and $t_{2g}$ states is the crystal field energy (Δ). In 1937, Hermann Jahn and Edward Teller reported a theorem about the stability and degeneracy of a molecule by using group theory and stated that "stability and degeneracy are not possible simultaneously unless the molecule is a linear one" [140]. In other words, the non-linear molecule with a spatially degenerated electronic ground state will undergo a geometrical distortion and decrease the degeneracy. One of the degenerate orbitals in $e_g$ or $t_{2g}$ will go down and another one go up in energy as shown in **Fig. 16c**. This results in an overall lowering of the electronic energy. The distorted $CoO_6$ octahedron is an example of this affect (see **Fig. 16a**).

The electronic configuration of $Co^{2+}$ in $Co(IO_3)_2$ is $3d^7$, wherein the electrons occupy the five *d* orbitals (**Fig. 16c**). At ambient conditions, the $CoO_6$ octahedron is distorted, the $e_g$ and $t_{2g}$ orbitals are doubly degenerate, and the splitting is described by $\Delta_t$ and $\Delta_e$. The *d-d* transition of the partially filled transition metal can be explained by the Tanabe-Sugano (TS) diagram [141], which is described by the crystal field energy Δ, and the inter-electronic repulsion interaction through Racah parameters *B* and *C*. According to the TS diagram for $3d^7$ ions, the ground state in the high spin is $^4T_{1g}(F)$, the spin allowed transition from the ground state to the other state are: $^4T_{1g}(F)\rightarrow{}^4T_{2g}(F)$, $^4T_{1g}(F)\rightarrow{}^4A_{2g}(F)$ and $^4T_{1g}(F)\rightarrow{}^4T_{1g}(P)$. The energy for the first spin allowed transition is lower than 1 eV and out of the experimental range in Ref. [58]. Furthermore, the transition from $^4T_{1g}(F)\rightarrow{}^4A_{2g}(F)$ is proportional to 2Δ and the band is too broad to be detected. Therefore, the transition $^4T_{1g}(F)\rightarrow{}^4T_{1g}(P)$ is the major feature in the multi-band spectra. In addition, the transition from ground state to the state $^2A_1(G)$ is observed over the whole pressure range and transition $^4T_{1g}(F)\rightarrow{}^2T_1(H)$ can be only observed at the pressure higher than 7.3 GPa.

Another interesting phenomenon in the multi-band spectra of $Co(IO_3)_2$ is the appearance of a new peak at pressure above 7.3 GPa and a gradual narrowing of the peak associated to the transition $^4T_{1g}(F)\rightarrow{}^4T_{1g}(P)$. This behavior can be explained by the pressure-induced symmetrization of the $CoO_6$ pyramid which is supported by the theoretical calculations [58]. Due to the Jahn-Teller effect, the Co atom is in a distorted 6-fold oxygen coordination environment, but the distortion index is only 0.06 at ambient conditions [58], so there is a slight splitting of the $e_g$ and $t_{2g}$ state (as shown in **Fig. 16c**). This splitting causes the broadening of the *d*-*d* transition absorption band. The distortion index of $CoO_6$ polyhedra decreases sharply with increasing pressure and reaches 0.01 at the pressures higher than 8 GPa [58]. Therefore, the splitting in the $e_g$ and $t_{2g}$ states decreases and the absorption bands become narrower, thus making the absorption peak, which was initially difficult to distinguish from neighboring peaks, more observable in the absorption spectra. The *d*-*d* transitions are only observed in the optical absorption spectra of $Co(IO_3)_2$ at HP. They provide another efficient and unique tool to investigate the pressure-induced symmetry change of $AO_6$ (A= metal) units now. One can expect to explore further interesting studies on the *d*-*d* transitions in partially filled transition metal iodates by the optical-absorption experiment.

### 6.3.3 Resistivity measurements for $Fe(IO_3)_3$ under compression

To the best of our knowledge, the HP resistivity has been reported in the literature only for $Fe(IO_3)_3$ [53]. As shown in **Fig. 17**, the resistivity exhibits a nonlinear behavior at pressures lower than 22 GPa, with slope changes at 6 and 15 GPa corresponding to the slope change observed in the bandgap energy under pressure (**Fig. 14**). The dramatic decrease of the resistance at pressures higher than 22 GPa could be correlated to the bandgap jump reported by experiments, which is a result of the pressure-induced first-order isostructural phase transition [29]. A simple Drude model can be used to fit the resistivity of $Fe(IO_3)_3$ at pressures higher than 22 GPa, with the resistivity being given by:

$$\rho(P) = [\sigma(P)] - 1 = [e(n+p)\mu_{aV}] - 1, where\ \ n = p = n_i$$

$$\rho(P) = [2en_i\mu_{aV}] - 1 = [2\mathrm{e}\sqrt{N_c N_V}\mu_{aV}]^{-1\frac{E_g(P)}{e^{2KT}}}$$

$$N_{C,V} = 2.51 \times 10^9 \left(\frac{m^*_{C,V}}{m_0}\right)\left(\frac{T}{300\ \ K}\right)^{\frac{3}{2}} cm^{-1}$$

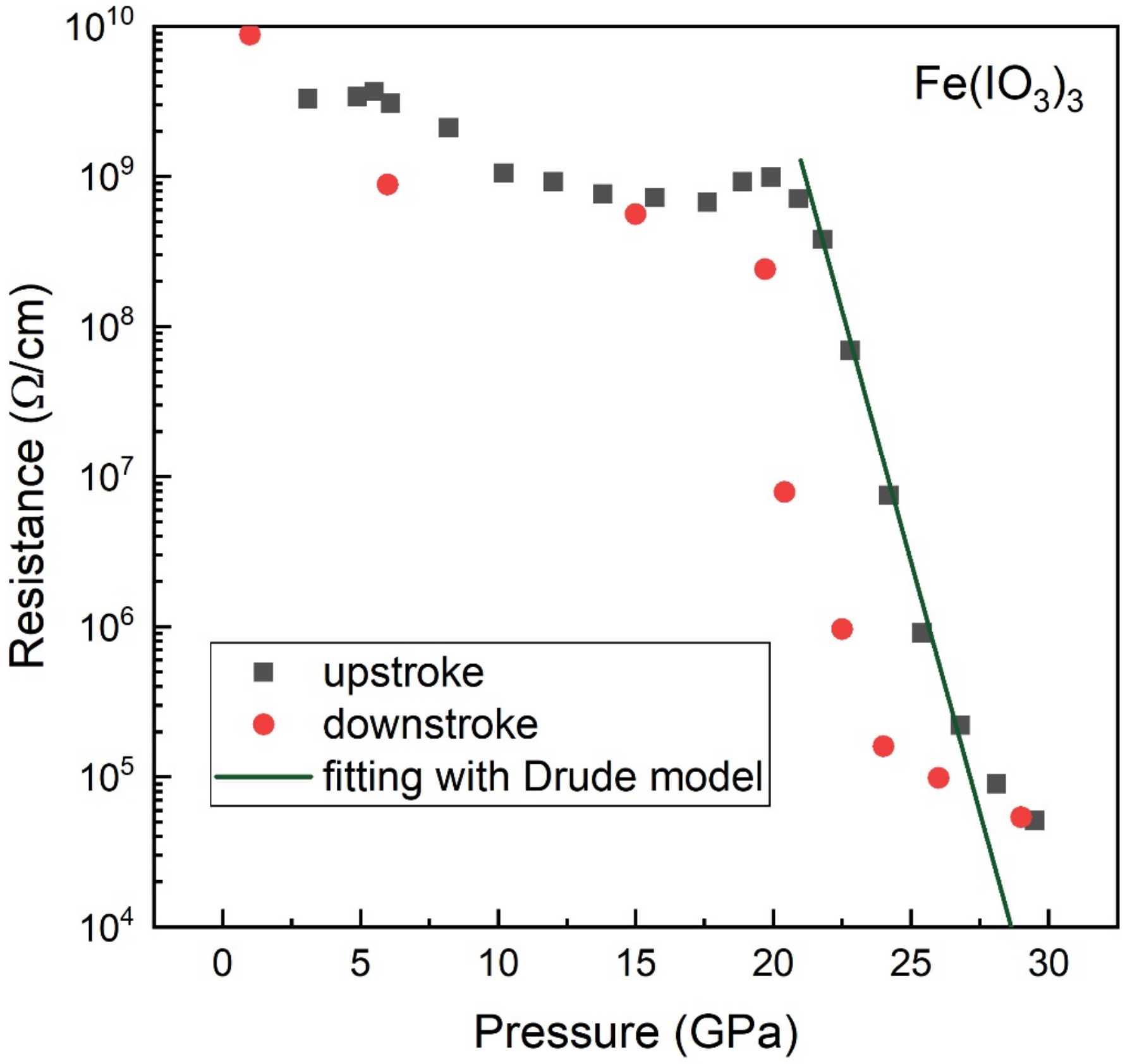

***Fig. 17.*** *Resistivity of Fe(IO₃)₃ powder measured under high pressure. The solid line is the fitting of the resistance collected at pressures higher than 22 GPa using the model discussed in the text.*

In the previous equations, $\rho$ and $P$ are the resistivity and pressure, respectively. $e$ is the electron charge, $n_i$ is the intrinsic carrier concentration, $\mu_{aV}$ is the average carrier mobility, $N_c$ and $N_v$ are the effective DOS in the conduction and valence band, $m^*_{C,V}$ is the electron effective mass in the conduction and valence band, $m_0$ is the electron mass, and $E_g(P)$ is the bandgap energy. The slope of the bandgap energy in the HP phase is $dE_g(P)/dP$= -60 meV/GPa. The green line in **Fig. 17** is produced by this model and the bandgap energy obtained at the HP phase and it matches well with the experimental resistivity. Consequently, the exponential decrease of the experimental resistivity of $Fe(IO_3)_3$ at pressures higher than 22 GPa is a consequence of the abrupt decrease of the bandgap energy.

### 6.3.4 Second-harmonic generation measurements under compression

The SHG signal of three metal iodates has been investigated under high pressure. They are $KIO_3$ [28], $BiOIO_3$ [31] and $Li_2Ti(IO_3)_6$ [32]. The SHG of these three materials shows a very similar behavior, that is, a pressure-induced nonlinear decrease in SHG intensity. The SHG response is strongly dependent on the polarity in the structure and it is only active when the crystal structure of the iodates is non-centrosymmetric. In metal iodates, the cation $I^{5+}$ with a LEP is normally in an asymmetric coordination environment and shows considerable polarity [11,62,70]. The SHG response can therefore be improved if the polar $[IO_3]^-$ pyramid is carefully aligned [70].

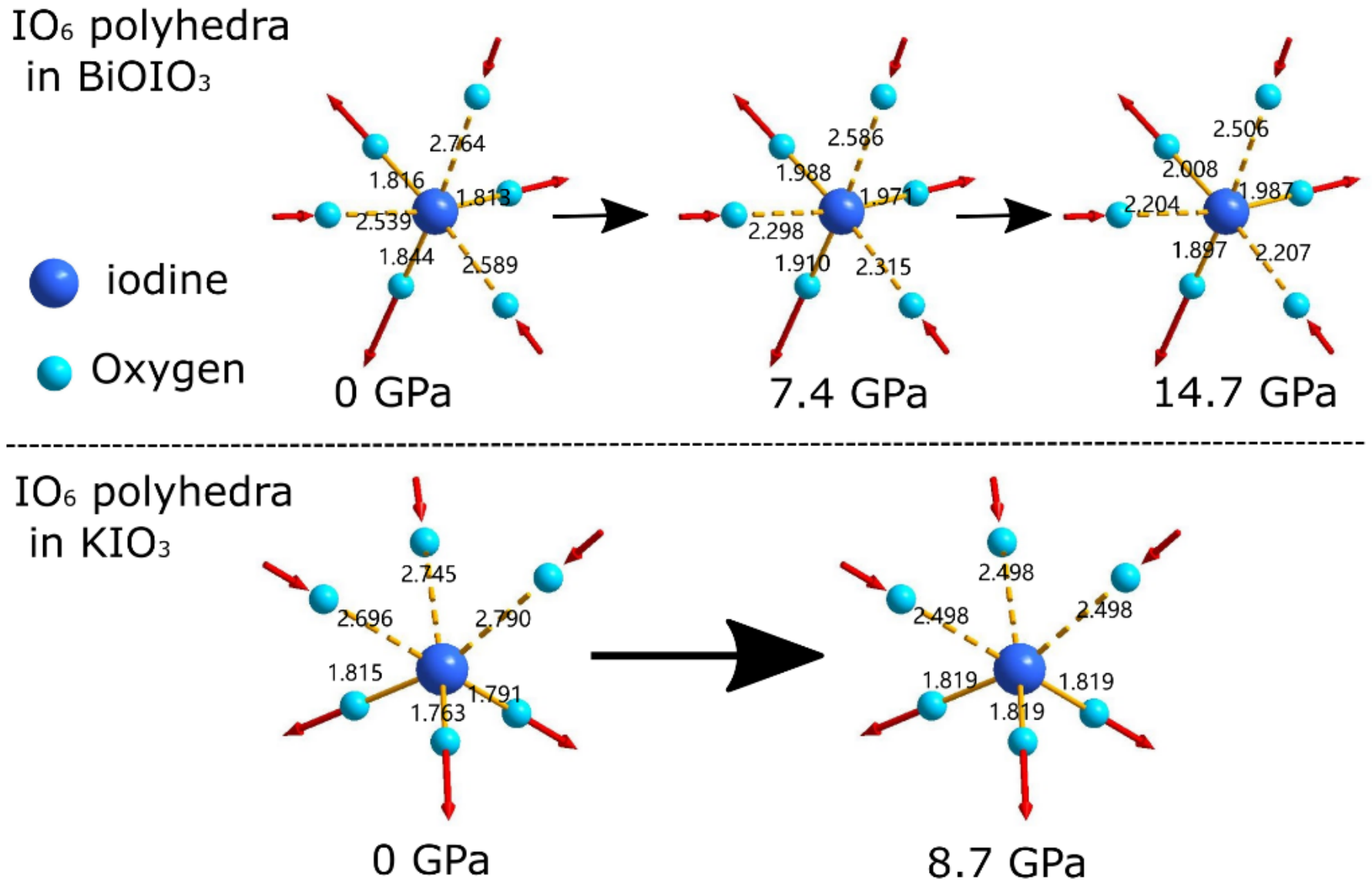

***Fig. 18.*** *Upper part: the bond distance change of the I-O bond in* $BiOIO_3$ *at high pressure* [31]. *Lower part: the bond distance change of I-O bond in* $KIO_3$ *reported at 0 GPa* [101] *and 8.7 GPa* [28], *respectively. The units for the bond distance is "Å". Solid line is the short I-O covalent bond which is existed at ambient condition, and the dash line is the weak long I···O bond.*

$BiOIO_3$ crystallizes at ambient conditions in an orthorhombic crystal structure described by the non-centrosymmetric space group $Pca2_1$ [70]. It undergoes two phase transition at around 5 and 10 GPa, respectively [31]. The first phase transition is an isostructural phase transition and the second HP phase is assumed to be described by space group $P2/n$ [31]. The SHG response at ambient conditions is about 12.5 times higher than that of the commonly used NLO material potassium dihydrogen phosphate (KDP). There is a pressure-induced two-step quenching of the SHG intensity. Two sharp decreases of the SHG intensity happened at the phase transition pressures. In the second HP phase, the SHG intensity reaches zero and could not be detected. The second HP phase of $BiOIO_3$ is SHG-inactive, in agreement with the space group assignment as $P2/n$ (No. 13) which is centrosymmetric. The HP behavior of the SHG signal observed in $BiOIO_3$ is caused by the pressure-induced suppression of the LEP and the oxygen coordination increase of iodine [31], which is a common behavior of metal iodates and has been observed in several metal iodates, including $Fe(IO_3)_3$ [29,52], $Zn(IO_3)_2$ [54], $Mg(IO_3)_2$ [30], $Co(IO_3)_2$ [57], and $Na_3Bi(IO_3)_6$ [59]. The oxygen coordination for iodine is 3-fold at ambient pressure, due to the existence of the LEP in iodine and the pressure-induced shortening of the bond distance between iodine and the second three nearest oxygen atoms. On the other hand, the three original oxygen atoms will be slightly pushed away from iodine to accommodate the new bonding oxygen atoms. Indeed, in the HP crystal structure of $BiOIO_3$, the bond distance between iodine and original three atoms (or the neighboring oxygen atoms) are larger (or shorter) than that of the bond distance at ambient conditions (**Fig. 18**), gradually transform the long I···O halogen bond into metavalent bonds (see **Section 8.2**) and induced a charge transfer from the iodine to the halogen I···O bond. Furthermore, the suppression of the LEP in iodine and coordination changes are supported by X-ray absorption spectra [31].

$KIO_3$ crystallizes at ambient conditions in a triclinic crystal structure described by the space group *P*1 [68]. It undergoes two phase transitions at 7 and 14 GPa [28]. The space group of the first HP phase is non-centrosymmetric and is described by space group *R*3, and the structure of the second HP phase remains unsolved. The SHG response of $KIO_3$ at ambient conditions is 0.815 times of that of KDP [142]. The SHG intensity of $KIO_3$ showed a pressure independent behavior from ambient pressure up to 7 GPa when nitrogen was used as pressure-transmitting medium. The SHG generation decreases sharply from 7 GPa to 14 GPa, after that the intensity continuously decreases but with a smaller slope. The intensity did not reach zero at the highest pressure (around 30 GPa) [28], indicating that the crystal structure of the second HP phase is non-centrosymmetric. The two changes of the slope of the SHG intensity at HP are believed to be triggered by the pressure-induced phase transition. In fact, the HP behavior of the SHG intensity observed in $KIO_3$ can be explained following the same rationale as in $BiOIO_3$. It was also found that the bond distance between iodine and nearest oxygen (ranging from 1.78-1.79 Å at ambient conditions) increased in the first HP phase to 1.819 Å, and, at the same time, the bond distance between iodine and the three next nearest oxygen atoms decreases, from 2.7-2.8 Å to 2.498 Å (**Fig. 18**), thereby indicating a pressure-induced oxygen coordination increase and transforming the halogen I···O bond into a metavalent bond as a consequence of suppression of the iodine LEP.

$Li_2Ti(IO_3)_6$ crystallizes in space group *P*6$_3$ at ambient conditions [10]. Its SHG intensity shows a pressure-induced decrease up to 40 GPa but the compound does not become SHG-inactive [32]. The SHG change is consistent with the crystal structure change, since the crystal structure of $Li_2Ti(IO_3)_6$ can be described by non-centrosymmetric space group $P6_3$ in the whole studied pressure range [32]. In addition, the SHG intensity changes in a similar fashion to the change of the *c*-axis of the crystal structure. This is the direction along which the LEPs of iodine are aligned. The *c*-axis in the crystal structure shows a large compressibility compared with the other two axes [32] and it is believed that the LEPs in iodine are suppressed under compression, which causes the decrease of the SHG intensity in $Li_2Ti(IO_3)_6$.

## 6.4 Theoretical studies

Advances in the accuracy and efficiency of first-principles calculations have allowed for the realization of detailed studies of the energetics of materials under HP as well as studies of the electronic structure, phonons, and elastic constants [36,143,144]. Usually, DFT calculations are accurate enough to explain experimental phenomena and/or predict the influence of the HP on the crystal structure and the occurrence of sudden changes in the arrangement of the atoms, i.e., the occurrence of structural phase transitions. The Gibbs free energy (at finite temperature) or the enthalpy (at 0 K) of the different possible crystal structures are the thermodynamic potentials used to determine the most stable phase. For different polymorphs these potentials vary under compression, and at some stage it becomes favorable for the material to change the type of atomic arrangement, which corresponds to the occurrence of a phase transition. This is exemplified by **Fig. 19** where the enthalpy of different potential polymorphs of $Cd_2V_2O_7$ are shown as function of pressure [145]. The polymorph of minimum enthalpy changes from $\alpha$ to $\gamma$ and then to $\tau$ at the pressures indicated by the blue arrows, which corresponds to the transition pressures of the predicted phase transitions.

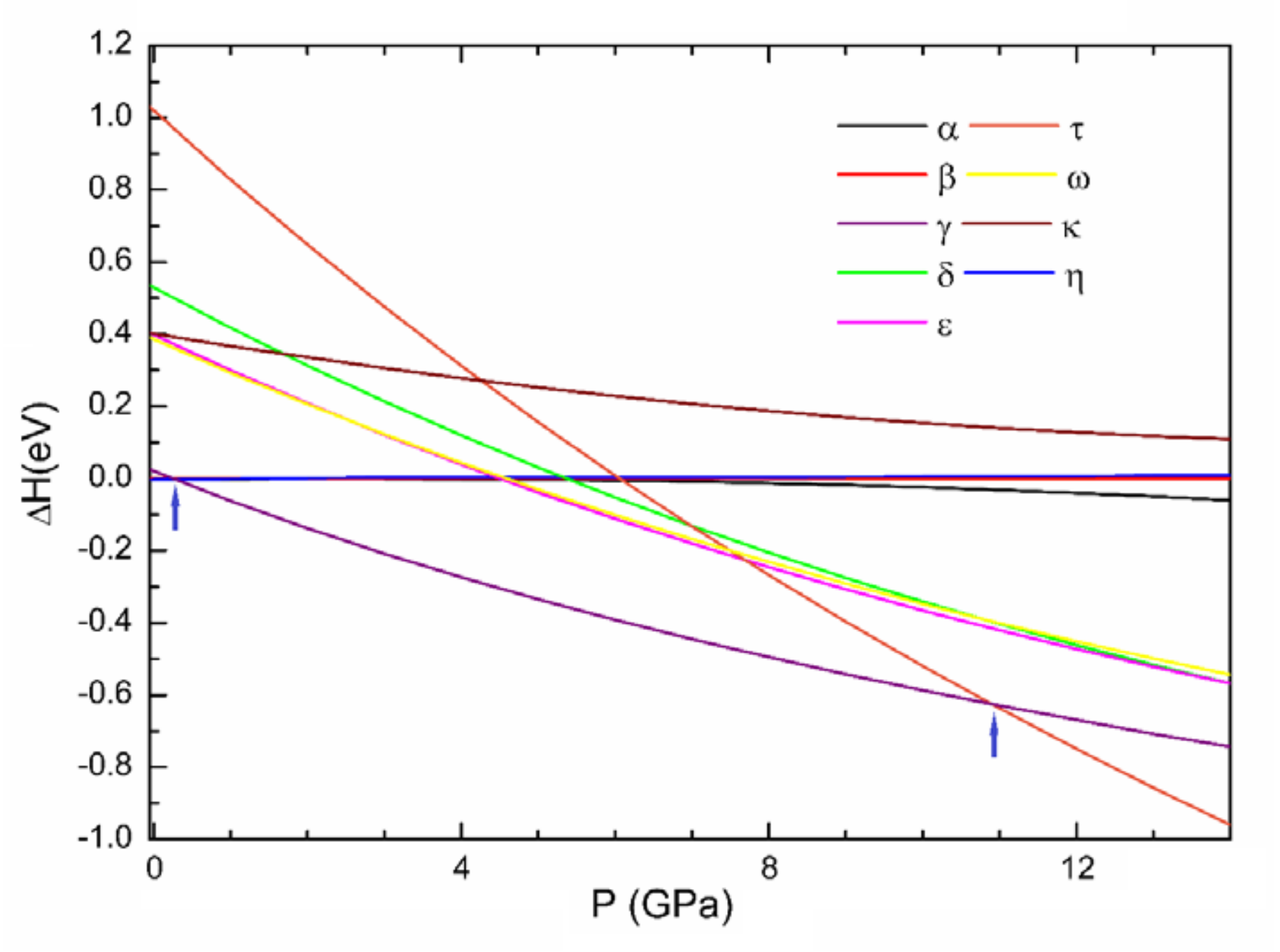

***Fig. 19.*** *Difference of enthalpy versus pressure between potential crystal structures of* $Cd_2V_2O_7$ *with respect to the to β-phase. For more information, see Ref.* [145].

In the case of metal iodates, the situation is more complex regarding theoretical predictions. In these compounds, isostructural phase transitions take place at relatively low pressures. These subtle phase transitions involve only continuous changes in the crystal structure, with no discontinuities in the unit-cell parameters or atomic positions. Due to the lack of symmetry breaking, isostructural transitions are extremely difficult to identify via DFT calculations even though they can lead to considerable changes in material properties [146]. One of the problems surrounding the identification of phase transitions comes from the fact that the two phases involved in the phase transition are indistinguishable regarding their thermodynamic potentials [29,30,52–58], at least within the precision of DFT calculations. If the isostructural phase transitions are first-order transitions, computer simulations find the isostructural transition during optimization of the crystal structure. For instance, in the case of $Fe(IO_3)_3$ DFT calculations found that the crystal structure spontaneously undergoes an abrupt change of the *c/a* axial ratio and a discontinuous decrease of the unit-cell volume. These results are shown in **Fig. 20**. These results were in agreement with XRD experiments [29] playing DFT calculations a crucial role in the interpretation of experiments.

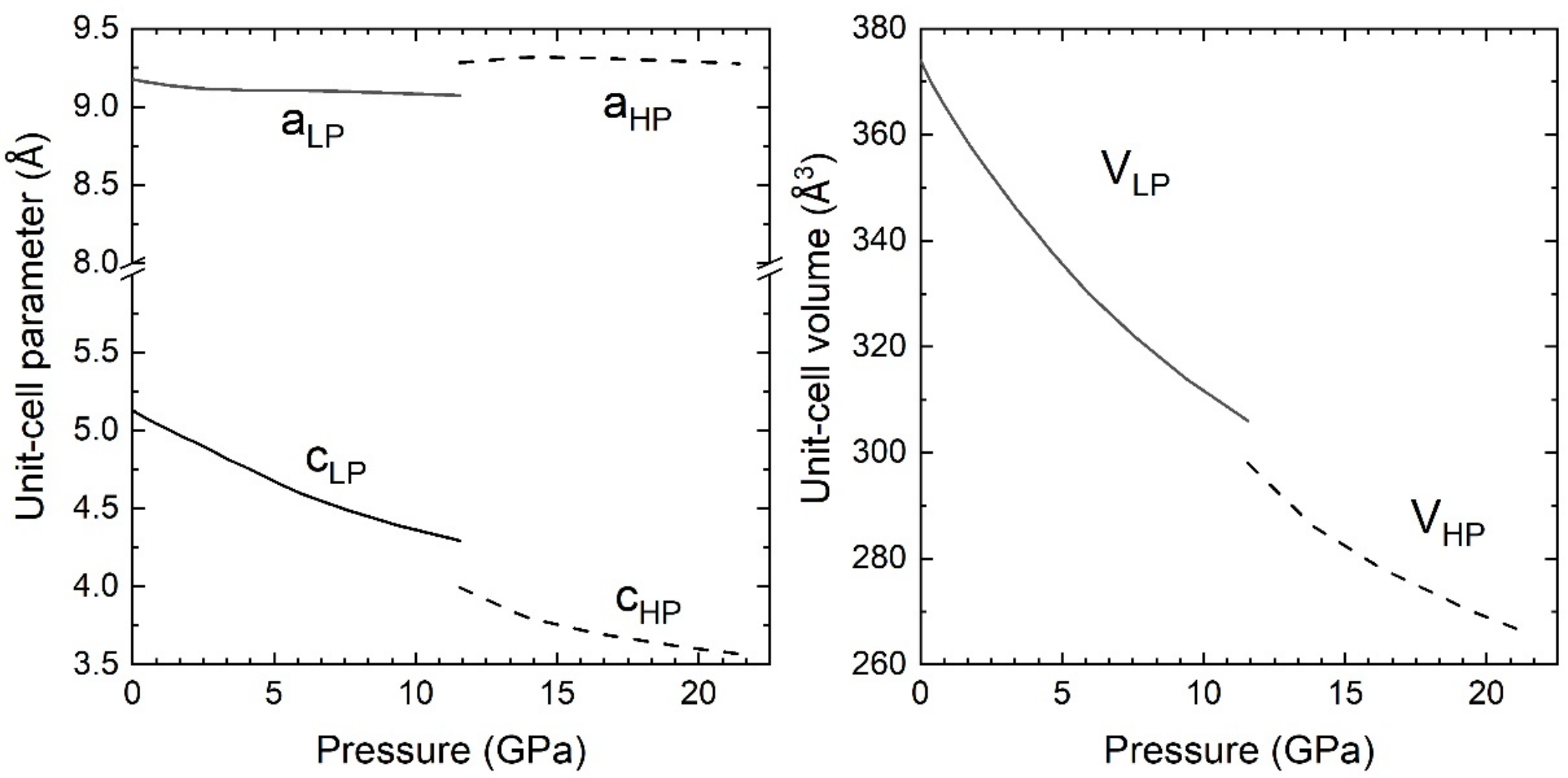

***Fig. 20.*** *Calculated pressure dependence of unit-cell parameters and volume for* $Fe(IO_3)_3$ [29]. *At 12 GPa assuming the low-pressure phase as starting model after optimization and structure involving discontinuities in the parameters is obtained.*

In the cases where there are no discontinuities in the volume at the phase transition, symmetry-preserving phase transitions are very subtle and difficult to detect. Usually, calculations identify such transitions by detecting changes in the pressure dependence of phonons and bond distances [29,30,52,54,57]. However, the elastic constants are the parameters most sensitive to isostructural phase transitions. **Fig. 21** shows the elastic constants calculated for $Fe(IO_3)_3$ [52]. There are clear discontinuities in $C_{33}$, $C_{66}$, $C_{12}$, and $C_{13}$ at 1.7 GPa which are a fingerprint of the occurrence of a phase transition. At 5.7 GPa, there are also changes in the elastic constants, especially in $C_{13}$ and $C_{33}$. They are less evident than at 1.7 GPa; however, the slope changes at 5.7 GPa indicate the existence of a second symmetry-preserving phase transition. DFT calculations have provided an insightful input to understand isomorphic phase transitions and for the interpretation of the changes induced by them in physical properties as discussed in different section of this review. DFT calculations also contribute to the understanding of the possible mechanisms driving the observed phase transitions. Additionally, DFT calculations provided crucial information for the interpretation of the behavior of phonon modes and changes in the electronic properties as discussed in previous sections.

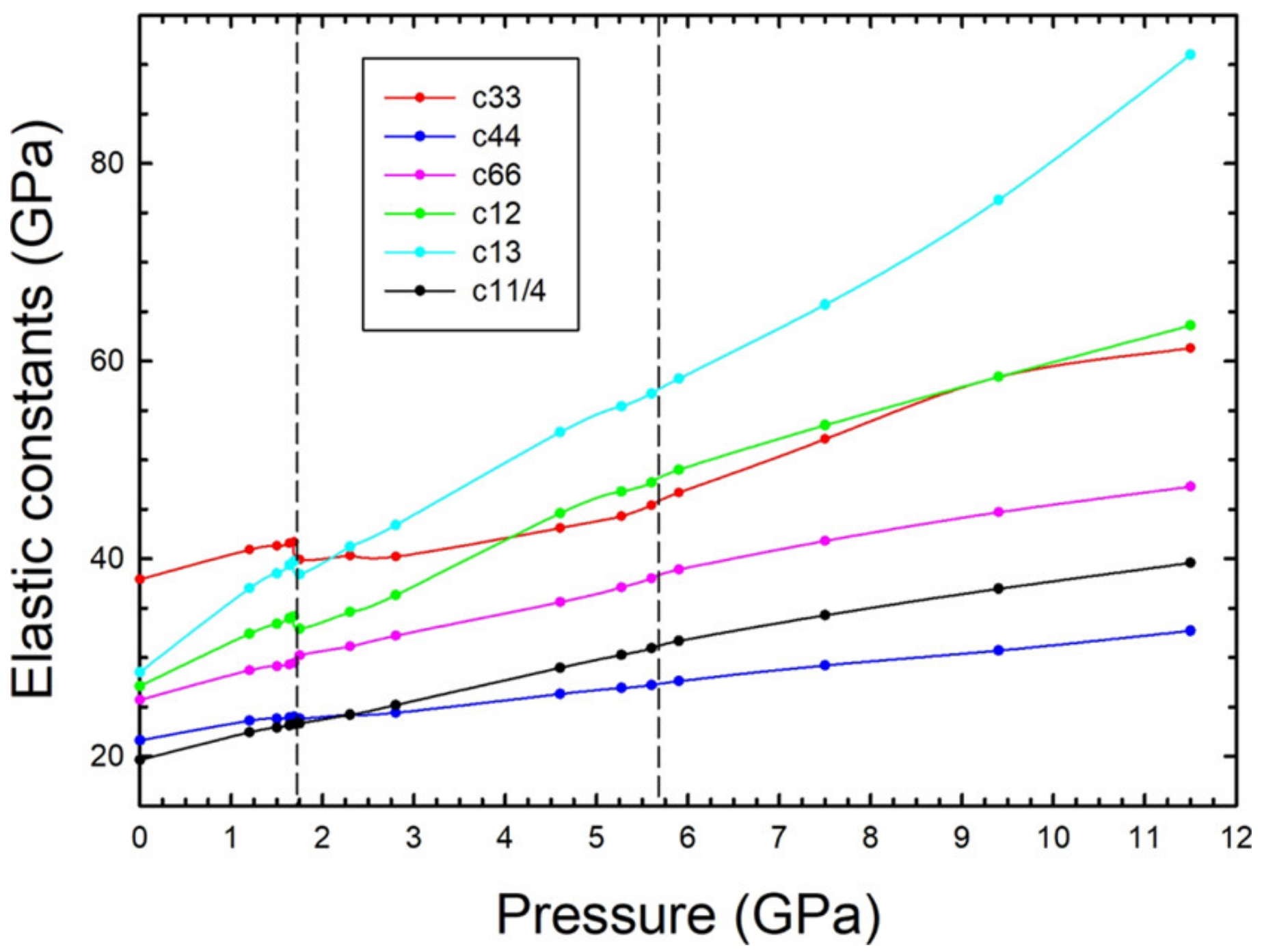

***Fig. 21.*** *Calculated pressure dependence of elastic constants in Fe(IO*$_3$*)*$_3$*. The vertical dashed lines indicate the pressures where isostructural transitions take place. Figure reproduced from Ref.* [52].

# 7. Phase transitions

Many of the studied metal iodates undergo isostructural phase transitions. The existence of these phase transitions is related to the enhancement of the intensity of lone pair stereochemical activity under compression. As a consequence, the interaction between iodine and oxygen atoms is boosted, gradually transforming the I···O halogen bonds into metavalent I-O bonds. This is evidenced by the fact that $Mg(IO_3)_2$ [30], $KIO_3$ [28], $BiOIO_3$ [31], $Zn(IO_3)_2$ [54], $Co(IO_3)_2$ [57], and $Fe(IO_3)_3$ [29] feature short I···O bonds ranging from 2.35 to 2.45 Å at high pressure. These bonds are extremely short for halogen bonds. Indeed, they are much shorter than the shortest I···O bonds observed in nature; e.g. α-$HIO_3$ (2.48 Å) [14,147] and $La(IO_3)_3(HIO_3)$ (2.47 Å) [148]. The modification of the iodine coordination sphere has important consequences of the physical properties of iodates as described in previous sections. For instance, the pressure dependence of the bandgap will be affected showing a non-linear pressure dependence. The enhancement of the I···O interactions occurs as charge transfer from LEPs to weak I···O halogen bonds between second-neighbor oxygen atoms of iodine. This is illustrated in **Fig. 22** by two-dimensional contour plots of the electron-localization function (ELF) of $Fe(IO_3)_3$ at selected pressures [53]. In **Fig. 22(a)** it can be seen that there is a are of charge depletion (blue) separating $IO_3$ molecules from the next neighboring O atoms. In **Fig. 22(b)**, after the first isostructural phase transition (predicted in DFT calculations to be around 2 GPa), charge begins to accumulate in the charge depletion zone. In **Fig. 22(c)**, after the second isostructural phase transition (~5.8 GPa predicted in calculation), there is an increase of charge accumulation. Finally, in **Fig. 22(d)**, after the first-order isostructural phase transition (~12 GPa predicted in calculation), additional bonds are formed between iodine and oxygen atoms. In the case of $Fe(IO_3)_3$ changes induced by pressure could modify the macroscopic electronic polarization along the *c*-axis related to the co-alignment of the stereoactive lone-pairs of the $[IO_3]^-$ and the magnetic $Fe^{3+}$ promoting an asymmetric exchange coupling making $Fe(IO_3)_3$ an ideal host for magnetic skyrmions [149].

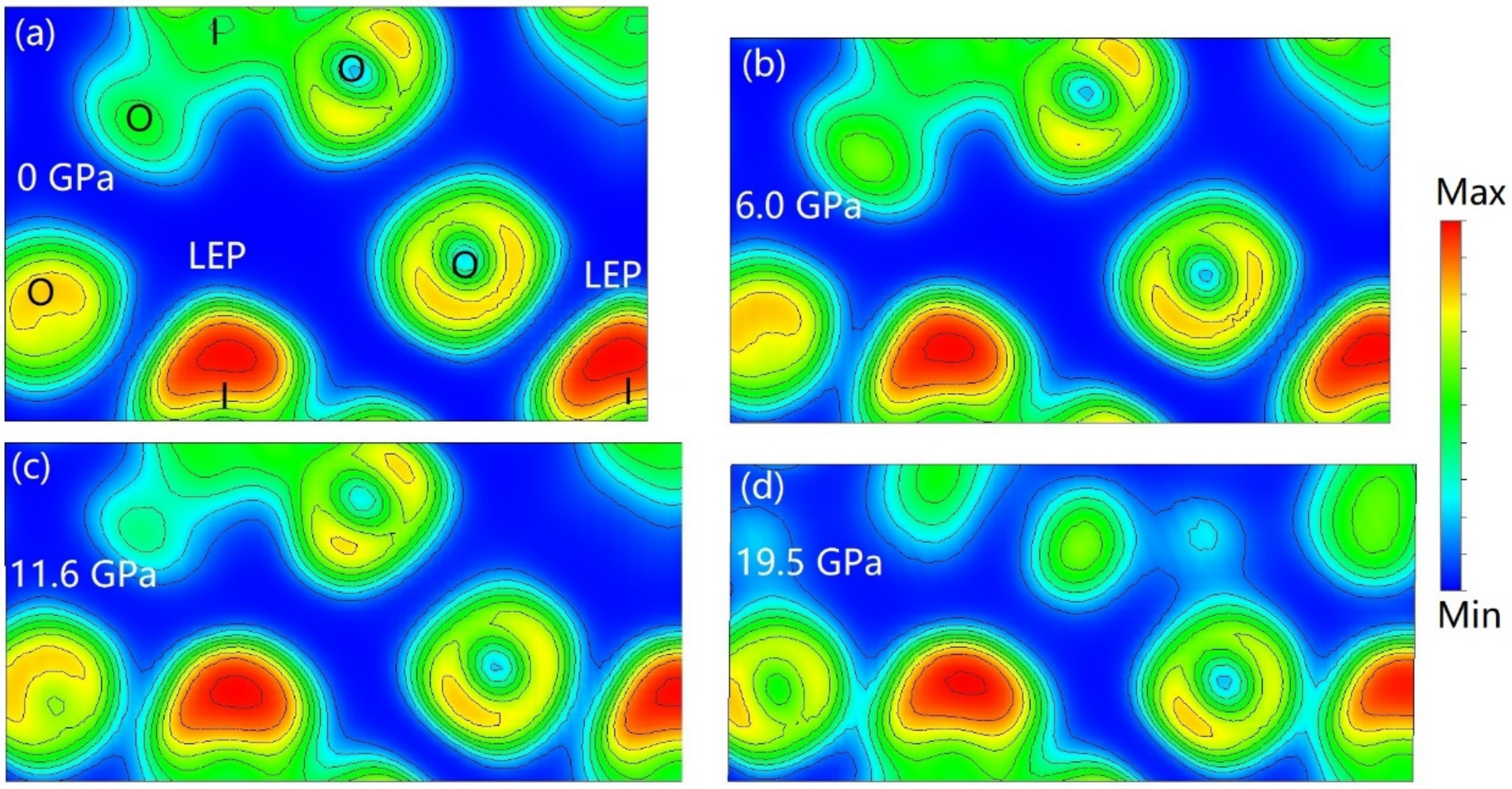

***Fig. 22.*** *Contour plot of the calculated electron-localization function of Fe(IO$_3$)$_3$ at different pressures* [53]. *(a), (b), (c), and (d) show the ELF of the ambit-pressure phase and the three high-pressure phases, respectively.*

The importance of the LEPs in phase transitions is not unique to iodates materials. In fact, it is a phenomenon observed under compression in multiple chalcogenide compounds [137]. In these compounds several isostructural phase transitions have been reported and have been connected to a change of the nature of bonding as described above in iodates. Given the fact that there are many iodates with structural frameworks similar to those of Zn(IO$_3$)$_2$, Co(IO$_3$)$_2$, Fe(IO$_3$)$_3$, and Na$_3$Bi(IO$_3$)$_6$, where the IO$_3$ pyramid has three single bonds with oxygen atoms and a LEP is a main building block, it is not unreasonable to speculate that many of them would undergo isostructural phase transitions under compression. In this regard it is very surprising that α-LIO$_3$ has been reported to remain stable in the ambient-pressure phase up to 75 GPa [27]. It is possible that previous studies, where low-resolution energy-dispersive XRD was performed and pressure steps of approximately 5 GPa were used, could have missed the existence of isostructural phase transitions in α-LiIO$_3$. This suggests that the HP behavior of this material should be revisited.

Mg(IO$_3$)$_2$ and KIO$_3$ are the only iodates studied recently where the isostructural phase transitions have not been reported [28,30]. In both compounds there are first order transitions which take place at around 7 GPa and they involve the enhancement of the crystal symmetry. In many regards the transitions induced by pressure in both materials resemble the transition induced by temperature in In(IO$_3$)$_3$ at 365 ºC from α-In(IO$_3$)$_3$ (space group *P*6$_3$) to β-In(IO$_3$)$_3$ (space group *R*-3) [150] and in LiO$_3$, α-β-γ structural sequence [151]. In magnesium iodate the transition transforms the material from monoclinic (space group $P2_1$) to trigonal (space group $P3$) [30]. In $KIO_3$ from triclinic (space group $P1$) to trigonal ($R3$) [28]. In both cases the transitions little affect the $IO_3$ pyramids. However, a substantial reduction of the long I···O halogen bond distances happens at the phase transition. In the case of $Mg(IO_3)_2$, the difference between short and long I-O bonds is less than 20% after the phase transition [30], therefore, in many respects, iodine atoms can be considered as 3 + 3 coordinated. In **Fig. 23** the low-pressure and high-pressure structures of $Mg(IO_3)_2$ are compared to illustrate the changes induced at the transitions. The similarities between both phases can be seen when comparing **Figs. 23(a)** and **23(b)**. Both structures are formed by $MgO_6$ octahedra corner linked by $IO_3$ pyramids. In **Fig. 23(c)** the three I···O halogen bonds are substantially reduced under compression. At the highest pressure covered by the studies [30], the distortion index of $IO_6$ octahedron is 0.05 if a 3+3 coordination is adopted for iodine which is consistent with the gradual transformation of halogen I···O bonds into metavalent I-O bonds.

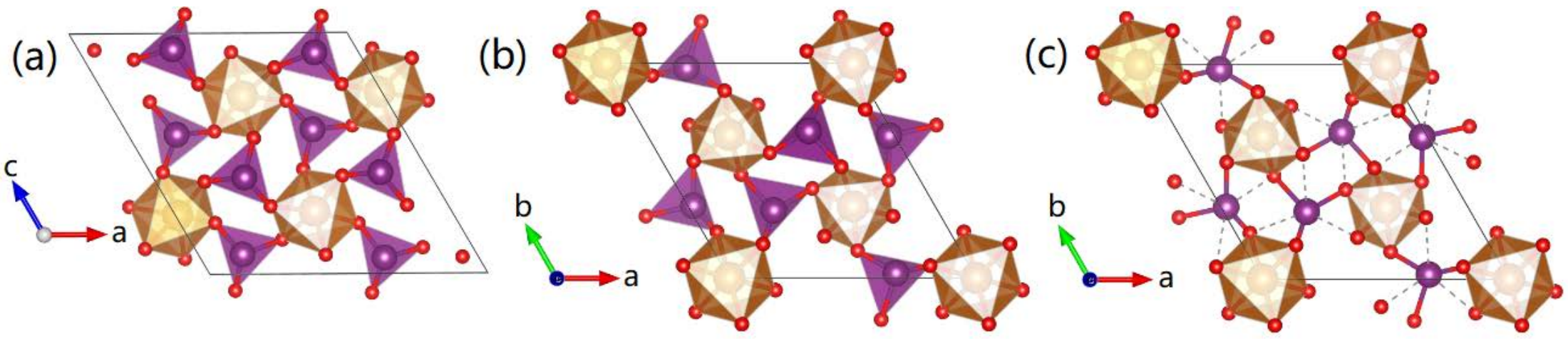

***Fig. 23.*** *(a) Crystal structure of $Mg(IO_3)_2$ at ambient pressure. (b) Crystal structure of the HP phase. (c) Crystal structure of the HP phase showing the short I-O bonds with solid lines and the long I-O bonds with dashed lines. $IO_3$ pyramids are shown in purple and $MgO_6$ octahedra in brown. Red (purple) circles are oxygen (iodine) atoms.*

# 8. Systematics of the high-pressure behavior of metal iodates

## 8.1 Common trends of pressure-induced phase transitions

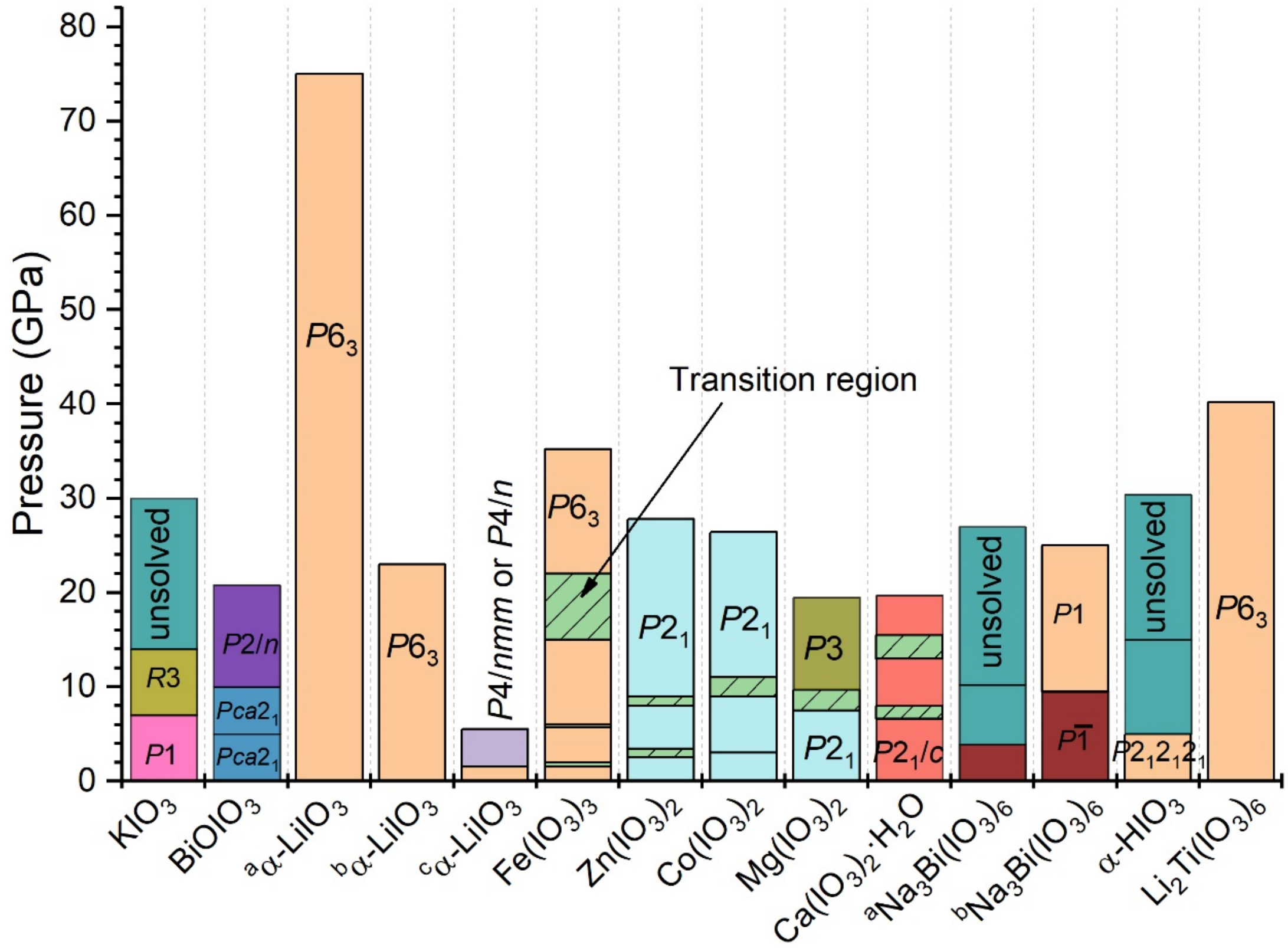

***Fig. 24.*** *A summary of the crystal structure evolution of several iodates under high pressure and room temperature, the different space group of the structure are shown in different color. The iodate in this figure including* $KIO_3$ [28], $BiOIO_3$ [31], *α-*$LiIO_3$ [27,38,42], $Fe(IO_3)_3$ [29,52], $Zn(IO_3)_2$ [54], $Co(IO_3)_2$ [57], $Mg(IO_3)_2$ [30], $Ca(IO_3)_2{\cdot}H_2O$ [129], $Na_3Bi(IO_3)_6$ [59,60], $HIO_3$ [61] *and* $Li_2Ti(IO_3)_6$ [32].

The pressure-driven structure evolution of several iodates at room temperature is shown in **Fig. 24**. The first two isostructural phase transitions in $Fe(IO_3)_3$ have been identified in experiment by the nonlinear-behavior of the Raman modes and by changes in the full-width at the half maximum (FWHM) of several Raman peaks under compression. They have been also evidenced, in theoretical calculations, by discontinuities observed in the I-O bond distance and elastic constants, as well as by instabilities found in the calculated phonon dispersion at the phase transition pressure [52]. The third isostructural phase in $Fe(IO_3)_3$ is accompanied by a 5% collapse in the unit-cell volume and can be assigned to a first-order transition. It is evidenced by the appearance of several new peaks in the XRD patterns and in infrared spectroscopy at high pressure. The pressure-induced isostructural phase transitions have been also found in $Zn(IO_3)_2$ and $Co(IO_3)_2$ which crystallize in the $P2_1$ space group. The isostructural phase transitions have been evidenced by the nonlinear behavior of the lattice parameters and the pressure dependence of the Raman and infrared modes [54,55,57]. In fact, the lattice parameters of $Li_2Ti(IO_3)_6$, which crystalizes in $P6_3$ space group at ambient conditions, also show the same nonlinear behavior. The mechanism behind the change is explained by the gear-spring model and no phase transition has been claimed in the literature [32]. However, a detailed analysis of the elastic constants and bond distances behavior is missing for this material. The isostructural phase transition found in hydrated $Ca(IO_3)_2$ is evidenced by the nonlinear behavior of bandgap energy and lattice parameters, including the monoclinic angle $\beta$ [129]. The isostructural phase transition found in $BiOIO_3$ is evidenced by peaks emerging in the XRD pattern, a nonlinear behavior of the SHG intensity, the appearance of a new peak found in Raman spectra and a bandgap discontinuous change. Additionally, there is a pressure-induced non-centrosymmetric ($Pca2_1$) to centrosymmetric ($P2/n$) structure transition at around 10 GPa [31]. The isostructural phase transitions observed in many metal iodates are believed to be triggered by the oxygen coordination increase in iodine atoms which will be discussed in the next section in detail.

The structure evolution of α-$LiIO_3$ is under debate. The pressure-temperature phase diagram established by Liu *et al.* suggested a phase transition at around 1.5 GPa at ambient temperature [38]. In contrast, the following HP studies up to 23 and 75 GPa did not find any evidence of a pressure-induced phase transition [27,42]. However, the limitations of the experimental techniques used to study $LiO_3$ could easily have led to the non-identification of isostructural phase transitions. Clearly the HP behavior of α-$LiO_3$ needs to be revisited. Based on the structure evolution of $Fe(IO_3)_3$, $Zn(IO_3)_2$, $Co(IO_3)_2$, and $Li_2Ti(IO_3)_6$. α-$LiIO_3$ is expected to undergo several isostructural phase transitions under pressure. Unfortunately, the pressure steps of experiments reported in Ref. [27] is as large as 5 GPa precluding the observation of nonlinear behaviors in the lattice parameters.

The lattice parameters in Zn and Co iodates, which exhibit nonlinear behavior under compression, become closer to each other in value and the value of their monoclinic angle becomes nearly 120 degrees at high pressure [54,57]. All of these changes in the structure indicate a gradual pressure-induced enhancement of the symmetry and the possible existence monoclinic-hexagonal phase transition under higher pressure. This has previously been reported in $Mg(IO_3)_2$, whose crystal structure transforms from space group $P2_1$ (monoclinic crystal system) to space group $P3$ (hexagonal crystal system) at the pressure region of 7.5-9.7 GPa [30].

$KIO_3$ undergoes a non-centrosymmetric to non-centrosymmetric pressure-induced phase transition. The structure of the second HP phase remains unsolved, however, it must be non-centrosymmetric because it is SHG-active [28]. In the HP Raman experiments of $Na_3Bi(IO_3)_6$ two phase transitions have been evidenced by the appearance of extra Raman peaks [59] at 1.9 and 8.2 GPa. The crystal structure of the new phase was not solved in that work because only Raman experiments were performed, however, the softening behavior of several Raman peaks in the high-frequency region can be correlated to the pressure-induced enlargement of the short I-O covalent bonds. Recently, a HP single-crystal XRD experiment revealed that $Na_3Bi(IO_3)_6$ undergoes a centrosymmetric-to-non-centrosymmetric space group phase transition at around 9.5 GPa [60], indicating an off-on switch in the SHG activity under pressure. The phase transitions of $HIO_3$ are observed in the Raman spectra at around 0.5, 5 and 15 GPa, the first transition could be assigned to isostructural phase transition and the second/third HP phases are remain unsolved [61]. It is interesting to note that out of all of the pressure induced phase transitions reported in metal iodates, all of them are reversible on decompression [28,29,129,30–32,52,54,57,60,61].

## 8.2 Pressure-induced coordination changes

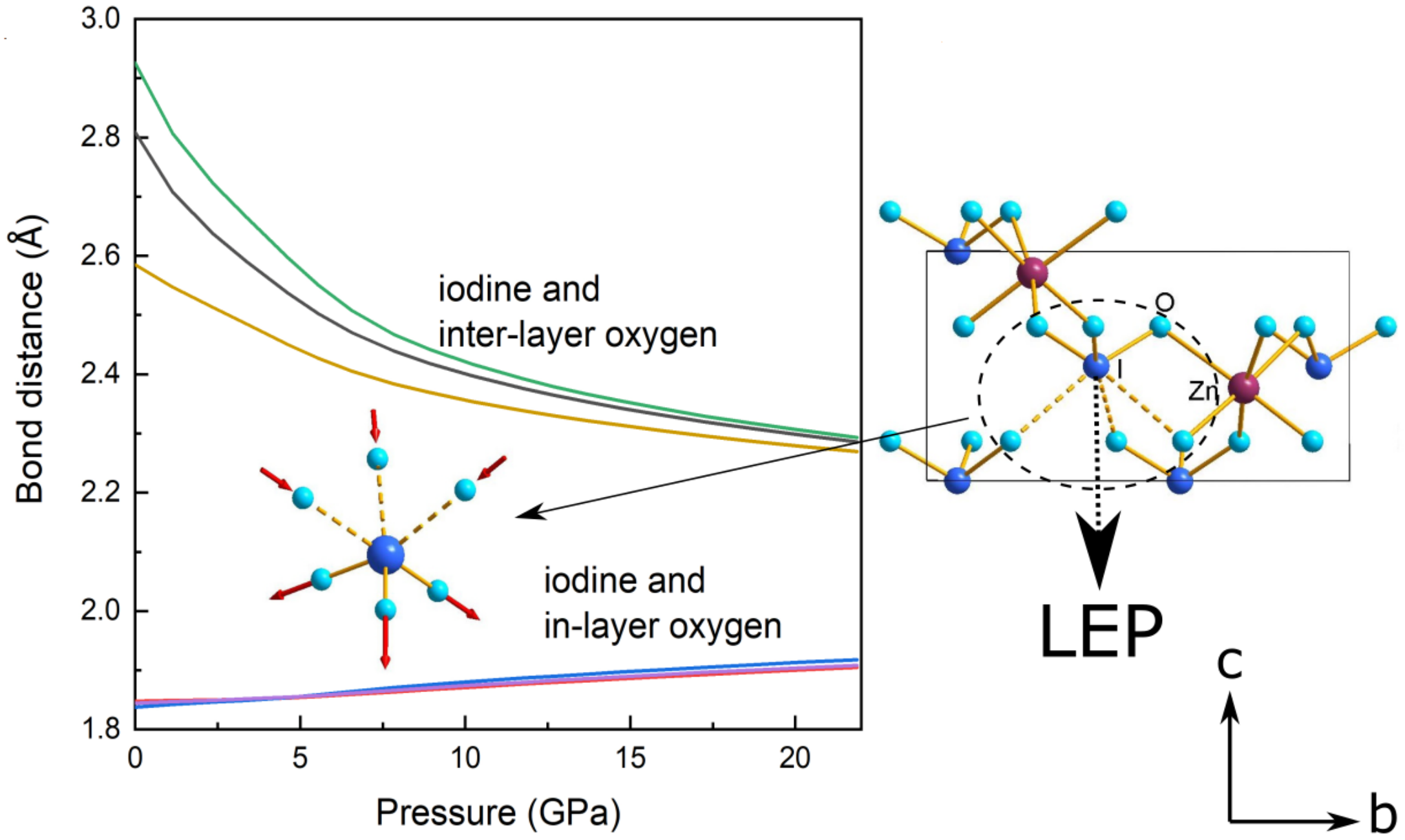

***Fig. 25.*** *(a). Calculated I-O bond distance in Zn(IO$_3$)$_2$ as a function of pressure* [54]. *The insert is a IO$_6$ polyhedron. The blue and light blue balls are iodine and oxygen atoms, respectively. The three oxygen atoms located below the iodine is the inlayer oxygen and the three oxygen atoms located above the iodine is the interlayer oxygen atoms. (b). The crystals structure of Zn(IO$_3$)$_2$ at ambient conditions* [78], *Zinc, iodine and oxygen atoms are represented by brown, blue and light blue balls, respectively. The direction of the LEP in iodine are indicated by black arrows.*

The pressure dependence of the I-O bond distances of Fe(IO$_3$)$_3$ [52], Zn(IO$_3$)$_2$ [54], Mg(IO$_3$)$_2$ [30], Co(IO$_3$)$_2$ [57], HIO$_3$ [61], KIO$_3$ [28], BiOIO$_3$ [31], hydrated Ca(IO$_3$)$_2$ [129], Na$_3$Bi(IO$_3$)$_6$ [60] share a similar characteristic. For example, the calculated pressure dependence of the I-O distances of Zn(IO$_3$)$_2$ [54] is shown in **Fig. 25**. At ambient pressure, there are two types of I-O distances, one is the covalent bond between iodine and inlayer oxygen atoms, the bond distances between them range from 1.7 to 1.9 Å (solid line in **Fig. 25**). Another I-O distance is between iodine and oxygen in the neighboring IO$_3$ units (interlayer oxygen atoms in **Fig. 25**), the distances between them, corresponding to the I···O halogen bonds, usually range from 2.6 to 3.2 Å (dashed lines in **Fig. 25**). The interaction between them is weak at ambient conditions. Under compression, the distance between iodine and interlayer oxygen atoms is shortened as shown in **Fig. 25**. In contrast, the distances between iodine and inlayer oxygen atoms will slightly expand, in order to accommodate the three extra oxygen atoms that approach iodine under pressure. Furthermore, it is reasonable to predicted that the same I-O distance change could be observed in other metal iodates under pressure, like α-$LiIO_3$, $Li_3Ti(IO_3)_6$.

In the study about the nature and energy of the iodine-oxygen interaction in α-$HIO_3$ [14] the I-O bonds have been divided into two types according to the topological analysis of the calculated electron density distribution. The first type is the covalent bond (bond between iodine and inlayer oxygen in $Zn(IO_3)_2$), the distance is short and ranges from 1.78 to 1.90 Å. The second type is the intermolecular interaction, or in other word, halogen bond (between iodine and the interlayer oxygen atoms in the case of $Zn(IO_3)_2$), the distances range from 2.48 to 3.2 Å. Furthermore, the halogen bond with a distance of 2.48 Å is claimed to be the extremely short halogen bond found in iodate.

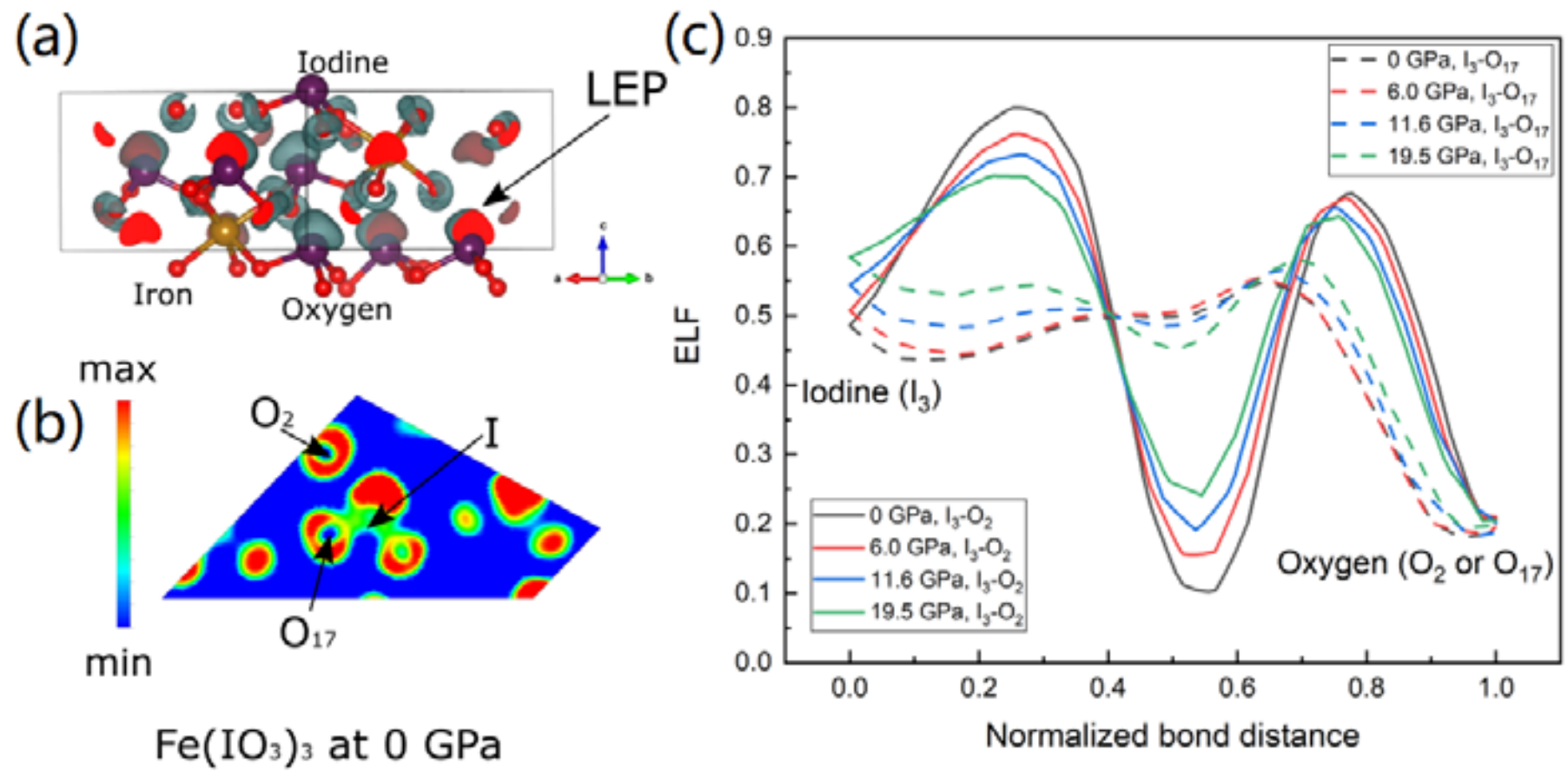

***Fig. 26**. (a) Theoretical calculated electron-localization function (ELF) of $Fe(IO_3)_3$ at ambient pressure, the atom iron, iodine and oxygen are shown in yellow, pink and red, respectively. (b) A plane has been chosen from **Fig. 26a** where include the iodine and both the inlayer (O17) and interlayer (O2) oxygen atoms. (c) The value of the ELF between the iodine and inlayer/interlayer oxygen atoms as a function of the normalized bond distance at different pressure.*

The theoretical calculated ELF of $Fe(IO_3)_3$ at ambient pressure is shown in **Fig. 26a**. The LEP of iodine is clearly observed at the head of iodine. The plane which contains the inlayer (O17) and interlayer (O2) oxygen atoms (**Fig. 26b**) corresponds to the structure shown in **Fig. 26a**. Clearly the interaction between the iodine and inlayer oxygen (I-O covalent bond) is much stronger than that between iodine and interlayer oxygen (weak I···O halogen bond) at ambient conditions. However, with increasing pressure, the interaction between the iodine and interlayer oxygen is enhanced, indicating by the increase of the ELF value between these two atoms (**Fig. 26c**).

Furthermore, the ELF value close to the iodine in the halogen bond decreases with the increasing of pressure, indicating a charge transfer from the iodine to the center of the halogen bond, or dislocation of the electrons in the iodine. Additionally, in the center of the I-O covalent bonds, the ELF shows a slight decrease at higher pressures. In fact, the weak I···O halogen bond is already shorter than 2.48 Å above 6 GPa [53]. Therefore, at high pressure, the bond between iodine and the interlayer oxygen atom O2 can no longer be considered as weak intermolecular interaction and there is possibly a pressure-induced transformation of the weak I···O halogen bond to an intermediate I-O metavalent bond. Metavalent bonding is a new categorization of bond and usually involves a medium ELF value between the weak metallic or Van der Waals bond and the strong ionic or covalent bonds [16]. The experimental evidence of the formation of metavalent bonds is the pressure-induced decrease of the resistance of $Fe(IO_3)_3$ at the pressure below of 23 GPa (**Section 6.3.3**). Since the electron will be shared between several metavalent bonds, so the mobility of the electrons is enhanced after the transformation, as well as the conductivity of the iodates.

Therefore, if 2.48 Å is set as the critical distance for the formation of metavalent bonds, then there is a gradual increase of the oxygen coordination in the aforementioned iodate under high pressure. The newly formed bonds will continuously shorten under compression but show a large resistance to the external pressure as shown in **Fig. 25**. At the highest pressure in the case of $Zn(IO_3)_2$, the metavalent bond is as short as 2.3 Å [54].

The formation of the metavalent bond between iodine and the interlayer oxygen atoms under compression is believed to be the main cause of the observed high pressure behaviors of several metal iodates. These behaviors include: **i)** The pressure-induced isostructural phase transitions in $Fe(IO_3)_3$ [52], $Zn(IO_3)_2$ [54], $Co(IO_3)_2$ [57], hydrated $Ca(IO_3)_2$ [129], $BiOIO_3$ [31] as well as the nonlinear behavior of the lattice parameter of $Li_2Ti(IO_3)_6$ under compression; **ii)** The formation of extra bonds which increase the bulk modulus of metal iodates at high pressure as shown in next section, due to the new formed bonds which enhance the repulsion force between the iodine and interlayer oxygen atoms; **iii)** The formation of the metavalent bonds which causes the nonlinear behavior of the Raman and infrared modes under compression [30–32,52,54,57,59,61]. The pressure coefficient change of the Raman modes arises from the interaction between iodine and interlayer oxygen atoms. This effect also includes the pressure-induced softening of the Raman modes in the high-frequency region, which is a consequence of the enlargement of the covalent I-O bonds; **iv)** The formation of metavalent bonds also causes the nonlinear change of the bandgap energy for several metal iodates plotted in **Fig. 14** in **Section 6.3.1**. The shortening of the distance between iodine and interlayer oxygen atoms favors increasing the interaction between them, thereby enlarging the difference between the bonding and anti-bonding states of the *p*-*p* interaction in the MO diagram around the Fermi level, thus leading to an opening of the bandgap; **v)** The formation of metavalent bonds which suppresses the LEP in iodine. This processes always involves a charge transfer from iodine and the extra oxygen atoms [14]. As a consequence, the polarity in $[IO_3]^-$ pyramids is reduced, thereby reducing the intensity of the SHG signal; **vi)** The transformation from the halogen bond to metavalent bonds, which also causes the decrease of the resistivity in iodates, as the electrons will be shared in several metavalent bond and the mobility of the electrons are improved.

## 8.3 Equation of states, axial and bulk compressibility

| Compounds | Phase | Method | $V_0$ (Å$^3$) | $B_0$ (GPa) | $B_0$' | Ref. |
|---|---|---|---|---|---|---|
| $LiIO_3$ | α ($P6_3$) | Exp. | 134.5 | 55(3) | 2.9(0.4) | [27] |
| | α ($P6_3$) | Exp. | 134.5 | 34(3) | 5.0(0.8) | [42] |
| | α ($P6_3$) | Theo. | 134.4 | 34(1) | 4.2(0.2) | [29] |
| | α ($P6_3$) | Theo. | 138.38 | 38.9(13) | 2.84 | [44] |
| | β($P4_2$/n) | Theo. | 592.65 | 67.4 | 4.62 | [49] |
| $KIO_3$ | phase III (P1) | Exp. | 355.3 | 24.3(5) | 4.0 | [28] |
| | HP I (*R*3) | Exp. | 152.3(4) | 67(3) | 4.0 | |
| | HP I | Theo. | 150.8 | 70.9 | 4.0 | |
| | HP I | Theo. | 150.8 | 67.9 | 5.9 | |
| $Fe(IO_3)_3$ | LP ($P6_3$) | Exp. | 385 | 55(2) | 4.3(0.3) | [29] |
| | LP ($P6_3$) | Theo. | 374 | 36(1) | 4.6(0.1) | |
| | HP ($P6_3$) | Exp. | 353(9) | 73(9) | 4.0 | |
| | HP ($P6_3$) | Theo. | 346(5) | 48(3) | 4.0 | |
| $Zn(IO_3)_2$ | Ambient+I1 +I2 ($P2_1$) | Exp. | 264 | 21.6(0.7) | 7.0(0.3) | [54] |
| | | Theo. | 258 | 18.4(1.5) | 6.7(0.7) | |
| $Co(IO_3)_2$ | Ambient+I1 ($P2_1$) | Exp. | 529 | 29.8(1) | 3.5(0.3) | [57] |
| | | Theo. | 505 | 32.1(1.1) | 2.3(0.3) | |
| | I2 ($P2_1$) | Exp. | 462.9(5.6) | 70.8(3.6) | 5.2(0.6) | |
| | | Theo. | 451.1(2.6) | 50.7(1.6) | 5.5(0.3) | |
| $Mg(IO_3)_2$ | LP ($P2_1$) | Exp. | 552.8 | 22.2(0.8) | 4.2(0.4) | [30] |
| | | Theo. | 540 | 26.4(0.8) | 2.9(0.2) | |
| | HP ($P3$) | Exp. | 369.6(3) | 63.6(0.4) | 3.3(0.1) | |
| | | Theo. | 363.9(5) | 44.4(2.4) | 3.6(0.4) | |
| $BiOIO_3$ | $BiOIO_3$-I (Pca21) | Exp. | 357.3(3) | 63.6(8) | 4.0 | [31] |
| | $BiOIO_3$-II (Pca21) | Exp. | 356.2(4) | 65.6(2) | 4.0 | |
| | $BiOIO_3$-III ($P2/n$) | Exp. | 336.5(12) | 88.0(3) | 4.0 | |
| $Li_2Ti(IO_3)_6$ | below 7 GPa ($P6_3$) | Exp. | 385.0(1) | 59.0(1) | 4.0 | [32] |
| $Ca(IO_3)_2{\cdot}H_2O$ | Phase I+II+III ($P2_1/c$) | Exp. | 640.0 | 39.0(7) | 4.0(1) | [129] |
| $Na_3Bi(IO_3)_6$ | α ($P\bar{1}$) | Exp. | 437.4(4) | 30.4(7) | 6.3(2) | [60] |
| $Hg(IO_3)_2$ | $P2_1$ | Theo. | 305.4 | 108.4 | 4.83 | [48] |

***Table 3.*** *A summary of the experiment (Exp.) and theoretical calculation (Theo.) determined equations of state of several metal iodates reported in the literature. Including zero-pressure unit-cell volume ($V_0$), bulk modulus ($B_0$) and the related pressure derivate ($B_0$'). Reference are shown in the right-hand column.*

The bulk modulus is a physical quantity that measures the compressibility of a crystal structure. The larger the bulk modulus, the smaller the change in unit-cell volume under pressure. The bulk moduli of several metal iodates in different phases reported in the literature are listed in **Table 3**, as well as the zero-pressure unit-cell volume and pressure derivate of the bulk modulus. The bulk modulus and pressure derivate are usually obtained by fitting the unit-cell volume under pressure with a second or third order Birch-Murnaghan equation of states (BM EOS) [152,153]. The third-order BM EOS is as follows:

$$P(V) = \frac{3}{2}B_0\left(\left(\frac{V_0}{V}\right)^{7/3} - \left(\frac{V_0}{V}\right)^{5/3}\right) \times \left(1 + \frac{3}{4}(B_0' - 4)\left(\left(\frac{V_0}{V}\right)^{2/3} - 1\right)\right)$$

where $P$ is the pressure, $V$ is the unit-cell volume at different pressure, $V_0$ is the zero-pressure volume, $B_0$ is the bulk modulus and $B_0$' is the pressure derivate. When the EOS is truncated at the second-order $B_0$'=4.0.

In experiment, for the ambient-pressure phase, the bulk moduli of $Zn(IO_3)_2$ [54], $Mg(IO_3)_2$ [30], $Co(IO_3)_2$ [57], $KIO_3$ [28], and $Na_3Bi(IO_3)_6$ [60] are comparable, ranging from 20 GPa to 30 GPa. The bulk modulus of α-$LiIO_3$ is under debate, with the two different experimentally determined bulk modulus values being 34(3) GPa [42] and 55(3) GPa [27]. The bulk modulus predicted from theoretical calculations is 34(1) GPa [29] or 38.9 GPa [44]. The bulk moduli of $Fe(IO_3)_3$ [29], β-$LiIO_3$ [49], $BiOIO_3$ [31] and $Li_2Ti(IO_3)_6$ [32] are much larger than those in the other metal iodates shown in **Table 3**, being around 60 GPa for the ambient-pressure phases.

The bulk moduli of the HP phases of the metal iodates which have been listed in **Table 3** are all higher than those of the ambient-pressure or low-pressure phases according to both experiments and calculations. In experiments, the bulk modulus of the HP phase in $Co(IO_3)_2$ [57] is twice as large as that of ambient-pressure phase, and it is three times larger in the HP phases of $Mg(IO_3)_2$ [30] and $KIO_3$ [28] compared to their respective low-pressure phases. The bulk moduli of the $BiOIO_3$-I and $BiOIO_3$-II phases are almost equal, as the observed transformation is an isostructural phase transition and there is no abrupt discontinuity in the unit-cell volume [30]. The increase of the bulk modulus at HP could be attributed to the decrease of the unit-cell volume. Furthermore, the pressure-induced transformation of the weak I···O halogen bond into metavalent bonds between iodine and oxygen atoms in the neighboring $[IO_3]^-$ pyramids also contributes to the increase of the bulk modulus at HP, since the atomic force of the metavalent bond is stronger than that in halogen intermolecular interaction as **Fig. 25** shown. After the formation of the extra bonds, the distance shortening between iodine and interlayer oxygen atoms becomes significantly slower due to change of stiffness.

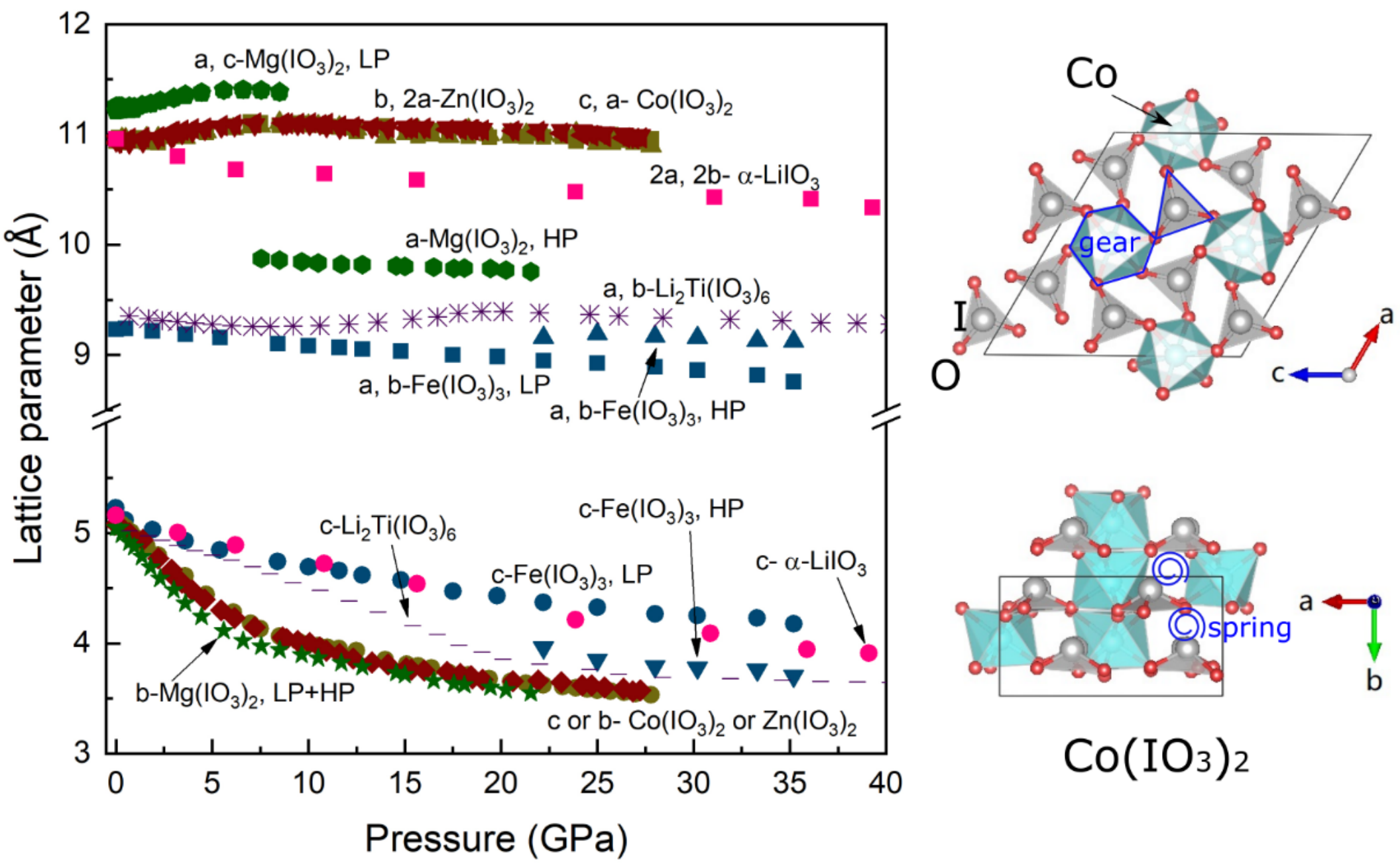

***Fig. 27.*** *Left: the lattice parameter for several metal iodates, crystalizing in the* $P6_3$ *and* $P2_1$ *space groups, as a function of pressure. In space group* $P6_3$*, it has α-*$LiIO_3$ [27], $Fe(IO_3)_3$ [29] *and* $Li_2Ti(IO_3)_6$ [32]. *In space group* $P2_1$*, it includes* $Zn(IO_3)_2$ [54], $Co(IO_3)_2$ [57], *and* $Mg(IO_3)_2$

[30]. *For a better comparison, the a-axis of $Zn(IO_3)_2$, and the a and b axes of α-$LiIO_3$ have been doubled in length. Right: the crystal structure of $Co(IO_3)_2$ at ambient pressure as a representation of the gear-spring model, atoms are shown in different color and labeled besides.*

There is a group of metal iodates whose crystal structure is described by $P6_3$ (No. 173) or $P2_1$(No. 4) space group. They are summarized in **Table 1** in **Section 3**, and the representation of the crystal structure can be found in **Fig. 2** in the same section. The crystal structure exhibits a layered structure along one direction. Within the layer, it contains the $IO_3$ pyramids which are aligned in a parallel manner, with each $IO_3$ layer is connected by $AO_6$ (A=metal) polyhedra. Therefore, the metal iodates with this type of crystal structure usually show a huge anisotropic behavior under pressure, as shown in **Fig. 27** which shows the pressure dependence of lattice parameters of three metal iodates crystallizing in space group $P6_3$ (α-$LiIO_3$ [27], $Fe(IO_3)_3$ [29], and $Li_2Ti(IO_3)_6$ [32]) and three metal iodates crystallizing in space group $P2_1$ ($Zn(IO_3)_2$ [54], $Mg(IO_3)_2$ [30] and $Co(IO_3)_2$ [57]). Their lattice parameters show the same behavior under pressure. The axis along which the $IO_3$ pyramids are aligned is the most compressible. Additionally, the other two axes show a nonlinear behavior under compression and expand in some pressure regions. The lengths of those two axes usually vary less under compression. In the HP study of $Li_2Ti(IO_3)_6$, a gear-spring model has been adopted to explain this kind of anisotropic behavior [32]. Since all the metal iodates in $P6_3$ and $P2_1$ share the same behavior it is reasonable to use this model to explain the crystal structure behavior of those two types of metal iodates under pressure. Here the crystal structure of $Co(IO_3)_2$ is used as an example (the right hand of **Fig. 27**). The rigid $CoO_6$ is the gear, as well as the $IO_3$ pyramid, the two-gears are alternately arranged along the *a*- and *c*-axes, the big gap and weak interaction between the $IO_3$ layer is the spring, it is easier to be compressed. In addition, the *b*-axis shows two different pressure coefficients at HP, which is a consequence of the formation of new metavalent bond between iodine and the oxygen atoms in the neighboring $IO_3$ layer, in agreement with the pressure dependence of I-O bonds distance (**Fig. 25**).

There is another family of metal iodates which crystallize in triclinic or monoclinic crystal system at ambient conditions and the $IO_3$ pyramid is randomly oriented relative the crystal axes. This family contains: $KIO_3$ (space group $P1$) [28], $BiOIO_3$ (space group $P\bar{1}$) [31], hydrated $Ca(IO_3)_2$ (space group $P2_1/c$) [129] and $Na_3Bi(IO_3)_6$ (space group $P\bar{1}$) [60]. The pressure-response of these iodates is less anisotropic due to the absence of the layered structure.

## 8.4 Polyhedral compressibility

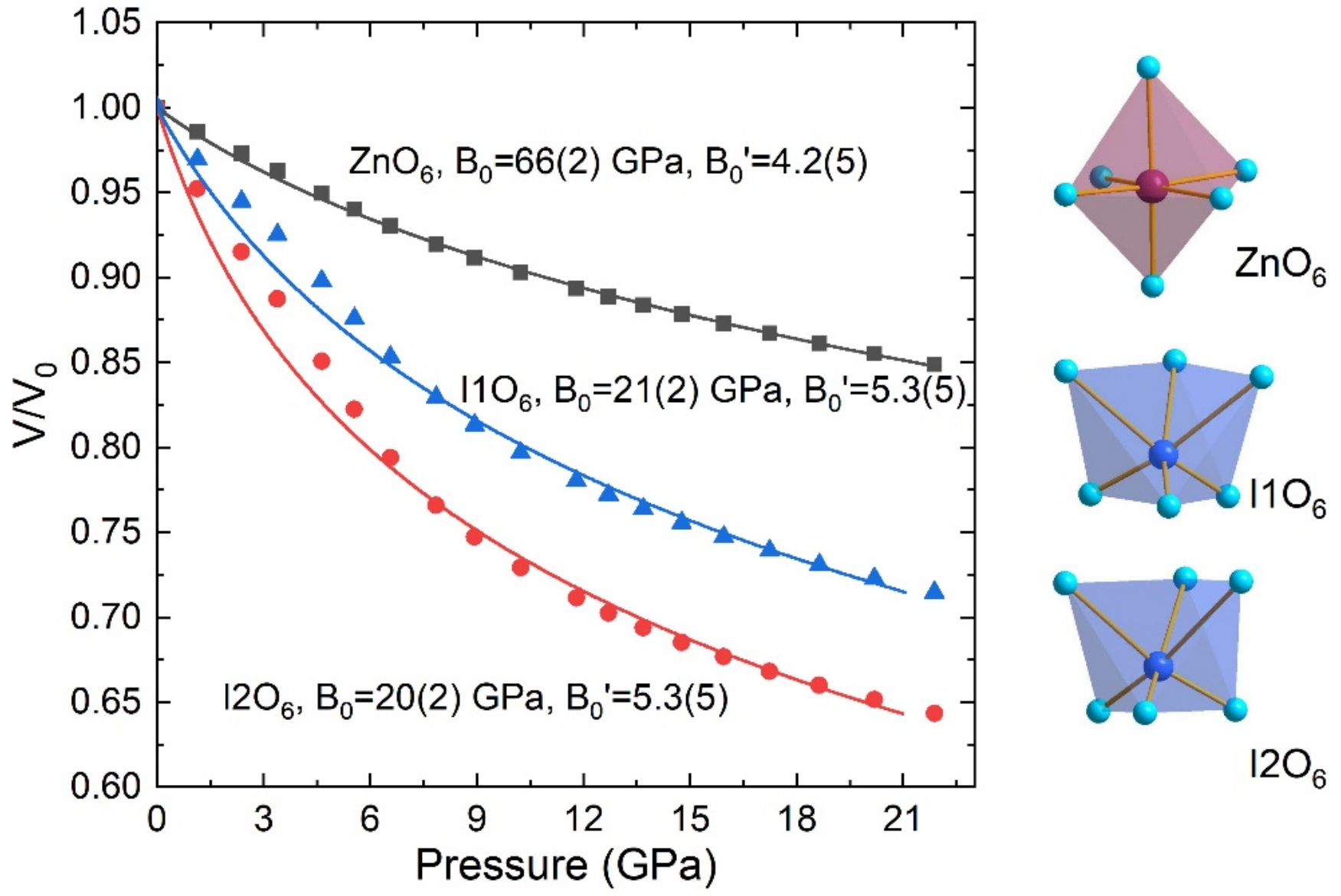

***Fig. 28.*** *Normalized volumes of $ZnO_6$, $I1O_6$ and $I2O_6$ polyhedra in the crystal structure of $Zn(IO_3)_2$ as a function of pressure (scattered symbols)* [54]. *The solid lines are the fits of the volume using a third-order BM EOS. The representation of the three polyhedra are shown in the right-hand of the figure.*

If a 3+3 coordination of iodine is assumed at ambient and high pressure, then $IO_6$ and $AO_6$ polyhedra are present in most of the metal iodates, even though there are some cases where the coordination of the metal is more than 6 [69]. In the crystal structure of $Zn(IO_3)_2$, zinc is located in one Wyckoff position and iodine is located in two different Wyckoff positions [78], so there is one kind of $ZnO_6$ and two kinds of $IO_6$ polyhedra (**Fig. 28**). By adopting a third-order BM EOS, the bulk moduli of $ZnO_6$, $I1O_6$ and $I2O_6$ are determined to be 66(2), 21(2) and 20(2) GPa, respectively (**Fig. 28**) [54]. The bulk modulus of $IO_6$ is comparable with that of the bulk modulus of $Zn(IO_3)_2$ (**Table 3**). Additionally, the bulk modulus of $ZnO_6$ is three times larger than that of $Zn(IO_3)_2$. Therefore, the compressibility of $Zn(IO_3)_2$ is dominated by that of $IO_6$ units rather than $ZnO_6$. This is due to the fact that the $IO_6$ polyhedra contain the three weak long I···O halogen bonds, and those three halogen bonds are more susceptible to compression [54]. This same feature also has been observed in the following metal iodates: $Fe(IO_3)_3$ [29], $Co(IO_3)_2$ [57] and $Mg(IO_3)_2$ [30]. It is reasonable therefore to predict that this is a general feature of all metal iodates, since all of them contain the iodine LEP and compressible weak I···O halogen bonds.

# 9. Future targets for the high-pressure studies of iodates

This section presents potential avenues for future research pertaining to high pressure iodate studies. The first obvious avenue is the study of the many metal iodates which have not been yet studied, including, but not limited to: $Cd(IO_3)_2$ [154], $RbIO_3$ [155], $Zr(IO_3)_4$ [156], $Ba(IO_3)_2$ [82], $Al(IO_3)_3$ [90], and $La(IO_3)_3$ [12]. All of these compounds share a common feature which is that the $IO_3$ groups form trigonal pyramids wherein iodine is closely bonded to three O atoms at an average distance of approximately 1.8 Å. The iodine atom is always at the vertex of the pyramid with the LEP pointing in the direction opposing the base of the pyramid. The O atoms which are bonded to I are also strongly bonded to the metal cation [Cd, Rb, Ba, Zr, Al, or La] with the $IO_3$ groups being corner linked to the coordination polyhedra of the metal cations. This fact strongly suggests that most metal iodates are likely to undergo isostructural phase transitions at relative low pressures below 10 GPa.

An area that needs to be systematically explored in the future is the transformation of halogen bonds into metavalent bonds. As described in this article, there is a lot of indirect evidence supporting the formation of new I-O bonds in iodates which has been tentatively assigned to the probable formation of pressure-induced metavalent bonds. This is a phenomenon well documented in chalcogenide compounds [16,157]. To fully prove the presence of metavalent bonding in iodates at HP, systematic calculations of the electronic density of I−O bonds, Born effective charges of I atoms, and optical dielectric constants under pressure are required. Such calculations as the target of forthcoming comprehensive DFT studies which should include calculations of the ELF and the reduced density gradient of the electron density [158]. X-ray absorption spectroscopy could only contribute to deepening the understanding of the pressure evolution of I-O and the coordination changes induced by pressure. These measured could be performed at the iodine K-edge (33.1 keV) using a diamond anvil cell or a Paris-Edinburgh press [157].

Studies at low-temperature and high-temperature could be one very interesting subject for future studies involving metal iodates with magnetic cations. In $Fe(IO_3)_3$ the existence of magnetic skyrmions has been found, which represent peculiar magnetic configurations with unique physical properties and potential applications in spintronic devices [149]. Their existence is favored by the coalignment of the stereoactive LEP of the $IO_3$ groups and the magnetic $Fe^{3+}$ cations. Similar phenomena have been reported for $Cu(IO_3)_2$ and $Mn(IO_3)_2$ [159] and would potentially exist in $Ni(IO_3)_2$ and $Co(IO_3)_2$. The study of the magnetic behavior of magnetic iodates at HP and low temperatures would be not only useful for understanding their magnetic properties but could also trigger the observation of new magnetic features due to the influence of pressure in magnetic coupling [160].

The discovery that $Li_2Ti(IO_3)_6$ is a material with zero-linear and zero-area compressibility makes this material a promising one for specific applications retaining constancy in specific directions or planes under external impacts [32]. A close examination of compressibility results of other iodates already studied and discussed in this review suggest that zero-linear and zero-area compressibility is a common characteristic of many metal iodates which deserves to be studied in detail. In particular, the possibility to use iodates as protective overcoats to improve the service life of substrate materials should be explored [161].

In $BiOIO_3$ the application of pressure lead to a two-step switch-off of the SHG signal [31]. Based on this result it has been proposed that $BiOIO_3$ could be used for high-capacity information storage and switch devices controllable by external pressure. For such applications it would be interesting to employ materials which are not SHG active where compression or stresses could switch-on of SHG activity. Good candidates for this are materials which undergo phase transitions from centrosymmetric to non-centrosymmetric structures. An excellent candidate for it is centrosymmetric $ZnIO_3F$ which under the application of pressure could probably take the crystal structure of non-centrosymmetric $CdIO_3F$ [162]. Another candidate could be $Na_3Bi(IO_3)_6$, since the single-crystal XRD results have revealed a centrosymmetric structure ($P\bar{1}$) to non-centrosymmetric structure ($P1$) transition at around 9.5 GPa [60].

Another interesting issue to explore in the future would be revisiting the HP behavior of α-$LiIO_3$ to search for missing phase transitions and to study the elastic constants by DFT calculations and Brillouin scattering. Elastic constants have been studied only for $Fe(IO_3)_3$ up to now [52]. It is also of interest the performance of HP-HT studies with the goal of creating new phases which could be recovered as metastable phases at ambient conditions as reported for $AgIO_3$ [47]. HP-HT could be also important for exploring the potential of some iodates as potential barocaloric materials. In particular, $KIO_3$ has been reported to become a superionic material at high temperature [163] therefore becoming a potential candidate to study large barocaloric effects [164].

To conclude, it should be added that the study of iodates under compression gives important hints for understanding the HP behavior of related compounds such as bromates which have crystal structures containing $BrO_3^-$ where one lone pair of electrons is present. In contrast to iodates, up to now only KBrO3 has been studied under compression [164].

# 10. Concluding remarks

This review systematically presents and examines the progress made during recent years on the study of high-pressure (HP) effects on the structural, vibrational, and electronic properties of metal iodates. The review focusses on the occurrence of pressure-induced phase transitions, which have been discussed comprehensively. Results obtained from X-ray diffraction (XRD), Raman and infrared spectroscopy, optical-absorption, resistivity measurements, and second-harmonic generation (SHG) measurements have been reported and discussed. Recent results obtained from density-functional theory (DFT) calculations which have been used for the interpretation of experiments are also described.

The systematics of the HP behavior of iodates are discussed in connection with the role played by the lone electrons pair (LEP) of iodine. The influence of the LEP in the formation of new bonds under compression is a crucial factor in the HP behavior of iodates. The presence of the LEP gives unique characteristics to the response of iodates to compression leading to unusual behaviors of the physical properties which have been discussed in detail throughout the review.

A systematic understanding of the band structure of iodates and their HP behavior has been discussed and, based upon the discussion of the already existing results, two rules for designing wide bandgap iodates have been proposed. Finally, possible implications of the reviewed results for technological applications have been commented on and possible directions for the future research on iodates have been proposed and discussed.

## Acknowledgements

This work was supported by the Generalitat Valenciana under Grants No. PROMETEO CIPROM/2021/075-GREENMAT and MFA/2022/007 and by the Spanish Ministerio de Ciencia e Innovación, Agencia Estatal de Investigación, and the European Union under (MCIN/AEI/10.13039/501100011033) under grants PID2019-106383GB-41 and RED2018-102612-T (MALTA Consolider Team). A.L. and D.E. thank the Generalitat Valenciana for the Ph.D. Fellowship No. GRISOLIAP/2019/025. R.T. and D.E. thank the Generalitat Valenciana for the postdoctoral Fellowship No. CIAPOS/2021/20.

# References

[1] S.P. Guo, Y. Chi, G.C. Guo, Recent achievements on middle and far-infrared second-order nonlinear optical materials, Coord. Chem. Rev. 335 (2017) 44.

[2] L. Xiao, Z. Cao, J. Yao, Z. Lin, Z. Hu, A new cerium iodate infrared nonlinear optical material with a large second-harmonic generation response, J. Mater. Chem. C. 5 (2017) 2130.

[3] A. You, M.A.Y. Be, I. In, Paramagnetic $Gd(IO_3)_3$. Crystal structure of the transition metal iodates. IV, J. Chem. Phys. 67 (1977) 1015.

[4] R. Liminga, S.C. Abrahams, J.L. Bernstein, Pyroelectric α-$Cu(IO_3)_2$. Crystal structure of the transition metal iodates. III, J. Chem. Phys. 62 (1975) 4388.

[5] P.S. Halasyamani, K.R. Poeppelmeier, Noncentrosymmetric Oxides, Chem. Mater. 10 (1998) 2753.

[6] R.E. Sykora, M.O. Kang, P.S. Halasyamani, T.E. Albrecht-Schmitt, Structural modulation of molybdenyl iodate architectures by alkali metal cations in $AMoO_3(IO_3)$ (A = K, RB, Cs): A facile route to new polar materials with large SHG responses, J. Am. Chem. Soc. 124 (2002) 1951.

[7] R.E. Sykora, K.M. Ok, P.S. Halasyamani, D.M. Wells, T.E. Albrecht-schmitt, New One-Dimensional Vanadyl Iodates: Hydrothermal Preparation, Structures, and NLO Properties of $A[VO_2(IO_3)_2]$ (A = K, Rb) and $A(VO)_2(IO_3)_3O_2]$ (A = $NH_4$, Rb, Cs), Chem. Mater. 14 (2002) 2741.

[8] M.O. Kang, P.S. Halasyamani, New metal iodates: Syntheses, structures, and characterizations of noncentrosymmetric $La(IO_3)_3$ and $NaYI_4O_{12}$ and centrosymmetric β-$Cs_2I_4O_{11}$ and $Rb_2I_6O_{15}(OH)_2{\cdot}H_2O$, Inorg. Chem. 44 (2005) 9353.

[9] D. Phanon, I. Gautier-Luneau, Promising material for infrared nonlinear optics: $NaI_3O_8$ salt containing an octaoxotriiodate(V) anion formed from condensation of $[IO_3]^-$ ions, Angew. Chem. Int. Ed. 46 (2007) 8488.

[10] H.Y. Chang, S.H. Kim, P.S. Halasyamani, K.M. Ok, Alignment of lone pairs in a new polar material: Synthesis, characterization, and functional properties of $Li_2Ti(IO_3)_6$, J. Am. Chem. Soc. 131 (2009) 2426.

[11] C.F. Sun, C.L. Hu, X. Xu, J.B. Ling, T. Hu, F. Kong, X.F. Long, J.G. Mao, $BaNbO(IO_3)_5$: A new polar material with a very large SHG response, J. Am. Chem. Soc. 131 (2009) 9486.

[12] Z. Hebboul, A. Ghozlane, R. Turnbull, A. Benghia, S. Allaoui, A. Liang, D. Errandonea, A. Touhami, A. Rahmani, I.K. Lefkaier, Simple New Method for the Preparation of $La(IO_3)_3$ Nanoparticles, Nanomaterials. 10 (2020) 2400.

[13] Y.J. Jia, Y.G. Chen, Y. Guo, X.F. Guan, C. Li, B. Li, M.M. Liu, X.M. Zhang, $LiM^{II}(IO_3)_3$ ($M^{II}$=Zn and Cd): Two Promising Nonlinear Optical Crystals Derived from a Tunable Structure Model of α-$LiIO_3$, Angew. Chem. Int. Ed. 58 (2019) 17194.

[14] Y. V. Nelyubina, M.Y. Antipin, K.A. Lyssenko, Extremely short halogen bond: The nature and energy of iodine-oxygen interactions in crystalline iodic acid,

Mendeleev Commun. 21 (2011) 250.
[15] J. Ling, T.E. Albrecht-Schmitt, Intercalation of lodic acid into the layered uranyl iodate, $UO_2(IO_3)_2(H_2O)$, Inorg. Chem. 46 (2007) 346.
[16] J.A. Sans, R. Vilaplana, E.L. Da Silva, C. Popescu, V.P. Cuenca-Gotor, A. Andrada-Chacón, J. Sánchez-Benitez, O. Gomis, A.L.J. Pereira, P. Rodríguez-Hernández, A. Muñoz, D. Daisenberger, B. García-Domene, A. Segura, D. Errandonea, R.S. Kumar, O. Oeckler, P. Urban, J. Contreras-García, F.J. Manjón, Characterization and Decomposition of the Natural van der Waals $SnSb_2Te_4$ under Compression, Inorg. Chem. 59 (2020) 9900.
[17] H.K. Mao, X.J. Chen, Y. Ding, B. Li, L. Wang, Solids, liquids, and gases under high pressure, Rev. Mod. Phys. 90 (2018) 15007.
[18] N. Dubrovinskaia, L. Dubrovinsky, N.A. Solopova, A. Abakumov, S. Turner, M. Hanfland, E. Bykova, M. Bykov, C. Prescher, V.B. Prakapenka, S. Petitgirard, I. Chuvashova, B. Gasharova, Y.L. Mathis, P. Ershov, I. Snigireva, A. Snigirev, Terapascal static pressure generation with ultrahigh yield strength nanodiamond, Sci. Adv. 2 (2016) e1600341.
[19] D. Errandonea, R.S. Kumar, F.J. Manjón, V. V. Ursaki, E. V. Rusu, Post-spinel transformations and equation of state in $ZnGa_2O_4$: Determination at high pressure by in situ x-ray diffraction, Phys. Rev. B. 79 (2009) 1.
[20] E. Bandiello, D. Errandonea, D. Martinez-Garcia, D. Santamaria-Perez, F.J. Manjón, Effects of high-pressure on the structural, vibrational, and electronic properties of monazite-type $PbCrO_4$, Phys. Rev. B. 85 (2012) 024108.
[21] R.J. Hemley, A.P. Jephcoat, H.K. Mao, L.C. Ming, M.H. Manghnani, Pressure-induced amorphization of crystalline silica, Nature. 334 (1988) 52.
[22] M.S. Torikachvili, S.L. Bud'ko, N. Ni, P.C. Canfield, Pressure induced superconductivity in $CaFe_2As_2$, Phys. Rev. Lett. 101 (2008) 057006.
[23] L. Zhang, Y. Wang, J. Lv, Y. Ma, Materials discovery at high pressures, Nat. Rev. Mater. 2 (2017) 1.
[24] P.F. McMillan, Pressing on: The legacy of Percy W. Bridgman, Nat. Mater. 4 (2005) 715.
[25] W.A. Bassett, Diamond anvil cell, 50th birthday, High. Press. Res. 29 (2009) 163.
[26] S. Klotz, J.C. Chervin, P. Munsch, G. Le Marchand, Hydrostatic limits of 11 pressure transmitting media, J. Phys. D. Appl. Phys. 42 (2009) 075413.
[27] W.W. Zhang, Q.L. Cui, Y.W. Pan, S.S. Dong, J. Liu, G.T. Zou, High-pressure x-ray diffraction study of $LiIO_3$ to 75 GPa, J. Phys. Condens. Matter. 14 (2002) 10579.
[28] L. Bayarjargal, L. Wiehl, A. Friedrich, B. Winkler, E.A. Juarez-Arellano, W. Morgenroth, E. Haussühl, Phase transitions in $KIO_3$, J. Phys. Condens. Matter. 24 (2012) 325401.
[29] A. Liang, S. Rahman, H. Saqib, P. Rodriguez-Hernandez, A. Munoz, G. Nenert, I. Yousef, C. Popescu, D. Errandonea, First-Order Isostructural Phase Transition Induced by High Pressure in $Fe(IO_3)_3$, J. Phys. Chem. C. 124 (2020) 8669.

[30] A. Liang, R. Turnbull, C. Popescu, F.J. Manjón, E. Bandiello, A. Muñoz, I. Yousef, Z. Hebboul, D. Errandonea, Pressure-induced phase transition and increase of oxygen-iodine coordination in magnesium iodate, Phys. Rev. B. 105 (2022) 054105.
[31] D. Jiang, H. Song, T. Wen, Z. Jiang, C. Li, K. Liu, W. Yang, H. Huang, Y. Wang, Pressure-Driven Two-Step Second-Harmonic-Generation Switching in $BiOIO_3$, Angew. Chem. Int. Ed. 61 (2022) e202116656.
[32] D. Jiang, T. Wen, H. Song, Z. Jiang, C. Li, K. Liu, W. Yang, H. Mao, Y. Wang, Intrinsic Zero-Linear and Zero-Area Compressibilities over an Ultrawide Pressure Range within a Gear-Spring Structure, CCS Chem. 4 (2022) 3246.
[33] M.I. Eremets, High Pressure Experimental Methods, Oxford University Press, 1996.
[34] H.K. Mao, B. Chen, J. Chen, K. Li, J.F. Lin, W. Yang, H. Zheng, Recent advances in high-pressure science and technology, Matter Radiat. Extrem. 1 (2016) 59.
[35] B. Winkler, V. Milman, Density functional theory based calculations for high pressure research, Zeitschrift Fur Krist. 229 (2014) 112.
[36] A. Mujica, A. Rubio, A. Muñoz, R.J. Needs, High-pressure phases of group-IV, III-V, and II-VI compounds, Rev. Mod. Phys. 75 (2003) 863.
[37] J. Liebertz, Dimorphie von Lithiumjodat ($LiJO_3$), Zeitschrift Für Phys. Chemie. 67 (1969) 94.
[38] J. Liu, Z. Shen, Y. Zhang, X. Yin, S. He, The P-T phase diagram of lithium iodate ($LiIO_3$) up to 40 Kbars, Acta Phys. Sin. 32 (1983) 118.
[39] F. Cerdeira, F.E.A. Melo, V. Lemos, Raman study of anharmonic effects in α-$LiIO_3$, 27 (1983) 7716.
[40] N. Choudhury, S.L. Chaplot, Ab initio studies of phonon softening and high-pressure phase transitions of α-quartz $SiO_2$, Phys. Rev. B. 73 (2006) 1.
[41] D. Errandonea, F.J. Manjón, On the ferroelastic nature of the scheelite-to-fergusonite phase transition in orthotungstates and orthomolybdates, Mater. Res. Bull. 44 (2009) 807.
[42] J. Hu, L. Chen, L. Wang, R. Tang, R. Che, Isothermal compression of α-$LiIO_3$ and its phase transition under high pressure and high temperature, Acta Phys, Sin. 36 (1987) 1099.
[43] P. Hermet, First-principles based analysis of the piezoelectric response in α-$LiIO_3$, Comput. Mater. Sci. 138 (2017) 199.
[44] G. Yao, Y. Chen, X.Y. An, Z.Q. Jiang, L.H. Cao, W.D. Wu, Y. Zhao, First-principles study of the structural, electronic and optical properties of hexagonal $LiIO_3$, Chinese Phys. Lett. 30 (2013) 067101.
[45] J. Mendes Fílho, V. Lemos, F. Cerdeira, R.S. Katiyar, Raman and x-ray studies of a high-pressure phase transition in β-$LiIO_3$ and the study of anharmonic effects, Phys. Rev. B. 30 (1984) 7212.
[46] Z.X. Shen, X.B. Wang, S.H. Tang, H.P. Li, F. Zhou, High Pressure Raman Study and Phase Transitions of $KIO_3$ Non-Linear Optical Single Crystals, Rev.

High Press. Sci. Technol. 7 (1998) 751.
[47] Y. Suffren, I. Gautier-Luneau, C. Darie, C. Goujon, M. Legendre, O. Leynaud, First evidence of a phase transition in a high-pressure metal iodate: Structural and thermal studies of $AgIO_3$ polymorphs, Eur. J. Inorg. Chem. 2013 (2013) 3526.
[48] B. Lagoun, B. Bentria, I.K. Lefkaier, Ab initio calculation of structural, electronic and optical properties of $Hg(IO_3)_2$, Phys. B Condens. Matter. 433 (2014) 117.
[49] G. Yao, X. An, Y. Chen, Y. Fu, Z. Jiang, Y. Liu, First-principles study of the structural, electronic and optical properties of tetragonal $LiIO_3$, Comput. Mater. Sci. 84 (2014) 350.
[50] D. Errandonea, C. Popescu, S.N. Achary, A.K. Tyagi, M. Bettinelli, In situ high-pressure synchrotron X-ray diffraction study of the structural stability in $NdVO_4$ and $LaVO_4$, Mater. Res. Bull. 50 (2014) 279.
[51] D. Errandonea, O. Gomis, D. Santamaría-Perez, B. García-Domene, A. Muñoz, P. Rodríguez-Hernández, S.N. Achary, A.K. Tyagi, C. Popescu, Exploring the high-pressure behavior of the three known polymorphs of $BiPO_4$: Discovery of a new polymorph, J. Appl. Phys. 117 (2015) 105902.
[52] A. Liang, S. Rahman, P. Rodriguez-Hernandez, A. Muñoz, F.J. Manjón, G. Nenert, D. Errandonea, High-pressure Raman study of $Fe(IO_3)_3$: Soft-mode behavior driven by coordination changes of iodine atoms, J. Phys. Chem. C. 124 (2020) 21329.
[53] A. Liang, P. Rodríguez-Hernandez, A. Munoz, S. Raman, A. Segura, D. Errandonea, Pressure-dependent modifications in the optical and electronic properties of $Fe(IO_3)_3$: The role of Fe 3*d* and I 5*p* lone-pair electrons, Inorg. Chem. Front. 8 (2021) 4780.
[54] A. Liang, C. Popescu, F.J. Manjon, A. Muñoz, Z. Hebboul, D. Errandonea, Structural and vibrational study of $Zn(IO_3)_2$ combining high-pressure experiments and density-functional theory, Phys. Rev. B. 103 (2021) 054102.
[55] A. Liang, R. Turnbull, E. Bandiello, I. Yousef, C. Popescu, Z. Hebboul, D. Errandonea, High-Pressure Spectroscopy Study of $Zn(IO_3)_2$ Using Far-Infrared Synchrotron Radiation, Crystals. 11 (2021) 34.
[56] A. Liang, R. Turnbull, P. Rodríguez-hernandez, A. Muñoz, M. Jasmin, L. Shi, D. Errandonea, General relationship between the band-gap energy and iodine-oxygen bond distance in metal iodates, Phys. Rev. Mater. 6 (2022) 044603.
[57] A. Liang, C. Popescu, F.J. Manjon, R. Turnbull, E. Bandiello, P. Rodriguez-Hernandez, A. Muñoz, I. Yousef, Z. Hebboul, D. Errandonea, Pressure-Driven Symmetry-Preserving Phase Transitions in $Co(IO_3)_2$, J. Phys. Chem. C. 125 (2021) 17448.
[58] A. Liang, F. Rodríguez, Rodríguez-Hernandez，P, A. Muñoz, R. Turnbull, D. Errandonea, High-pressure tuning of *d–d* crystal-field electronic transitions and electronic band gap in $Co(IO_3)_2$, Phys. Rev. B. 105 (2022) 115204.
[59] H. Song, D. Jiang, N. Wang, W. Xing, R. Guo, Z. Lin, J. Yao, Y. Wang, H. Tu, G. Zhang, $Na_3Bi(IO_3)_6$: An Alkali-Metal Bismuth Iodate with Intriguing One-

Dimensional [$BiI_6O_{18}$] Chains and Pressure-Induced Structural Transition, Inorg. Chem. 60 (2021) 2893.
[60] R. Turnbull, J. González-Platas, A. Liang, D. Jiang, Y. Wang, C. Popescu, P. Rodriguez-Hernandez, A. Munoz, J. Ibáñez, D. Errandonea, Pressure-Induced Phase Transition and Band-Gap Decrease in Semiconducting $Na_3Bi(IO_3)_6$, Results Phys. 44 (2023) 106156.
[61] B.B. Sharma, P.S. Ghosh, A.K. Mishra, H.K. Poswal, Hyper-coordinated iodine in $HIO_3$ under pressure, Vib. Spectrosc. 117 (2021) 103318.
[62] H.Y. Chang, S.H. Kim, M.O. Kang, P.S. Halasyamani, Polar or nonpolar? $A^+$ cation polarity control in $A_2Ti(IO_3)_6$ (A = Li, Na, K, Rb, Cs, Tl), J. Am. Chem. Soc. 131 (2009) 6865.
[63] F.F. Mao, C.L. Hu, J. Chen, B.L. Wu, J.G. Mao, $HBa_{2.5}(IO_3)_6(I_2O_5)$ and $HBa(IO_3)(I_4O_{11})$: Explorations of Second-Order Nonlinear Optical Materials in the Alkali-Earth Polyiodate System, Inorg. Chem. 58 (2019) 3982.
[64] X. Xu, B. Yang, C. Huang, J. Mao, Explorations of New Second-Order Nonlinear Optical Materials in the Ternary Rubidium Iodate System: Noncentrosymmetric β-$RbIO_3(HIO_3)_2$ and Centrosymmetric $Rb_3(IO_3)_3(I_2O_5)(HIO_3)_4(H_2O)$, Inorg. Chem. 53 (2014) 1756.
[65] X. Xu, C. Hu, B. Li, B. Yang, J. Mao, α-$AgI_3O_8$ and β-$AgI_3O_8$ with Large SHG Responses: Polymerization of $IO_3$ Groups into the $I_3O_8$ Polyiodate Anion, Chem. Mater. 26 (2014) 3219.
[66] K.M. Ok, P.S. Halasyamani, The Lone-Pair Cation $I^{5+}$ in a Hexagonal Tungsten Oxide-Like Framework: Synthesis, Structure, and Second-Harmonic Generating Properties of $Cs_2I_4O_{11}$, Angew. Chem. 116 (2004) 5605.
[67] D. Phanon, B. Bentria, E. Jeanneau, D. Benbertal, A. Mosset, I. Gautier-Luneau, Crystal structure of $M(IO_3)_2$ metal iodates, twinned by pseudo-merohedry, with $M^{II}$: $Mg^{II}$, $Mn^{II}$, $Co^{II}$, $Ni^{II}$ and $Zn^{II}$, Zeitschrift Fur Krist. 221 (2006) 635.
[68] S.A. Hamid, Symmetrie von $KJO_3$ und die Struktur der Zimmertemperaturphase, Zeitschrift Fur Krist. - New Cryst. Struct. 137 (1973) 412.
[69] J. Yeon, S.H. Kim, P.S. Halasyamani, New thallium iodates-Synthesis, characterization, and calculations of $Tl(IO_3)_3$ and $Tl_4(IO_3)_6$, [$Tl^+{}_3Tl^{3+}(IO_3)_6$], J. Solid State Chem. 182 (2009) 3269.
[70] S.D. Nguyen, J. Yeon, S.H. Kim, P.S. Halasyamani, $BiO(IO_3)$: A new polar iodate that exhibits an Aurivillius-type $(Bi_2O_2)^{2+}$ layer and a large SHG response, J. Am. Chem. Soc. 133 (2011) 12422.
[71] R. Mass, J. Claude Guitel, Préparation chimique et structure cristalline de I´iodate d´ argent $AgIO_3$, J. Solid State Chem. 32 (1980) 177.
[72] J.L. De Boer, F. van Bolhuis, R.V. Olthof-Hazekamp, Re-investigation of the crystal structure of lithium iodate, Acta Crystallogr. 21 (1966) 841.
[73] C. Huang, C.L. Hu, X. Xu, B.P. Yang, J.G. Mao, Explorations of a series of second order nonlinear optical materials based on monovalent metal gold(III) iodates, Inorg. Chem. 52 (2013) 11551.

[74] M.T. Rogers, L. Helmholz, The crystal structure of iodic acid, J. Am. Chem. Soc. 63 (1941) 278.
[75] M. Zhang, C. Hu, T. Abudouwufu, Z. Yang, S. Pan, Functional Materials Design via Structural Regulation Originated from Ions Introduction: A Study Case in Cesium Iodate System, Chem. Mater. 30 (2018) 1136.
[76] Q. Wu, H. Liu, F. Jiang, L. Kang, L. Yang, Z. Lin, Z. Hu, X. Chen, X. Meng, J. Qin, $RbIO_3$ and $RbIO_2F_2$: Two Promising Nonlinear Optical Materials in Mid-IR Region and Influence of Partially Replacing Oxygen with Fluorine for Improving Laser Damage Threshold, Chem. Mater. 28 (2016) 1413.
[77] M. Luo, F. Liang, X. Hao, D. Lin, B. Li, Z. Lin, N. Ye, Rational Design of the Nonlinear Optical Response in a Tin Iodate Fluoride $Sn(IO_3)_2F_2$, Chem. Mater. 32 (2020) 2615.
[78] J.K. Liang, C.G. Wang, The structure of $Zn(IO_3)_2$ Crystal, Acta Chim. Sin. 40 (1982) 985.
[79] Y.J. Jia, Y.G. Chen, T. Wang, Y. Guo, X.F. Guan, X.M. Zhang, $KBi(IO_3)_3(OH)$ and $NaBi(IO_3)_4$: From the centrosymmetric chain to a noncentrosymmetric double layer, Dalt. Trans. 48 (2019) 10320.
[80] S.J. Oh, H.G. Kim, H. Jo, T.G. Lim, J.S. Yoo, K.M. Ok, Photoconversion Mechanisms and the Origin of Second-Harmonic Generation in Metal Iodates with Wide Transparency, $NaLn(IO_3)_4$ (Ln = La, Ce, Sm, and Eu) and $NaLa(IO_3)_4{:}Ln^{3+}$ (Ln = Sm and Eu), Inorg. Chem. 56 (2017) 6973.
[81] S. Ghose, C. Wan, O. Wittke, The crystal structure of synthetic lautarite, $Ca(IO_3)_2$, Acta Crystallogr. Sect. B. B34 (1978) 84.
[82] V. Petříček, K. Malý, B. Kratochvíl, J. Podlahová, J. Loub, Barium diiodate, Acta Crystallogr. Sect. B. B36 (1980) 2130.
[83] E.E. Oyeka, M.J. Winiarski, T.T. Tran, Study of integer spin s = 1 in the polar magnet $\beta$-$Ni(IO_3)_2$, Molecules. 26 (2021) 7210.
[84] C.F. Sun, T. Hu, X. Xu, J.G. Mao, Syntheses, crystal structures, and properties of three new lanthanum(iii) vanadium iodates, Dalt. Trans. 39 (2010) 7960.
[85] T. Kellersohn, E. Alici, D. Eßer, H.D. Lutz, $Pb(IO_3)_2$ I – Das erste Halogenat eines zweiwertigen Hauptgruppenmetalls mit Schichtenstruktur – Kristallstruktur, IR- und Ramanspektren, Zeitschrift Für Krist. 203 (1993) 225.
[86] G. Park, H.R. Byun, J.I. Jang, K.M. Ok, Dimensionality-Band Gap-Third-Harmonic Generation Property Relationship in Novel Main-Group Metal Iodates, Chem. Mater. 32 (2020) 3621.
[87] R.E. Sykora, Z. Assefa, R.G. Haire, T.E. Albrecht-Schmitt, Hydrothermal synthesis, structure, Raman spectroscopy, and self-irradiation studies of $^{248}Cm(IO_3)_3$, J. Solid State Chem. 177 (2004) 4413.
[88] M. Jansen, Zur Kistallstruktur von $FeJ_3O_9$, J. Solid State Chem. 17 (1976) 1.
[89] Y.H. Kim, T.T. Tran, P.S. Halasyamani, K.M. Ok, Macroscopic polarity control with alkali metal cation size and coordination environment in a series of tin iodates, Inorg. Chem. Front. 2 (2015) 361.
[90] X. Chen, H. Xue, X. Chang, H. Zang, W. Xiao, Hydrothermal synthesis and crystal structures of $Nd(IO_3)_3$ and $Al(IO_3)_3$, J. Alloys Compd. 398 (2005) 173.

[91] B. Bentria, D. Benbertal, M. Bagieu-Beucher, R. Masse, A. Mosset, Crystal structure of anhydrous bismuth iodate, $Bi(IO_3)_3$, J. Chem. Crystallogr. 33 (2003) 867.
[92] D. Phanon, A. Mosset, I. Gautier-Luneau, New iodate materials as potential laser matrices. Preparation and characterisation of α-$M(IO_3)_3$ (M = Y, Dy) and β-$M(IO_3)_3$ (M = Y, Ce, Pr, Nd, Eu, Gd, Tb, Dy, Ho, Er). Structural evolution as a function of the $Ln^{3+}$ cationic radius, Solid State Sci. 9 (2007) 496.
[93] Z. Assefa, J. Ling, R.G. Haire, T.E. Albrecht-Schmitt, R.E. Sykora, Syntheses, structures, and vibrational spectroscopy of the two-dimensional iodates $Ln(IO_3)_3$ and $Ln(IO_3)_3(H_2O)$ (Ln=Yb, Lu), J. Solid State Chem. 179 (2006) 3653.
[94] W. Runde, A.C. Bean, L.F. Brodnax, B.L. Scott, Synthesis and characterization of f-element iodate architectures with variable dimensionality, α- and β-$Am(IO_3)_3$, Inorg. Chem. 45 (2006) 2479.
[95] K. Liu, J. Han, J. Huang, Z. Wei, Z. Yang, S. Pan, $SrTi(IO_3)_6{\cdot}2H_2O$ and $SrSn(IO_3)_6$: distinct arrangements of lone pair electrons leading to large birefringences, RSC Adv. 11 (2021) 10309.
[96] D. Phanon, A. Mosset, I. Gautier-Luneau, New materials for infrared non-linear optics. Syntheses, structural characterisations, second harmonic generation and optical transparency of $M(IO_3)_3$ metallic iodates, J. Mater. Chem. 17 (2007) 1123.
[97] F. Mao, J. Liu, J. Hu, H. Wu, From $Ag_2Zr(IO_3)_6$ to $LaZr(IO_3)_5F_2$: A Case of Constructing Wide-band-gap Birefringent Materials through Chemical Cosubstitution, Chem. - An Asian J. 15 (2020) 3487.
[98] B.P. Yang, C.L. Hu, X. Xu, J.G. Mao, New Series of Polar and Nonpolar Platinum Iodates $A_2Pt(IO_3)_6$ (A = $H_3O$, Na, K, Rb, Cs), Inorg. Chem. 55 (2016) 2481.
[99] H. Jo, H.G. Kim, H.R. Byun, J.I. Jang, K.M. Ok, Synthesis, structure, and third-harmonic generation measurements of a mixed alkali metal iodate, $KLi_2(IO_3)_3$, J. Solid State Chem. 282 (2020) 121120.
[100] J. Chen, C.L. Hu, F.F. Mao, X.H. Zhang, B.P. Yang, J.G. Mao, $LiMg(IO_3)_3$: An excellent SHG material designed by single-site aliovalent substitution, Chem. Sci. 10 (2019) 10870.
[101] V.R. Kalinin, V. V Ilyukhin, N. V Belov, On crystal structure of triclinic modification of potassium iodate, Dokl. Akad. Nauk SSSR. 239 (1978) 590.
[102] B.Y.B.W. Lucas, Structure (Neutron) of Room-Temperature Phase III Potassium Iodate, $KIO_3$, Acta Crystallogr. Sect. C. C40 (1984) 1989.
[103] A.L. Hector, S.J. Henderson, W. Levason, M. Webster, Hydrothermal synthesis of rare earth iodates from the corresponding periodates: Structures of $Sc(IO_3)_3$, $Y(IO_3)_3{\cdot}2H_2O$, $La(IO_3)_3{\cdot}1/2H_2O$ and $Lu(IO_3)_3{\cdot}2H_2O$, Zeitschrift Fur Anorg. Und Allg. Chemie. 628 (2002) 198.
[104] D. Tuschel, Practical Group Theory and Raman Spectroscopy, Part I: Normal Vibrational Modes, Spectroscopy. 29 (2014) 14.
[105] J.S. Ogden, Introduction to Molecular Symmetry, Oxford University Press,

Oxford, 2001.
[106] A. Vincent, Molecular Symmetry and Group Theory, 2nd Ed., John Wiley & Sons, Chichester, 2001.
[107] R.L. Carter, Molecular Symmetry and Group Theory, John Wiley & Sons, New York, 1998.
[108] P.H. Walton, Beginning Group Theory for Chemistry, Oxford University Press, Oxford, 1998.
[109] D.C. Harris, M.D. Bertolucci, Symmetry and Spectroscopy: An Introduction to Vibrational and Electronic Spectroscopy, Dover, New York, 1989.
[110] T. Iwasita, A. Rodes, E. Pastor, Vibrational spectroscopy of carbonate adsorbed on Pt(111) and Pt(110) single-crystal electrodes, J. Electroanal. Chem. 383 (1995) 181.
[111] S.T. Shen, Y.T. Yao, T.Y. Wu, Depolarization of Raman lines and structure of chlorate, bromate and iodate ions, Phys. Rev. 51 (1937) 235.
[112] D.T. Cromer, A.C. Larson, The crystal structure of $Ce(IO_3)_4$, Acta Crystallogr. 9 (1956) 1015.
[113] T.C. Kochuthresia, I. Gautier-Luneau, V.K. Vaidyan, M.J. Bushiri, Raman and Ftir Spectral Investigations of Twinned $M(IO_3)_2$ (M = Mn, Ni, Co, and Zn) Crystals, J. Appl. Spectrosc. 82 (2016) 941.
[114] Z.X. Shen, X.B. Wang, S.H. Tang, H.P. Li, F. Zhou, High Pressure Raman Study and Phase Transitions of $KIO_3$ Non-Linear Optical Single Crystals, Rev. High Press. Sci. Technol. 7 (1998) 751.
[115] W.E. Dasent, T.C. Waddington, 491. Iodine-oxygen compounds. Part I. Infrared spectra and structure of iodates, J. Chem. Soc. 491 (1960) 2429.
[116] A.F. Wells, Bifurcated hydrogen bonds, Acta Crystallogr. 2 (1949) 128.
[117] L. Lin, X. Jiang, C. Wu, Z. Lin, Z. Huang, M.G. Humphrey, C. Zhang, $CsZrF_4(IO_3)$: The First Polar Zirconium Iodate with cis-$[ZrO_2F_6]$ Polyhedra Inducing Optimized Balance of Large Band Gap and Second Harmonic Generation, Chem. Mater. 33 (2021) 5555.
[118] M. Ding, H. Yu, Z. Hu, J. Wang, Y. Wu, $Na_7(IO_3)(SO_4)_3$: The first noncentrosymmetric alkaline-metal iodate-sulfate with isolated $[IO_3]$ and $[SO_4]$ units, Chem. Commun. 57 (2021) 9598.
[119] C.K. Aslani, V. V. Klepov, M.A.A. Aslani, H.C. Zur Loye, Hydrothermal Synthesis of New Iodates $Ln_2(IO_3)_3(IO_4)$ (Ln = La, Nd, Pr) Containing the Tetraoxoiodate(V) Anion: Creation of Luminescence Properties by Doping with Eu, Dy, and Tb, Cryst. Growth Des. 21 (2021) 4707.
[120] Q.M. Huang, C.L. Hu, B.P. Yang, Z. Fang, Y. Lin, J. Chen, B.X. Li, J.G. Mao, $[GaF(H_2O)][IO_3F]$: a promising NLO material obtained by anisotropic polycation substitution, Chem. Sci. 12 (2021) 9333.
[121] D. Wang, Z. Bai, L. Liu, L. Hu, Y. Huang, F. Yuan, D. Wei, Z. Lin, L. Zhang, A Polar Material with Facility of Crystal Growth and a Large Second-Harmonic Generation Response, Cryst. Growth Des. 21 (2021) 1734.
[122] F.F. Mao, J.Y. Hu, B.X. Li, H. Wu, $Bi_4O(I_3O_{10})(IO_3)_3(SeO_4)$: Trimeric condensation of $IO_4^{3-}$monomers into the $I_3O_{10}^{5-}$polymeric anion observed in a

three-component mixed-anion NLO material, Dalt. Trans. 49 (2020) 15597.
[123] J. Tauc, Optical properties and electronic structure of amorphous Ge and Si, Mater. Res. Bull. 3 (1968) 37.
[124] J. Chen, C.L. Hu, F.F. Mao, B.P. Yang, X.H. Zhang, J.G. Mao, $REI_5O_{14}$ (RE=Y and Gd): Promising SHG Materials Featuring the Semicircle-Shaped $I_5O_{14}^{3-}$ Polyiodate Anion, Angew. Chem. Int. Ed. 58 (2019) 11666.
[125] M.K.Y. Chan, G. Ceder, Efficient band gap prediction for solids, Phys. Rev. Lett. 105 (2010) 196403.
[126] P. Borlido, J. Schmidt, A.W. Huran, F. Tran, M.A.L. Marques, S. Botti, Exchange-correlation functionals for band gaps of solids: benchmark, reparametrization and machine learning, Npj Comput. Mater. 6 (2020) 96.
[127] F.F. Mao, C.L. Hu, J. Chen, J.G. Mao, A Series of Mixed-Metal Germanium Iodates as Second-Order Nonlinear Optical Materials, Chem. Mater. 30 (2018) 2443.
[128] W. Zhou, N. Umezawa, Band gap engineering of bulk and nanosheet SnO: An insight into the interlayer Sn-Sn lone pair interactions, Phys. Chem. Chem. Phys. 17 (2015) 17816.
[129] A. Liang, L. Shi, R. Turnbull, F.J. Manjón, J. Ibáñez, C. Popescu, M. Jasmin, J. Singh, K. Venkatakrishnan, G. Vaitheeswaran, D. Errandonea, Pressure-induced band-gap energy increase in a metal iodate, Phys. Rev. B. 106 (2022) 235203.
[130] D.N. Nikogosyan, Rarely Used and Archive Crystals, in: Nonlinear Opt. Cryst. A Complet. Surv., Springer, New York, 2005: pp. 319–398.
[131] Y. Huang, X. Meng, P. Gong, L. Yang, Z. Lin, X. Chen, J. Qin, $A_2BiI_5O_{15}$ (A = $K^+$ or $Rb^+$): Two new promising nonlinear optical materials containing $[I_3O_9]^{3-}$ bridging anionic groups, J. Mater. Chem. C. 2 (2014) 4057.
[132] R. Wu, X. Jiang, M. Xia, L. Liu, X. Wang, Z. Lin, C. Chen, $K_8Ce_2I_{18}O_{53}$: A novel potassium cerium(IV) iodate with enhanced visible light driven photocatalytic activity resulting from polar zero dimensional $[Ce(IO_3)_8]^{4-}$ units, Dalt. Trans. 46 (2017) 4170.
[133] L. Xiao, F. You, P. Gong, Z. Hu, Z. Lin, Synthesis and structure of a new mixed metal iodate $Ba_3Ga_2(IO_3)_{12}$, CrystEngComm. 21 (2019) 4981.
[134] F.-F. Mao, C.-L. Hu, X. Xu, D. Yan, B.-P. Yang, J.-G. Mao, $Bi(IO_3)F_2$: The First Metal Iodate Fluoride with a Very Strong Second Harmonic Generation Effect , Angew. Chem. 129 (2017) 2183.
[135] H. Liu, Q. Wu, X. Jiang, Z. Lin, X. Meng, X. Chen, J. Qin, $ABi_2(IO_3)_2F_5$ (A=K, Rb, and Cs): A Combination of Halide and Oxide Anionic Units To Create a Large Second-Harmonic Generation Response with a Wide Bandgap, Angew. Chem. Int. Ed. 56 (2017) 9492.
[136] B.P. Yang, C.L. Hu, X. Xu, C. Huang, J.G. Mao, $Zn_2(VO_4)(IO_3)$: A novel polar zinc(II) vanadium(V) iodate with a large SHG response, Inorg. Chem. 52 (2013) 5378.
[137] H.H.H. Osman, F.J. Manjón, Metavalent bonding in chalcogenides: DFT-chemical pressure approach, Phys. Chem. Chem. Phys. 24 (2022) 9936.

[138] A. Jayaraman, D.L. Wood, R.G. Maines, High-pressure Raman study of the vibrational modes in $AlPO_4$ and $SiO_2$ (α-quartz), Phys. Rev. B. 35 (1987) 8316.
[139] F.J. Manjón, J. Serrano, I. Loa, K. Syassen, C.T. Lin, M. Cardona, Effect of pressure on the Raman anomaly of zinc-blende CuBr and Raman spectra of high-pressure phases, Phys. Rev. B. 64 (2001) 643011.
[140] H.A. Jahn and E. Teller, Stability of Polyatomic Molecules in Degenerate Electronic States I-Orbital Degeneracy, Proc. R. Soc. A. 161 (1937) 220.
[141] Y. Tanabe, S. Sugano, On the Absorption Spectra of Complex Ions. I, J. Phys. Soc. Japan. 9 (1954) 753.
[142] P.G.S. Roche, A.L. Fathima, P. Selvarajan, CRYSTAL NUCLEATION, CRYSTAL GROWTH, SPECTRAL AND SHG STUDIES OF POTASSIUM IODATE CRYSTAL, Int. J. Adv. Trends Eng. Technol. 3 (2018) 70.
[143] D. Errandonea, R.S. Kumar, L. Gracia, A. Beltrán, S.N. Achary, A.K. Tyagi, Experimental and theoretical investigation of $ThGeO_4$ at high pressure, Phys. Rev. B. 80 (2009) 1.
[144] F.J. Manjón, J.Á. Sans, P. Rodríguez-hernández, A. Muñoz, Combined experimental and theoretical studies: Lattice-dynamical studies at high pressures with the help of ab initio calculations, Minerals. 11 (2021) 1283.
[145] D. Díaz-Anichtchenko, D. Errandonea, Pressure-induced phase transitions and electronic properties of $Cd_2V_2O_7$, RSC Adv. 12 (2022) 14827.
[146] R. Debord, H. Euchner, V. Pischedda, M. Hanfland, A. San-Miguel, P. Mélinon, S. Pailhès, D. Machon, Isostructural phase transition by point defect reorganization in the binary type-I clathrate $Ba_{7.5}Si_{45}$, Acta Mater. 210 (2021) 116824.
[147] K. Ståhl, M. Szafranski, A Single-Crystal Neutron Diffraction Study of $HIO_3$ at 295 and 30 K and of $DIO_3$ at 295 K, Acta Chem. Scand. 46 (1992) 1146.
[148] M.B. Taouti, A. Gacemi, D. Benbertal, I. Gautier-Luneau, Crystal structure of lanthanum triiodate iodic acid, $La(IO_3)_3(HIO_3)$, Zeitschrift Fur Krist. - New Cryst. Struct. 223 (2008) 179.
[149] E.E. Oyeka, M.J. Winiarski, A. Błachowski, K.M. Taddei, A. Scheie, T.T. Tran, Potential Skyrmion Host $Fe(IO_3)_3$: Connecting Stereoactive Lone-Pair Electron Effects to the Dzyaloshinskii-Moriya Interaction, Chem. Mater. 33 (2021) 4661.
[150] D. Phanon, A. Mosset, I. Gautier-Luneau, New materials for infrared non-linear optics. Syntheses, structural characterisations, second harmonic generation and optical transparency of $M(IO_3)_3$ metallic iodates, J. Mater. Chem. 17 (2007) 1123.
[151] J.-M. Crettez, E. Coquet, B. Michaux, J. Pannetier, J. Bouillot, P. Orlans, A. Nonat, J.-C. Mutin, Lithium Iodate: Phase Transitions Revisited, Phys. B+C. 144 (1987) 277.
[152] F.D. Murnaghan, The Compressibility of Media under Extreme Pressures, Proc. Natl. Acad. Sci. 30 (1944) 244.
[153] F. Birch, Finite elastic strain of cubic crystals, Phys. Rev. 71 (1947) 809.
[154] H. Bach, H. Küppers, Cadmium Diiodate, Acta Crystallogr. Sect. B. B34

(1978) 263.
[155] N.W. Alcock, The crystal structure of α-rubidium iodate, Acta Crystallogr. Sect. B. B28 (1972) 2783.
[156] A.C. Larson, D.T. Cromer, The crystal structure of $Zr(IO_3)_4$, Acta Crystallogr. 14 (1961) 128.
[157] Y. Cheng, S. Wahl, M. Wuttig, Metavalent Bonding in Solids: Characteristic Representatives, Their Properties, and Design Options, Phys. Status Solidi - Rapid Res. Lett. (2020).
[158] L.I. Karaouzène, T. Ouahrani, Á. Morales-García, D. Errandonea, Theoretical calculations of the effect of nitrogen substitution on the structural, vibrational, and electronic properties of wolframite-type $ScTaO_4$ at ambient conditions, Dalt. Trans. 51 (2022) 3642.
[159] E.E. Oyeka, M.J. Winiarski, M. Sorolla, K.M. Taddei, A. Scheie, T.T. Tran, Spin and Orbital Effects on Asymmetric Exchange Interaction in Polar Magnets: $M(IO_3)_2$ (M = Cu and Mn), Inorg. Chem. 60 (2021) 16544.
[160] D.P. Kozlenko, N.T. Dang, S.E. Kichanov, L.T.P. Thao, A. V. Rutkaukas, E. V. Lukin, B.N. Savenko, N. Tran, D.T. Khan, L. V. Truong-Son, L.H. Khiem, B.W. Lee, T.L. Phan, N.L. Phan, N. Truong-Tho, N.N. Hieu, T.A. Tran, M.H. Phan, High pressure enhanced magnetic ordering and magnetostructural coupling in the geometrically frustrated spinel $Mn_3O_4$, Phys. Rev. B. 105 (2022) 094430.
[161] X. Shi, B.D. Beake, T.W. Liskiewicz, J. Chen, Z. Sun, Failure mechanism and protective role of ultrathin ta-C films on Si (100) during cyclic nano-impact, Surf. Coatings Technol. 364 (2019) 32.
[162] L. Cao, M. Luo, C. Lin, Y. Zhou, D. Zhao, T. Yan, N. Ye, From centrosymmetric to noncentrosymmetric: Intriguing structure evolution in $d^{10}$-transition metal iodate fluorides, Chem. Commun. 56 (2020) 10734.
[163] M.M. Abdel Kader, F. El-Kabbany, H.M. Naguib, W.M. Gamal, Charge transport mechanism and low temperature phase transitions in $KIO_3$, J. Phys. Conf. Ser. 423 (2013) 012036.
[164] A.K. Sagotra, D. Errandonea, C. Cazorla, Mechanocaloric effects in superionic thin films from atomistic simulations, Nat. Commun. 8 (2017) 963.